\documentclass[aps,prd,twocolumn,superscriptaddress,nofootinbib,showpacs]{revtex4-2}
\usepackage{amsmath}
\usepackage{amssymb}
\usepackage{graphicx}
\usepackage{xcolor,color}
\usepackage{url}
\usepackage{bm}
\usepackage{mathrsfs}
\usepackage[utf8]{inputenc}
\usepackage{hyperref}
\usepackage{enumerate}
\usepackage{amsthm}
\usepackage{bbm}
\usepackage[normalem]{ulem}
\usepackage{upgreek}
\usepackage{tensor}
\usepackage{siunitx}
\usepackage{orcidlink}
\usepackage{float}
\usepackage{colortbl}
\usepackage{tikz}
\usepackage[caption=false]{subfig}

\definecolor{colour1}{HTML}{0571b0} 
\definecolor{colour2}{HTML}{92c5de} 
\definecolor{colour3}{HTML}{f4a582} 
\definecolor{colour4}{HTML}{ca0020} 
\definecolor{colour5}{HTML}{fe4a49} 
\definecolor{colour6}{HTML}{2d3092} 

\definecolor{colour99}{HTML}{0000FF} 

\hypersetup{colorlinks=true, linkcolor=colour99, citecolor=colour99,
filecolor=colour99, urlcolor=colour99}

\theoremstyle{definition}

\usepackage{color,xcolor}
\usepackage{ulem}
\usepackage{float}
\usepackage{cleveref}
\usepackage{multirow}
\usepackage{booktabs}
\usepackage{booktabs}
\usepackage{threeparttable}

\usepackage[export]{adjustbox}
\usepackage{times}
\usepackage{commath}
\usepackage{lipsum}

\def\HCa{{\bf{H}}_{\rm C1} }

\def\HCb{{\bf{H}}_{\rm C2} }

\crefname{equation}{Eq.}{Eq.}
\Crefname{equation}{Eqs.}{Eqs.}

\crefname{section}{Sec.}{Sec.}
\Crefname{section}{Secs.}{Secs.}

\crefname{table}{Table}{Table}
\Crefname{table}{Tables}{Tables}
\crefname{figure}{Figure}{Figure}
\Crefname{figure}{Figures.}{Figures.}

\crefname{appendix}{Appendix}{Appendix}

\usepackage{threeparttable}
\graphicspath{{Figures/}}

\begin{document}

\title{Efficient and stable computation of gravitational-wave fluxes from generic Kerr orbits via a unified Heun-function framework}

\author{Changkai Chen\,\orcidlink{0000-0002-4023-0682}}
\affiliation{School of Physics and Astronomy, Beijing Normal University, Beijing, 100875, China}
\affiliation{Department of Physics, Key Laboratory of Low Dimensional Quantum Structures and Quantum Control of Ministry of Education, Synergetic Innovation Center for Quantum Effects and Applications, Hunan Normal University, Changsha, 410081, Hunan, China}
\affiliation{Institute for Frontiers in Astronomy and Astrophysics, Beijing Normal University, Beijing, 102206, China}
\author{Zhoujian Cao\,\orcidlink{0000-0002-1932-7295} } \email[Contact author: ]{zjcao@bnu.edu.cn}
\affiliation{School of Physics and Astronomy, Beijing Normal University, Beijing, 100875, China}
\affiliation{Institute for Frontiers in Astronomy and Astrophysics, Beijing Normal University, Beijing, 102206, China}
\affiliation{School of Fundamental Physics and Mathematical Sciences, Hangzhou Institute for Advanced Study, UCAS, Hangzhou, 310024, China}

\author{Jiliang Jing\,\orcidlink{0000-0002-2803-7900} }\email[Contact author: ]{jljing@hunnu.edu.cn}
\affiliation{Department of Physics, Key Laboratory of Low Dimensional Quantum Structures and Quantum Control of Ministry of Education, Synergetic Innovation Center for Quantum Effects and Applications, Hunan Normal University, Changsha, 410081, Hunan, China}
\date{\today}
\begin{abstract}

Modeling extreme-mass-ratio inspirals hinges on the accurate and efficient computation of gravitational-wave fluxes from generic Kerr orbits.
Conventional frequency-domain techniques are often limited by costly auxiliary parameter searches and numerical instabilities in the strong-field or high-frequency regimes. We address these challenges by reformulating both the angular and radial Teukolsky equations in terms of confluent Heun functions.
Employing a hybrid analytic continuation algorithm to compute the connection coefficients eliminates the dependence on auxiliary parameters, directly yielding globally convergent solutions and scattering amplitudes. To resolve the highly oscillatory source integrands for generic orbits, we implement an adaptive bi-power mapping quadrature.
Comprehensive benchmarks under standard double-precision arithmetic demonstrate that, for the total radiative flux summed over 168 low-order modes, our method achieves relative errors of order $10^{-11}$, with computational costs typically reduced by factors of 3--13 compared to the state-of-the-art \texttt{GeneralizedSasakiNakamura.jl} and \texttt{pybhpt} packages.
Notably, for highly oscillatory high-order modes, our framework achieves a speedup of up to 60 times compared to specialized oscillatory integrators like \texttt{GeneralizedSasakiNakamura.jl}.
These demonstrated gains in precision and efficiency establish the framework as a robust tool for strong-field perturbation theory, providing the numerical foundation for high-order self-force calculations and rapid, high-precision waveform generation.

\end{abstract}

\maketitle
\newpage

\section{Introduction}

Over the past decade, ground-based interferometric detectors, operating within the LIGO-Virgo-KAGRA network~\cite{LIGOScientific:2016wkq,LIGOScientific:2016dsl,
LIGOScientific:2018mvr,LIGOScientific:2020ibl,KAGRA:2021duu}, have ushered in the era of gravitational-wave (GW) astronomy and substantially broadened our knowledge of compact astrophysical objects.
The next frontier lies in the millihertz band, where space-based observatories---LISA~\cite{LISA:2017pwj,LISA:2024hlh}, TianQin~\cite{TianQin:2015yph,TianQin:2020hid,Li:2024rnk}, and Taiji~\cite{Ruan:2018tsw,Gong:2021gvw}---are expected to reveal an entirely new class of sources. Chief among these are  extreme-mass-ratio inspirals (EMRIs, $q \in [10^{4}, 10^{6}]$) and intermediate-mass-ratio inspirals (IMRIs, $q \in [10^{2}, 10^{4}]$)~\cite{Amaro-Seoane:2012lgq,LISA:2022yao}, both of which are also targeted by proposed decihertz observatories such as DECIGO~\cite{Kawamura:2020pcg} and next-generation ground-based facilities including the Einstein Telescope~\cite{ET:2025xjr,Reitze:2019iox,Punturo:2010zz}.
These asymmetric-mass systems offer substantial scientific returns.
High-precision measurements of binary parameters---achievable at sub-percent levels for EMRIs~\cite{Babak:2017tow,Berry:2019wgg}---would enable strong-field tests of general relativity~\cite{Barack:2003fp,Gair:2012nm,Maselli:2021men,Speri:2024qak} and allow the spacetime geometry of the central massive black hole (MBH) to be mapped with unprecedented fidelity~\cite{Babak:2017tow}.
Beyond tests of fundamental physics, such observations would shed light on the accretion environments surrounding MBHs~\cite{Bonga:2019ycj,Speri:2022upm,Khalvati:2024tzz}, constrain the MBH mass function~\cite{Gair:2010yu,Chapman-Bird:2022tvu}, and probe the stellar dynamics in galactic nuclei~\cite{Amaro-Seoane:2007osp}.
IMRI and EMRI signals may further serve as independent standard sirens for measuring the cosmic expansion history~\cite{MacLeod:2007jd,Laghi:2021pqk,LISACosmologyWorkingGroup:2022jok}.

Realizing this scientific potential hinges critically on the availability of accurate theoretical waveform templates.
Since IMRI and EMRI systems accumulate $\sim 10^3$--$10^6$ orbital cycles within the detector band, sub-radian phase coherence must be maintained throughout, without which parameter estimation would be severely biased and search sensitivity degraded~\cite{Burke:2023lno,wanghe}.
The gravitational self-force framework, grounded in black hole perturbation theory~\cite{Barack:2009ux,Pound:2019lzj,Warburton:2021kwk,Wardell:2021fyy}, addresses these demands by systematically expanding the Einstein field equations in powers of the small mass ratio around the background spacetime of the primary; see Refs.~\cite{Poisson:2011nh,Barack:2018yvs,Pound:2021qin} for comprehensive reviews.
Within this framework, the GW-driven evolution of the binary's orbital constants is governed at leading (adiabatic) order by flux-balance relations~\cite{Mino:2003yg,Drasco:2005is,Sago:2005fn,Isoyama:2018sib,Grant:2024ivt}, which are evaluated through the separable Teukolsky equation~\cite{Teukolsky:1972my,Teukolsky:1973ha,Press:1973zz,Teukolsky:1974yv} for the Weyl scalars $\psi_0$ and $\psi_4$~\cite{Wald:1973wwa}.
This master equation naturally decomposes into coupled ordinary differential equations for the angular and radial coordinates---commonly referred to as the angular Teukolsky equation (ATE) and radial Teukolsky equation (RTE).
The multiscale expansion strategy~\cite{Hinderer:2008dm,Miller:2020bft,Pound:2021qin,Mathews:2025txc} then separates the problem into a computationally expensive offline data-generation step and a rapid online interpolation step, enabling efficient waveform construction.
Substantial efforts have been devoted to computing these GW fluxes for increasingly generic orbital geometries. On the numerical side, Hughes pioneered a framework that employs a spectral expansion for the ATE alongside direct integration of the RTE to obtain fluxes and inspiral trajectories for spherical orbits in Kerr spacetime~\cite{Hughes:1999bq,Hughes:2001jr}, a program later extended to fully generic eccentric and inclined orbits~\cite{Drasco:2003ky,Drasco:2005kz,Hughes:2021exa}.
In parallel, the Mano--Suzuki--Takasugi (MST) method~\cite{Mano:1996mf,Mano:1996vt} provides a semi-analytic framework by expanding RTE homogeneous solutions into convergent hypergeometric series. Fujita \textit{\textit{et al.}}~\cite{Fujita:2009us} later refined this formalism with efficient numerical algorithms for computing the renormalized angular momentum. These developments enable convergent RTE solutions and yield high-precision GW flux evaluations across a broad range of orbital parameters.
Within the MST framework, high-order post-Newtonian (PN) flux formulae have been systematically derived~\cite{Fujita:2012cm,Sago:2015rpa,Munna:2020iju,Sago:2024mgh,Castillo:2024isq,Sago:2026gxb}, providing analytical expressions that are particularly efficient in the weak-field regime.

Despite these advances, each existing method carries inherent limitations.
Traditional direct numerical integration of the Teukolsky equation~\cite{Hughes:1999bq,Drasco:2005kz}, while robust, encounters computational bottlenecks for orbits with large eccentricities or in the deep strong-field regime, where high-frequency modes proliferate and the radial potential demands increasingly fine resolution.
Similarly, PN expansions, by construction, degrade in accuracy as the binary approaches the innermost stable circular orbit.
Covering the full IMRI and EMRI parameter space therefore demands methods that are simultaneously accurate across both the weak- and strong-field extremes and computationally tractable for large-scale data generation.
The resulting flux data underpin waveform generation packages such as the \textsc{FastEMRIWaveforms} framework~\cite{Chua:2020stf,Katz:2021yft,Chapman-Bird:2025xtd}, which has rendered high-fidelity Bayesian parameter estimation of EMRI signals feasible on timescales of hours~\cite{Speri:2024qak,wanghe}.

Two recently developed codes have significantly advanced the state of the art in Teukolsky-based flux computation, addressing these computational challenges through distinct methodological innovations.
Lo revamped Hughes's numerical integration framework to the generalized Sasaki--Nakamura (GSN) formalism, implemented in the Julia package \texttt{GeneralizedSasakiNakamura.jl}~\cite{GeneralizedSasakiNakamura}.
This work establishes homogeneous solutions of the RTE~\cite{Lo:2023fvv}, constructs $s=-2$ inhomogeneous solutions for fluxes at infinity~\cite{Yin:2025kls}, and finally derives the nonsingular source term for $s=+2$ to enable horizon flux calculations without regularization~\cite{Lo:2025lpo}.
Independently, the \texttt{pybhpt} code~\cite{Nasipak:2022xjh,Nasipak:2023kuf,pybhpt} adopts a hybrid semi-analytical, semi-numerical strategy: MST series for the RTE and Frobenius expansions of the confluent Heun (HeunC) function provide accurate boundary data, while adaptive Runge--Kutta integration and spectral source methods handle the radial evolution and mode projection.
Its double-precision arithmetic (machine epsilon $\sim 10^{-16}$) delivers substantially higher computational speed and memory efficiency compared to other numerical schemes~\cite{BHPToolkit}.
Notably, both packages incorporate the highly efficient spectral eigenvalue method for the ATE pioneered by Cook and Zalutskiy~\cite{Cook:2014cta}, which ensures rapid and stable convergence for spin-weighted spheroidal harmonics (SWSHs).

The HeunC function has emerged as a particularly powerful and natural framework in black hole perturbation theory~\cite{slavyanov2000unified}.
The mathematical parameters of the HeunC function exhibit a precise correspondence with the global causal structure of the black hole spacetime---its horizons, null infinity, spatial infinity, and timelike infinity. This correspondence reveals that the confluent Heun equation (CHE) does not merely govern the evolution of perturbation fields but intrinsically encodes the global topology of the background geometry~\cite{Minucci:2024qrn}.
London~\cite{London:2023aeo,London:2023idh} established the orthogonality of solutions to the RTE, a result that enables quantitative predictions of the excitation amplitudes of individual GW modes and furnishes a precise diagnostic for probing how astrophysical environments---such as accretion disks or dark-matter halos---leave imprints on the quasinormal mode (QNM) spectrum.
A connection formula for the HeunC function between the regular singularity $x=0$ and the irregular singularity $x=\infty$ was recently derived within the MST framework~\cite{Chen:2023ese,Chen:2023lsa}, which provides a high-precision pathway to compute radiative fluxes for arbitrary perturbation fields.

Beyond the MST framework, several independent strategies for evaluating connection coefficients of Heun class functions have been developed.
Bonelli \textit{et al.} derived connection formulas for Heun class functions using the Liouville conformal field theory~\cite{Bonelli:2022ten}.
While powerful, this method shares a structural feature with the MST method: both require the determination of an auxiliary parameter, namely the renormalized angular momentum $\nu$ for the MST method and the parameter $\mathfrak{a}(m_i, u, L)$ for the Bonelli \textit{et al.} formulation. These two parameters are related by $\mathfrak{a} = \tfrac{1}{2} - \nu$~\cite{Bautista:2023sdf}.
Since both strategies are grounded in power-series expansions of special functions, a systematic comparison has been carried out in Ref.~\cite{Bautista:2023sdf}.
Furthermore, for QNM calculations using these two connection formulas, the QNM results derived from the Nekrasov--Shatashvili technology are documented in Refs.~\cite{Aminov:2020yma,Bonelli:2021uvf}, whereas those obtained through the MST method are reported in Ref.~\cite{Casals:2019vdb}.
A conceptually related but technically distinct alternative is the isomonodromic method~\cite{ablowitz1981solitons,its2006isomonodromic}, which Cavalcante \textit{et al.} applied to QNMs of massive scalar perturbations in near-extremal Kerr black holes~\cite{Cavalcante:2024swt,Cavalcante:2024kmy}.
This method, however, faces the same inherent difficulty as its counterparts: the calculation of auxiliary parameters playing a role analogous to $\nu$ remains computationally expensive, limiting the overall efficiency of the scheme.
Despite the high accuracy achieved by these methods, the calculation of connection coefficients via such series-based approaches constitutes the principal computational bottleneck and constrains the overall efficiency of flux calculations.
A more computationally tractable route was proposed by Motygin~\cite{Motygin2018}, who devised a numerical algorithm for the HeunC function in which connection coefficients are extracted by truncating local Frobenius series at a matching point between two singular points and solving the resulting linear system through matrix inversion.
Because no auxiliary parameters need to be determined, this strategy achieves high numerical precision at a fraction of the computational cost.
McMaken \textit{et al.} subsequently demonstrated its utility in a black hole context by applying Motygin's scheme to the Hawking radiation spectrum of Reissner--Nordstr\"{o}m black holes~\cite{McMaken:2023tft}.
Although the basic analytic continuation follows the scheme proposed by Motygin~\cite{Motygin2018}, we have enhanced the algorithm to handle the high-order centrifugal barriers inherent to Kerr perturbations. Specifically, the transition from fixed to adaptive matching is a critical modification required to maintain machine-precision accuracy for EMRI waveforms.

In this work, we reformulate our previous method~\cite{Chen:2023ese,Chen:2023lsa} by employing HeunC solutions for both the ATE and RTE, and utilizing Motygin's efficient numerical algorithm to calculate scattering amplitudes.
The resulting HeunC-based numerical framework delivers excellent numerical precision and superior computational efficiency under standard double-precision arithmetic.
Crucially, its performance advantages become increasingly prominent for high-order harmonic modes and extreme orbital parameter configurations, where conventional numerical schemes typically suffer from degraded accuracy and heavy computational overhead.

The remainder of this paper is organized as follows.
Section \ref{sec:GWFlux} introduces the novel HeunC framework for computing GW radiative fluxes from particles on generic orbits around a Kerr black hole. The correspondence between the Teukolsky equations and the confluent Heun equations is established in \cref{subsec:teukolsky_heun}. HeunC solutions for the angular and radial equations are detailed in \cref{subsec:heunc_ATE} and \cref{subsec:heunc_RTE}, respectively. The methodology for calculating radiative fluxes is described in \cref{subsec:fluxes}. Section \ref{sec:comparisons} presents comprehensive benchmarks against established methods, covering spin-weighted spheroidal harmonics (\cref{subsec:spheroidal}), inspiral fluxes (\cref{subsec:inspirals}), and QNM tests (\cref{subsec:ringdown}). Finally, we summarize our conclusions and outline future prospects in \cref{sec:conclusion}.
In this paper, we adopt geometrized units $c=G=1$, and $M=1$.


\section{GW Radiative Fluxes via the Teukolsky equation}\label{sec:GWFlux}

In this section, we reformulate our previous framework~\cite{Chen:2023ese,Chen:2023lsa} to enable accurate and efficient computation of GW radiative fluxes in Kerr spacetime.
This reformulation applies HeunC functions to both the angular and radial sectors of the Teukolsky equation.
By adopting Motygin's numerical algorithm~\cite{Motygin2018} for evaluating connection coefficients, we construct a streamlined Green's function formalism.
This unified treatment not only simplifies the mathematical structure but also significantly enhances computational throughput and numerical stability, particularly for high-order harmonic modes and extreme orbital configurations.

\subsection{Teukolsky and HeunC equations}\label{subsec:teukolsky_heun}
We begin by establishing the correspondence between the Teukolsky equation and the CHE.
Using the Newman--Penrose formalism, Teukolsky~\cite{Teukolsky:1973ha} showed that several types of perturbation fields in Kerr spacetime are governed by a single master equation
\begin{equation}\label{eq:Teuk-Master-Eq}
  {\cal O}\psi  = 4\pi \Sigma T,
\end{equation}
where $T$ is the source term and $\psi$ has spin weight $s$ (e.g., $s=0, \pm 1, \pm 2$ for scalar, electromagnetic, and gravitational fields, respectively). In vacuum ($T=0$), \cref{eq:Teuk-Master-Eq} separates into radial and angular parts via the ansatz
\begin{equation}\label{eq:separationEq-Teuko}
  \psi(t,r,\theta,\phi) = \mathrm{e}^{-i\omega t} \mathrm{e}^{im\phi}S_{\ell m}({\Theta }){R_{\ell m}}(r).
\end{equation}

Introducing the variable $\Theta = \cos\theta$, the angular function $S_{\ell m}(\Theta)$ satisfies the ATE,
\begin{equation}\label{eq:GFoATE}
  \left[ \frac{\rm d}{{\rm d}\Theta}\left( \nabla\frac{\rm d}{{\rm d}\Theta} \right) + U(\Theta) \right]S_{\ell m} = 0,
\end{equation}
where $\nabla = (1 - \Theta)(1 + \Theta)$,
and the potential $U(\Theta)$ is given by
\begin{equation}\label{eq:U_ATE_Kerr}
U(\Theta) = a^2\omega^2\Theta^2 - \frac{(m + s\Theta)^2}{1 - \Theta^2} - 2a\omega s\Theta + s + {}_sA_{\ell m}.
\end{equation}

Correspondingly, the radial function $R_{\ell m}(r)$ satisfies the inhomogeneous RTE,
\begin{equation}\label{eq:GFoRTE}
 \left[ \Delta^{-s+1}\frac{\rm d}{{\rm d}r}\left( \Delta^{s+1}\frac{\rm d}{{\rm d}r} \right) + V(r) \right]R_{\ell m} = T_{\ell m},
\end{equation}
where $\Delta = (r - r_-)(r - r_+)$ and ${r_ \pm} = M \pm \sqrt {{M^2} - {a^2}}$. The potential $V(r)$ for Kerr black holes is given by
\begin{equation}\label{eq:V_RTE_Kerr}
V(r) = K^2 - isK\Delta' + (2isK' - \hat{\lambda})\Delta,
\end{equation}
where $K = (r^2 + a^2)\omega - am$ and $\hat{\lambda} = {}_sA_{\ell m}(a\omega) + a^2\omega^2 - 2am\omega$.

Both the ATE \eqref{eq:GFoATE} and the homogeneous RTE (obtained by setting $T_{\ell m}=0$) can be systematically reduced to the standard form of the CHE. This reduction proceeds in two steps. First, a Möbius (linear fractional) transformation is applied to the independent variable,
\begin{equation}\label{eq:mtran0}
x = \frac{\hat{a}y +\hat{b}}{\hat{c}y+\hat{d}},
\end{equation}
where $y=\Theta$ for the ATE and $y=r$ for the homogeneous RTE, and $\{\hat{a}, \hat{b}, \hat{c}, \hat{d}\}$ are constants chosen to map the physical singularities of the Teukolsky equations to the canonical singular points of the CHE at $x=0$, $x=1$, and $x=\infty$. Second, an S-homotopic transformation is employed to absorb the singular behavior of the dependent variable, yielding exact solutions of the form
\begin{align}
  S_{\ell m}(\Theta) &= S_\theta  (x)\,\mathbf{H}(x), \label{eq:Sm-Sol}\\
  {R_{\ell m}}(r) &= S_r (x)\,\mathbf{H}(x), \label{eq:Rm-Sol}
\end{align}
where $\mathbf{H}(x)$ denotes the HeunC function. The S-homotopic prefactors in the angular and radial directions are explicitly given by
\begin{align}
S_\theta  (x) &= (-x)^{\tfrac{1}{2}\beta}(1-x)^{\tfrac{1}{2}\gamma}\mathrm{e}^{\tfrac{1}{2}\alpha x}, \label{eq:S-homotopic-theta-d2}\\
S_{r} (x) &= (-x)^{\tfrac{1}{2}(\beta-s)}(1-x)^{\tfrac{1}{2}(\gamma-s)}\mathrm{e}^{\tfrac{1}{2}\alpha x}. \label{eq:S-homotopic-r}
\end{align}

The CHE is obtained from the general Heun equation via a confluence of two regular singularities~\cite{Ronveaux:1995,slavyanov2000unified,olver2010nist}. Its standard form reads
\begin{equation}\label{eq:HeunC-Eq}
 {\mathbf{H}}''(x) +\left( \frac{\tilde{\gamma}}{x} + \frac{\tilde{\delta}}{x - 1} + \tilde{\epsilon} \right){\mathbf{H}}'(x) + \frac{\tilde{\alpha}x - \tilde{q}}{x(x - 1)}{\mathbf{H}}(x) = 0.
\end{equation}
The solution $\mathbf{H}(x)$ admits two widely used parametrizations. The first, $\mathbf{H}(x) = \mathrm{HeunC}(\alpha, \beta, \gamma, \delta, \eta; x)$, emphasizes the intrinsic symmetries of the CHE.
Specifically, the parameters $(\alpha, \beta, \gamma, \delta, \eta)$ encode transformation properties under sign reversals and parameter exchanges, as detailed in Appendix \ref{app:Symmetries_HC_TE}.
This parametrization is particularly advantageous for constructing local Frobenius solutions and analyzing their behavior near the singular points.
The second, $\mathbf{H}(x) = \mathrm{HeunC}(\tilde{q}, \tilde{\alpha}, \tilde{\gamma}, \tilde{\delta}, \tilde{\epsilon}; x)$, aligns with the conventions of most computational software packages and numerical libraries, facilitating direct implementation of Motygin's algorithm.
The correspondence between these two parameter sets is given by
\begin{subequations}\label{eq:HeunC_Para_Relat}
\begin{align}
  \tilde{q} &= \frac{1}{2}\alpha \beta  - \frac{1}{2}\beta \gamma  + \frac{1}{2}\alpha  - \frac{1}{2}\beta  - \eta  - \frac{1}{2}\gamma ,\\
  \tilde{\alpha} &= \frac{1}{2}\alpha \beta  + \frac{1}{2}\alpha \gamma  + \alpha  + \delta ,\\
  \tilde{\delta} &= \gamma  + 1,\\
  \tilde{\epsilon} &= \alpha ,\\
  \tilde{\gamma} &= \beta  + 1.
\end{align}
\end{subequations}
Based on these parametrizations, we define two fundamental sets of linearly independent solutions:
\begin{enumerate}
  \item {Local basis at the regular singularity $x=0$:}

Exploiting the $\beta \leftrightarrow -\beta$ symmetry inherent to the CHE in the first parametrization, we define two linearly independent Frobenius solutions, denoted $\HCa(x)$ and $\HCb(x)$:
\begin{subequations}\label{eq:HCSol-x0}
  \begin{align}
\HCa(x) &= \mathrm{HeunC}(\alpha ,\beta ,\gamma ,\delta ,\eta ;x), \label{eq:HCSol-x0-HCa}\\
\HCb(x) &=  x^{-\beta}\mathrm{HeunC}(\alpha , - \beta ,\gamma ,\delta ,\eta ;x).\label{eq:HCSol-x0-HCb}
  \end{align}
\end{subequations}
A crucial property for constructing physical solutions is the behavior of the HeunC function near the singular point $x=0$.
The asymptotic normalizations for $\HCa(x)$ and $\HCb(x)$ are prescribed as:
\begin{subequations}\label{eq:HCA-HCB-x0}
  \begin{align}
\lim_{x \to 0}& \HCa(x) = 1, \label{eq:Asy-HCA-x0} \\
\lim_{x \to 0}& x^{\beta}\HCb(x) = 1.
  \end{align}
\end{subequations}
This normalization \eqref{eq:Asy-HCA-x0} ensures that the singular behavior of the full angular and radial functions in \cref{eq:Sm-Sol,eq:Rm-Sol} is entirely captured by the S-homotopic prefactors $S_\theta$ and $S_r$, respectively. Consequently, physical boundary conditions at the event horizon or poles can be imposed directly on the prefactors without ambiguity.
  \item {Asymptotic basis at the irregular singularity $x=\infty$:}

  For the behavior at infinity, we employ the second parametrization. The two independent solutions, denoted $\HCa^\infty(x)$ and $\HCb^\infty(x)$, are constructed as follows:
\begin{subequations}
  \begin{align}
&\HCa^{\infty }(x)={\rm{HeunC}}_{A,\infty }({\tilde q},{\tilde \alpha},{\tilde \gamma},{\tilde \delta},{\tilde \epsilon};x),\label{eq:HCSol-xInf-a} \\
&\HCb^{\infty }(x)={\rm{HeunC}}_{B,\infty }({\tilde q},{\tilde \alpha},{\tilde \gamma},{\tilde \delta},{\tilde \epsilon};x), \\
&= {\mathrm{e}^{ - {\tilde \epsilon}x}}{\rm{HeunC}}_{A,\infty }({\tilde q} - {\tilde \gamma}{\tilde \epsilon},{\tilde \alpha} - {\tilde \epsilon}({\tilde \gamma} + {\tilde \delta}),{\tilde \gamma},{\tilde \delta}, - {\tilde \epsilon};x) \nonumber,\label{eq:HCSol-xInf-b}
  \end{align}
\end{subequations}
  where ${\rm{HeunC}}_{A,\infty}$ is evaluated numerically using Motygin's algorithm~\cite{Motygin2018}. The asymptotic normalizations for these solutions are prescribed as:
\begin{subequations}\label{eq:HCA-HCB-inf}
  \begin{align}
& {\lim}_{|x| \to \infty } {( - x)^{\frac{{{\tilde \alpha}}}{{{\tilde \epsilon}}}}}\HCa^{\infty }(x)= 1, \\
 & {\lim}_{|x| \to \infty }  {\mathrm{e}^{\tilde \epsilon x}}{( - x)^{\tilde \gamma  + \tilde \delta  - \frac{{\tilde \alpha }}{{\tilde \epsilon }}}}\HCb^{\infty }(x)= 1.
 \end{align}
\end{subequations}
\end{enumerate}
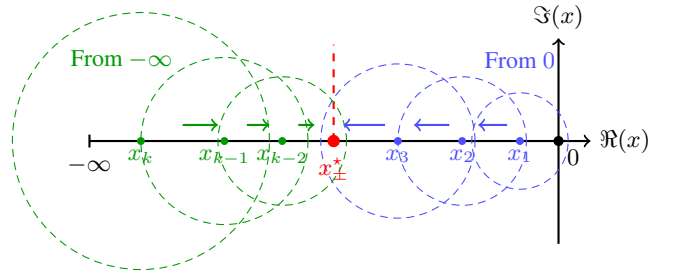
\begin{figure}[htbp]
    \centering
    \begin{tikzpicture}[scale=0.85, every node/.style={font=\small}]
    \draw[->, thick] (-7.3,0) -- (0.5,0) node[right] { $\Re(x)$};
    \draw[->, thick] (0,-1.6) -- (0,1.6) node[above] { $\Im(x)$};
    \filldraw[black] (0,0) circle (2pt) node[below right] {$0$};

    \coordinate (xb1) at (-0.6,0); \draw[densely dashed, blue!70] (xb1) circle [radius=0.75];
    \coordinate (xb2) at (-1.5,0); \draw[densely dashed, blue!70] (xb2) circle [radius=1.0];
    \coordinate (xb3) at (-2.5,0); \draw[densely dashed, blue!70] (xb3) circle [radius=1.2];
    \filldraw[blue!70] (xb1) circle (1.5pt) node[below] {$x_1$};
    \filldraw[blue!70] (xb2) circle (1.5pt) node[below] {$x_2$};
    \filldraw[blue!70] (xb3) circle (1.5pt) node[below] {$x_3$};

    \coordinate (xstar) at (-3.5,0);
    \filldraw[red] (xstar) circle (2.5pt) node[below=3pt] {$x_\pm^\star$};
    \draw[dashed, thick, red] (xstar) ++(0,0) -- ++(0,1.5);

    \coordinate (xg1) at (-4.3,0); \draw[densely dashed, green!60!black] (xg1) circle [radius=1.0];
    \coordinate (xg2) at (-5.2,0); \draw[densely dashed, green!60!black] (xg2) circle [radius=1.3];
    \coordinate (xg3) at (-6.5,0); \draw[densely dashed, green!60!black] (xg3) circle [radius=2.0];
    \filldraw[green!60!black] (xg1) circle (1.5pt) node[below] {$x_{k-2}$};
    \filldraw[green!60!black] (xg2) circle (1.5pt) node[below] {$x_{k-1}$};
    \filldraw[green!60!black] (xg3) circle (1.5pt) node[below] {$x_{k}$};

    \draw[->, thick, blue!70] (-0.8,0.25) -- (-1.25,0.25);
    \draw[->, thick, blue!70] (-1.7,0.25) -- (-2.25,0.25);
    \draw[->, thick, blue!70] (-2.7,0.25) -- (-3.35,0.25);
    \draw[<-, thick, green!60!black] (-3.8,0.25) -- (-4.05,0.25);
    \draw[<-, thick, green!60!black] (-4.5,0.25) -- (-4.85,0.25);
    \draw[<-, thick, green!60!black] (-5.3,0.25) -- (-5.85,0.25);

    \draw[thick] (-7.3,0.1) -- (-7.3,-0.1) node[below] {$-\infty$};
    \node[blue!70, above] at (-0.6, 1.0) {From $0$};
    \node[green!60!black, above] at (-6.8, 1.0) {From $-\infty$};
\end{tikzpicture}
    \caption{Schematic of the bidirectional analytic continuation scheme. The local solution from $x=0$ (blue chain) and the far-field asymptotic solution from $x \to -\infty$ (green chain) are propagated toward the intermediate matching point $x_\pm^\star$. The dashed circles represent effective convergence disks. Continuity of the function and its derivative at $x_\pm^\star$ yields the connection coefficients.}
    \label{fig:matching_scheme}
\end{figure}

A critical limitation arises in Kerr perturbation theory, where the physical radial coordinate maps to the semi-infinite interval $x \in (-\infty, 0]$. The Frobenius expansions $\HCa(x)$ and $\HCb(x)$ converge strictly for $|x|<1$, while the asymptotic expansions $\HCa^\infty(x)$ and $\HCb^\infty(x)$ are only meaningful for $|x|\gg 1$. Neither local basis alone suffices to cover the entire physical domain.
To bridge this gap, we adopt the analytic continuation framework developed by Motygin~\cite{Motygin2018}. Although the solutions defined above are initially local, Motygin's algorithm constructs globally valid, single-valued functions through a \emph{hybrid matching scheme} that combines medium-distance analytic continuation with far-field asymptotic expansions. As detailed in \S7 of Ref.~\cite{Motygin2018}, the algorithm switches from step-by-step continuation to the asymptotic representation once the radial coordinate satisfies $|\tilde{\epsilon} x| > \mathfrak{R}$. The threshold $\mathfrak{R}$, termed the ``far-field radius,'' is implicitly defined by the requirement that the minimal term of the asymptotic series drops below the machine precision $\varepsilon_{\mathrm{mach}}$:
\begin{equation}\label{eq:far_field_radius}
\min_{n \ge 0} \left| \frac{n!}{\mathfrak{R}^n} \right| < \varepsilon_{\mathrm{mach}}.
\end{equation}
When $|\tilde{\epsilon} x| > \mathfrak{R}$, the local HeunC function is evaluated efficiently via the precomputed connection coefficients and the asymptotic basis, bypassing further stepwise continuation.

The connection between the horizon basis and the far-field asymptotic basis is formally governed by the two-point connection problem. In each Stokes sector $S_\pm$ (defined by $\pm \mathrm{Im}(x) > 0$), the asymptotic solutions are expressed as linear combinations of the locally continued solutions:
\begin{subequations}\label{eq:connection_inf_0}
\begin{align}
\HCa^\infty(x) &= E_1^\pm \HCa(x) + E_2^\pm \HCb(x), \\
\HCb^\infty(x) &= D_1^\pm \HCa(x) + D_2^\pm \HCb(x),\label{eqb:connection_inf_0}
\end{align}
\end{subequations}
where $E_i^\pm$ and $D_i^\pm$ are the forward connection coefficients. The inverse relations,
\begin{subequations}\label{eq:connection_0_inf}
\begin{align}
\HCa(x) &= d_{11}^\pm \HCa^\infty(x) + d_{12}^\pm \HCb^\infty(x),\label{eqa:connection_0_inf} \\
\HCb(x) &= d_{21}^\pm \HCa^\infty(x) + d_{22}^\pm \HCb^\infty(x),
\end{align}
\end{subequations}
with $\mathbf{d}^\pm = \left[ \begin{smallmatrix} E_1^\pm & E_2^\pm \\ D_1^\pm & D_2^\pm \end{smallmatrix} \right]^{-1}$, enable the transformation of solutions between the event horizon ($x=0$) and infinity ($|x|\to\infty$) representations.
To compute these coefficients numerically, we employ a hybrid analytic continuation scheme that proceeds as follows (see also \cref{fig:matching_scheme}):
\begin{enumerate}[(i)]
    \item
     { Adaptive Matching Point Selection: Theoretically, the radial solutions $R_{\ell m}^{\text{in,up}}$ are analytically invariant with respect to the choice of the matching point $x_\pm^\star$, provided it lies within the overlapping convergence regions of the Frobenius and asymptotic expansions. However, the numerical stability of the connection coefficients is highly dependent on this choice. For high-order modes, the increased centrifugal potential barrier makes the radial Teukolsky equation significantly more stiff, necessitating that the Frobenius series and the asymptotic expansion be matched in a region where the latter is sufficiently convergent. Consequently, we employ an adaptive matching strategy: a fixed point $x_\pm^\star = \frac{5}{4}\mathrm{e}^{i\theta}$ is chosen on the anti-Stokes line. For real orbital frequencies, this line coincides with the negative real axis ($\theta = \pi$), yielding $x_\pm^\star = -1.25$.
This fixed-point strategy is utilized for low-order harmonics ($\ell \leq 3$) and moderate orbital parameters ($e = a \leq 0.7$).
 In all other regimes, we push $x_\pm^\star$ to the far-field radius $x_\infty = {\mathfrak{R}}\mathrm{e}^{i\theta}/{|\tilde{\epsilon}|}$ to maintain machine-precision accuracy. This ensures that the radial solutions remain robust across extreme orbital configurations and resolve the strong-field potential structure.}

    \item {Bidirectional computation:}
    From $x=0$, the solution is analytically continued outward to $x_\pm^\star$ via a chain of overlapping power-series expansions. The convergence radius at each center $x_k$ scales as $\mathfrak{R}_k \propto |x_k|$, so the disks increase in size as the chain moves away from the origin.
    From the far field $x_\infty = {\mathfrak{R}}\mathrm{e}^{i\theta}/{|\tilde{\epsilon}|}$, the asymptotic solution is continued inward to $x_\pm^\star$ using optimally truncated series. Since the distance to $x=0$ decreases along this path, the corresponding convergence disks decrease in size as they approach $x_\pm^\star$.
    \item {Connection coefficient extraction:} At $x_\pm^\star$, function values and derivatives from both chains are matched by solving a $2\times 2$ linear system, yielding the forward coefficients $E_i^\pm, D_i^\pm$ and their inverse matrix $\mathbf{d}^\pm$.

    \item {Efficient far-field evaluation:} For any subsequent point satisfying $|\tilde{\epsilon} x| > R$, the solution is computed directly via
    \begin{equation}\label{eq:far_field_eval}
    \HCa(x) = d_{11}^\pm \HCa^\infty(x) + d_{12}^\pm \HCb^\infty(x),
    \end{equation}
    without further analytic continuation.
\end{enumerate}

This hybrid scheme offers significant advantages over pure analytic continuation. It substantially enhances computational efficiency by computing the connection coefficients once and caching them, which reduces far-field evaluation to an $O(1)$ asymptotic series summation and accelerates the process by two to three orders of magnitude compared to step-by-step continuation. The approach also ensures high numerical precision, as the optimal truncation of asymptotic series exponentially suppresses remainder terms, while Wronskian-stabilized matching yields connection coefficients with errors on the order of $10^{-11}$--$10^{-14}$.

Finally, by substituting the ans\"{a}tze \eqref{eq:Sm-Sol} and \eqref{eq:Rm-Sol} into the ATE and RTE, and matching coefficients with \cref{eq:HeunC-Eq}, the physical parameters of the Kerr perturbations can be uniquely mapped to the CHE parameters. The connection machinery detailed above guarantees that this mapping remains valid across the entire radial domain, thereby enabling exact analytic treatments of horizon-to-infinity scattering and QNM boundary conditions.

\subsection{HeunC solutions of ATEs}\label{subsec:heunc_ATE}
Using the basis functions defined in \cref{eq:HCSol-x0}, the general solution $S_{\ell m}$ of the ATE \eqref{eq:GFoATE} is constructed as a linear combination of two linearly independent solutions:
\begin{equation}\label{eq:GSol_ATE}
S_{\ell m}(x) = \mathfrak{D}_1 S_\theta (x)\HCa(x) + \mathfrak{D}_2 S_\theta (x) \HCb(x),
\end{equation}
where $\mathfrak{D}_1$ and $\mathfrak{D}_2$ are arbitrary constants, and $S_\theta(x)$ is the S-homotopic prefactor defined in \cref{eq:S-homotopic-theta-d2}.

The ATE possesses regular singularities at the poles $\Theta = \pm 1$.
To construct solutions regular at each pole, we introduce local coordinates that map the respective singularity to the origin:
\begin{align}
x_{+1} &= \tfrac{1}{2}(1 - \Theta) \quad \text{for } \Theta = +1 \text{ (north pole)}, \\
x_{-1} &= \tfrac{1}{2}(1 + \Theta) \quad \text{for } \Theta = -1 \text{ (south pole)}.
\end{align}
Depending on the chosen coordinate, the CHE parameters $(\alpha, \beta, \gamma, \delta, \eta)$ adopt distinct forms. For Kerr black holes, these parameter sets are explicitly given by
\begin{subequations}\label{eq:ATE_HCprameters_Kerr}
 \begin{align}
 &\alpha_{-1} = 4a\omega, \quad\quad\quad\quad\; \alpha_{+1} = -4a\omega,\\
 &\beta_{-1} = |m - s|,\quad\quad\quad\;\; \beta_{+1} = |m + s|,\\
 &\gamma_{-1} = |m + s|,\quad\quad\quad \gamma_{+1} = |m - s|,\\
 &\delta_{-1} = 4sa\omega,\quad\quad\quad\quad \delta_{+1} = -4sa\omega,\\
 &\eta_{-1} = \tfrac{1}{2}m^2 - \tfrac{1}{2}s^2 - s - 2ma\omega - 2sa\omega - \hat{\lambda},\\
 &\eta_{+1} = \tfrac{1}{2}m^2 - \tfrac{1}{2}s^2 - s - 2ma\omega + 2sa\omega - \hat{\lambda}.
\end{align}
\end{subequations}
Substituting the parameter sets from \cref{eq:ATE_HCprameters_Kerr} into the general solution \eqref{eq:GSol_ATE} yields two linearly independent local solutions around each pole. To enforce global regularity across the entire angular domain, we construct the Wronskian using the particular solution regular at the north pole, $S_{\ell m}(x_{+1})$, and the one regular at the south pole, $S_{\ell m}(x_{-1})$. These are explicitly given by
\begin{widetext}
\begin{align}
 &{S_{\ell m}}({x_{ + 1}}) = {S_\theta }({\alpha _{ + 1}},\beta_{+1},\gamma_{+1},{\delta _{ + 1}},{\eta _{ + 1}};{x_{ - 1}}){\rm{HeunC}}({\alpha _{ + 1}},\beta_{+1},\gamma_{+1},{\delta _{ + 1}},{\eta _{ + 1}};{x_{ + 1}}),\\
   &{S_{\ell m}}({x_{ - 1}}) = {S_\theta }({\alpha _{ - 1}},\beta_{-1},\gamma_{-1},{\delta _{ - 1}},{\eta _{ - 1}};{x_{ - 1}}){\rm{HeunC}}({\alpha _{ - 1}},\beta_{-1},\gamma_{-1},{\delta _{ - 1}},{\eta _{ - 1}};{x_{ - 1}}).
   \label{eq:ATE_W2}
\end{align}
Using these solutions of the ATE \eqref{eq:GFoATE}, we can calculate the SWSHs and the angular separation constants ${}_s A_{\ell m}(a\omega)$.
The solutions are regular at both poles if and only if these two branches are linearly dependent, which is equivalent to the vanishing of their Wronskian~\cite{Chen:2025sbz}:
\begin{equation}
  W_\theta = S_{\ell m}(x_{-1})\frac{{\rm d}}{{\rm d}\Theta}S_{\ell m}(x_{+1}) - S_{\ell m}(x_{+1})\frac{{\rm d}}{{\rm d}\Theta}S_{\ell m}(x_{-1}).
\end{equation}
At the equatorial plane $\Theta=0$ (where $x_{+1}=x_{-1}=1/2$), the S-homotopic prefactors $S_\theta(x)$ become identical up to a constant phase and factor out of the determinant.
Consequently, the regularity condition $W_\theta = 0$ simplifies to a transcendental equation involving only the HeunC functions and their derivatives:
\begin{align}\label{eq:ATE_Wronskian}
 W_\theta &\propto \mathrm{HeunC}(\alpha_{-1},\beta_{-1},\gamma_{-1},\delta_{-1},\eta_{-1};\tfrac{1}{2})\,\mathrm{HeunC}'(\alpha_{+1},\beta_{+1},\gamma_{+1},\delta_{+1},\eta_{+1};\tfrac{1}{2}) \nonumber \\
 &\quad + \mathrm{HeunC}(\alpha_{+1},\beta_{+1},\gamma_{+1},\delta_{+1},\eta_{+1};\tfrac{1}{2})\,\mathrm{HeunC}'(\alpha_{-1},\beta_{-1},\gamma_{-1},\delta_{-1},\eta_{-1};\tfrac{1}{2}).
\end{align}
\end{widetext}
where $\mathrm{HeunC}'(\cdots;x)$ denotes the $x$-derivative.
The condition $W_\theta = 0$ yields a transcendental equation that uniquely determines the eigenvalue $\hat{\lambda}$ and the angular separation constants ${}_s A_{\ell m}(a\omega)$; further details on the extraction procedure are provided in our previous work~\cite{Chen:2025sbz}. Although this fixes the eigenvalue, the overall amplitude of the eigenfunction remains undetermined. To resolve this multiplicative freedom, we impose the standard orthonormality condition for SWSHs,
\begin{equation}\label{eq:Normalization}
  \int_0^{2\pi}\!\!\int_0^\pi \tilde{S}_{\ell m}(\theta ,\phi )\,\tilde{S}^*_{\ell m}(\theta ,\phi )\,{\rm d}\Omega = 1,
\end{equation}
which yields the normalized angular solution
\begin{equation}\label{eq:ATE_sol_phi}
  \tilde{S}_{\ell m}(\theta ,\phi ) = \frac{(-1)^\sigma}{\mathcal{N}}\,S_\theta(x)\,\HCa(x)\,\mathrm{e}^{im\phi}.
\end{equation}
Here, $(-1)^\sigma$ is a conventional phase factor chosen to match standard conventions: $\sigma = 1$ if (i) $\ell$ is odd and $m+s$ is even, (ii) $\ell$ and $m+s$ are both odd with $m \ge s$, or (iii) $\ell$ is even, $m-s$ is odd, and $m \le s$; otherwise, $\sigma = 0$.

The normalization factor $\mathcal{N}$ is defined by the integral
\begin{equation}\label{eq:Norm_Fac_0}
  \mathcal{N}^2 = \int_0^{2\pi}\!\!\int_0^\pi \left|S_\theta(x)\,\HCa(x) \right|^2 \sin\theta\,{\rm d}\theta\,{\rm d}\phi.
\end{equation}
The explicit evaluation of $\mathcal{N}$ is detailed in Appendix~\ref{app:normalization}. For practical implementation, we employ two complementary strategies tailored to different mode regimes. For low-order modes, a direct power-series summation of the HeunC expansion provides a straightforward evaluation,
\begin{equation}\label{eq:Norm_Fac_2}
  \mathcal{N} = \sqrt{4\pi \sum_{i=0}^{n_{\max}} \sum_{j=0}^i b_j b_{i-j} \, |I(i)|},
\end{equation}
where $b_j$ are the recurrence coefficients and $I(i)$ is an analytically integrable kernel.

However, for higher-order modes, the recurrence relations become increasingly ill-conditioned, making direct summation numerically unstable. In this regime, we exploit the Sturm--Liouville structure of the angular equation to express $\mathcal{N}$ in terms of the derivative of the Wronskian with respect to the eigenparameter $\tilde{q}$,
\begin{equation}\label{eq:Norm_DWdq}
  \mathcal{N} = \sqrt {\left|( - 1)^{\tilde\gamma }4\pi p(x)\frac{\partial W}{\partial \tilde{q}}\frac{f_0}{f_1} \right|},
\end{equation}
which bypasses explicit quadrature and maintains machine-precision stability even for large $\ell$.

\subsection{HeunC solutions of homogeneous RTEs}\label{subsec:heunc_RTE}
The radial function $R_{\ell m}(r)$ satisfies the homogeneous RTE. To analyze solutions in the vicinity of the event horizon $r=r_+$, we introduce the coordinate transformation
\begin{equation}\label{eq:mtran}
x_r = -\frac{r - r_+}{r_+ - r_-},
\end{equation}
which maps the event horizon to $x_r=0$ and spatial infinity to $x_r \to -\infty$.
The homogeneous RTE admits a general solution near the event horizon that can be expressed as a linear combination of two linearly independent solutions:
\begin{equation}\label{eq:GSol_RTE}
R_{\ell m}(x_r) = \mathfrak{C}_1 S_r(x_r)\,\HCa(x_r) + \mathfrak{C}_2 S_r(x_r)\,\HCb(x_r),
\end{equation}
where $\mathfrak{C}_1, \mathfrak{C}_2$ are constants determined by physical boundary conditions, and $S_r(x_r)$ is the radial S-homotopic prefactor given in \cref{eq:S-homotopic-r}. The full set of CHE parameters is explicitly given by
\begin{subequations}\label{eq:RTE_HCparameters_Kerr}
 \begin{align}
&{\alpha_r} = 2i\omega ({r_ + } - {r_ - }),\\
&{\beta_r} = -s - \frac{{2i\omega (r_ + ^2 + {a^2}) - 2iam}}{{{r_ + } - {r_ - }}},\\
&{\gamma_r} = s - \frac{{2i\omega (r_ - ^2 + {a^2}) - 2iam}}{{{r_ + } - {r_ - }}},\\
&{\delta_r} = 2is\omega ({r_ - } - {r_ + }) + 2{\omega ^2}\Delta, \\
&\eta_r  = 2is\omega {r_ + } - \tfrac{1}{2}{s^2} - s - \hat{\lambda} \\
& -\frac{{2\left[ {\omega (r_ + ^2 + {a^2}) - am} \right]\left[ { ({a^2} + 2{r_ - }{r_ + } - r_ + ^2)\omega - am} \right]}}{{{{({r_ + } - {r_ - })}^2}}}.\nonumber
\end{align}
\end{subequations}
For a detailed derivation, see our previous work~\cite{Chen:2023ese,Chen:2023lsa}.

Physical scattering problems require solutions satisfying specific asymptotic behaviors.
We define two fundamental solutions: the purely ingoing wave at the event horizon, $R_{\ell m}^{\rm in}$, and the purely outgoing wave at infinity, $R_{\ell m}^{\rm up}$. Their asymptotic forms are prescribed as
\begin{align}
    & R_{\ell m}^{{\rm{in}}}\to \left\{ {
    \begin{array}{*{20}{l}}
{{\Delta ^{ - s}}{{\rm{e}}^{ - iP{r^*}}},}&{r \to {r_{\rm{ + }}},}\\
{B_{\ell m}^{{\rm{ref}}}\frac{{{{\rm{e}}^{i\omega {r^*}}}}}{{{r^{1{\rm{ + }}2s}}}} + B_{\ell m}^{{\rm{inc}}}\frac{{{{\rm{e}}^{ - i\omega {r^*}}}}}{r},}
&{r \to  + \infty ,}
\end{array}
    } \right.\label{eq:boundary1}\\
 & R_{\ell m}^{{\rm{up}}} \to \left\{\begin{array}{*{20}{l}}
{C_{\ell m}^{{\rm{up}}}{{\rm{e}}^{iP{r^*}}} + C_{\ell m}^{{\rm{ref}}}\frac{{{{\rm{e}}^{ - iP{r^*}}}}}{\Delta^s },}&{r \to {r_ + },}\\
{{r^{ - 1 - 2s}}{{\rm{e}}^{i\omega {r^*}}},}&{r \to  + \infty ,}
\end{array} \right.\label{eq:boundary2}
\end{align}
where $P = \omega - m \Omega_{\rm H}$ and $\Omega_{\rm H} = a/(2Mr_+)$. The tortoise coordinate $r^*$ is defined by
\begin{equation}
 r^* = r + \frac{2Mr_+}{r_+ - r_-} \ln \frac{r - r_+}{2M} - \frac{2Mr_-}{r_+ - r_-} \ln \frac{r - r_-}{2M}.
\end{equation}

To construct these global solutions, we employ the hybrid analytic continuation framework established in \cref{subsec:teukolsky_heun}.
A central advantage of Motygin's \emph{hybrid matching scheme} is its automatic representation switching at a designated matching point $x_\pm^\star$:
for $|x_r| < |x_\pm^\star|$, the solution is evaluated via chained power-series expansions, while for $|x_r| > |x_\pm^\star|$,
it seamlessly transitions to the asymptotic basis using precomputed connection coefficients.
This design inherently avoids the numerical stiffness associated with direct continuation over large radial intervals.

Regularity of the ingoing mode at the event horizon requires the absence of the singular Frobenius branch, which fixes $\mathfrak{C}_2=0$ in \cref{eq:GSol_RTE}.
Thus, the ingoing wave solution is proportional to the regular basis function:
\begin{equation}\label{eq:uHor}
R_{\ell m}^{\rm in}(x_r) = S_r(x_r) \, \HCa(x_r).
\end{equation}
To extract the scattering amplitudes at infinity, we utilize this hybrid structure.
As the radial coordinate crosses the matching point into the far-field regime ($|x_r| > |x_\pm^\star|$), $\HCa(x_r)$ is automatically represented in terms of the asymptotic basis via the inverse connection formula \eqref{eqa:connection_0_inf}:
\begin{equation}\label{eq:uHor_inf}
R_{\ell m}^{\rm in}(x_r) = S_r(x_r) \left[ d_{11}^\pm \HCa^\infty(x_r) + d_{12}^\pm \HCb^\infty(x_r) \right].
\end{equation}
Substituting the asymptotic expansions \eqref{eq:HCA-HCB-inf} of $\HCa^\infty$ and $\HCb^\infty$ as $|x_r| \to \infty$ and matching the resulting expression to the physical boundary conditions \eqref{eq:boundary1}, we obtain the analytic expressions for the transmission, incident, and reflection amplitudes:
\begin{subequations}\label{eq:Rin_Amp}
\begin{align}
&B_{\ell m}^{\rm trans} = \mathrm{e}^{i P r_+}\frac{{{{({r_ + } - {r_ - })}^{2s + 2iMP}}}}{{{{({r_ + } + {r_ - })}^{2iMP}}}}, \\
&B_{\ell m}^{\rm inc} = \mathrm{e}^{i\omega r_+} \frac{(r_+ - r_-)^{1 + 2iM\omega}}{(r_+ + r_-)^{2iM\omega}} d_{11}^\pm, \\
&B_{\ell m}^{\rm ref} = \mathrm{e}^{-i\omega r_+} \frac{(r_+ - r_-)^{2s + 1 - 2iM\omega}}{(r_+ + r_-)^{-2iM\omega}} d_{12}^\pm.
\end{align}
\end{subequations}
Similarly, the outgoing wave solution $R_{\ell m}^{\rm up}$ is defined by its asymptotic behavior at spatial infinity. In the far-field region ($|x_r| > |x_\pm^\star|$), it corresponds directly to the outgoing asymptotic branch $\HCb^\infty(x_r)$:
\begin{equation}\label{eq:uOut1}
R_{\ell m}^{\rm up}(x_r) = S_r(x_r) \, \HCb^\infty(x_r).
\end{equation}
To determine the horizon reflection and transmission coefficients, we trace this solution inward.
As $x_r$ crosses the matching point back into the near-horizon domain ($|x_r| < |x_\pm^\star|$), the hybrid algorithm automatically maps the far-field basis to the Frobenius basis using the forward connection formula \eqref{eqb:connection_inf_0}:
\begin{equation}\label{eq:uOut_hor}
R_{\ell m}^{\rm up}(x_r) = S_r(x_r) \left[ D_1^\pm \HCa(x_r) + D_2^\pm \HCb(x_r) \right].
\end{equation}
Matching the asymptotic behavior of \cref{eq:uOut_hor} to the horizon boundary conditions \eqref{eq:boundary2} yields the corresponding amplitudes:
\begin{subequations}\label{eq:amp_C}
\begin{align}
&C_{\ell m}^{\rm trans} = \mathrm{e}^{-i \omega r_+} \frac{{{{({r_ + } - {r_ - })}^{2s + 1 - 2iM\omega }}}}{{{{({r_ + } + {r_ - })}^{{\rm{ - }}2i\omega M}}}}, \\
&C_{\ell m}^{\rm up} = (-1)^{s + \frac{4iMP r_+}{r_+ - r_-}} \mathrm{e}^{-i P r_+}\left( \frac{r_+ + r_-}{r_+ - r_-} \right)^{2iMP} D_2^\pm, \\
&C_{\ell m}^{\rm ref}  = \mathrm{e}^{i P r_+} \frac{{{{({r_ + } - {r_ - })}^{2s + 2iMP}}}}{{{{({r_ + } + {r_ - })}^{2iMP}}}} D_1^\pm.
\end{align}
\end{subequations}
These amplitudes, determined by the numerically computed connection coefficients $d_{ij}^\pm$ and $D_i^\pm$,
serve as essential building blocks for constructing the full radial solutions.
Together with the normalized angular eigenfunctions \eqref{eq:ATE_sol_phi},
they constitute the complete set of mode functions required to assemble the Green's function for Kerr perturbations.

\subsection{Radiative fluxes}\label{subsec:fluxes}
The radial function $R_{\ell m}(r)$, governed by the inhomogeneous RTE \eqref{eq:GFoRTE}, describes the gravitational field generated by a perturbing point particle.
Causality and the absorbing nature of the event horizon impose strict boundary conditions: the solution must be purely outgoing at spatial infinity and purely ingoing at the event horizon.
It is crucial to distinguish this source-driven inhomogeneous solution from the homogeneous basis functions $R_{\ell m}^{\rm in,up}$ constructed in \cref{subsec:heunc_RTE}, which are used as building blocks for the Green's function.
Focusing on the gravitational perturbation sector (spin weight $s=-2$, corresponding to the Weyl scalar $\psi_4$), we find that the asymptotic behavior of $R_{\ell m}(r)$ under these physical conditions is given by
\begin{equation}
  {R_{\ell m}}(r) \to \left\{ \begin{array}{l}
Z_{\ell m\omega}^\infty {r^3}{e^{i\omega {r^*}}}\;,\qquad r \to \infty \;,\\
Z_{\ell m\omega}^{\rm{H}}\Delta {e^{ - iP{r^*}}}\;,\qquad r \to {r_ + }\;.
\end{array} \right.
\end{equation}
where $Z_{\ell m\omega}^{\infty}$ and $Z_{\ell m\omega}^{\rm H}$ are the complex scattering amplitudes at infinity and the event horizon, respectively.
Unlike the homogeneous modes, these amplitudes encode the strength and phase of the gravitational radiation excited by the particle source.
They are determined via the Green's function formalism, in which the particular solution is constructed by integrating the homogeneous basis $R_{\ell m}^{\rm in, up}$ against the source term $T_{\ell m}$~\cite{Chen:2023ese,Chen:2023lsa,Hughes:2021exa}.

For a test particle on a generic bound geodesic, the source term is periodic in the radial and polar motions.
Consequently, the radiation spectrum is discrete, consisting of harmonics of the fundamental orbital frequencies.
The orbit is fully characterized by the parameter set $(a, p, e, x_I)$, where $a$ is the spin, $p$ the semi-latus rectum, $e$ the eccentricity, and $x_I = \cos\iota$ the inclination parameter.
Serving as initial inputs, these parameters uniquely determine the fundamental Mino--time frequencies and the explicit Boyer--Lindquist coordinate evolution $x^\mu(\lambda)$, enabling the direct mapping of the particle trajectory onto the phase-space integrals~\cite{Fujita:2009bp,Hughes:2021exa}.
In this parametrization, the amplitudes $Z_{\ell m \omega}^{\infty ,{\rm{H}}}$ can be expressed as a series sum involving the product of a double integral $J^{\infty,{\rm H}}_{\ell mnk}$ and a single integral,
\begin{equation}
Z_{\ell m \omega}^{\infty ,{\rm{H}}} = \sum\limits_{k =  - \infty }^\infty  \sum\limits_{n =  - \infty }^\infty {J_{\ell mnk}^{\infty ,{\rm{H}}}}  \int_{ - \infty }^\infty  {{{\rm{e}}^{i{\Upsilon _t}(\omega - {\omega _{mnk}})\lambda }}} {\kern 1pt} \;{\rm{d}}\lambda,
\label{eq:Zlmw}
\end{equation}
where ${\omega _{mnk}} = m{\Omega _\phi } + k{\Omega _\theta } + n{\Omega _r}$ is the Doppler-shifted harmonic frequency observed at infinity, with $\Omega_i = \Upsilon_i/\Upsilon_t$ being the coordinate-time frequencies.
The double integral
\begin{equation}
  J^{\infty,{\rm H}}_{\ell mnk} = \frac{1}{(2 \pi)^2}\int_0^{2 \pi} \int_0^{2 \pi}{\bf G}_{\ell m}^{\infty ,{\rm{H}}}{\rm d}\lambda_\theta\,{\rm d}\lambda_r,
\label{eq:Jlmnk_def}
\end{equation}
captures the spatial overlap between the source trajectory and the homogeneous wave modes. The integrand is given by
\begin{equation}\label{eq:integrand}
 {\bf G}_{\ell m \omega}^{\infty ,{\rm{H}}} = {\cal I}_{\ell m}^{\infty ,{\rm{H}}}(r,\theta ){\rm{exp}}\left( \begin{array}{l}
i\omega (\Delta {t_r} + \Delta {t_\theta })\\
 - im(\Delta {\phi _r} + \Delta {\phi _\theta })\\
 + in{\Upsilon _r}{\lambda _r} + ik{\Upsilon _\theta }{\lambda _\theta }
\end{array} \right).
\end{equation}
A detailed discussion of the kernel function ${\cal I}^{\infty,{\rm H}}_{\ell m\omega}(r,\theta)$ can be found in our previous work~\cite{Chen:2023ese,Chen:2023lsa}.
This function is schematically a Green's function used to obtain homogeneous radial solutions ${R_{\ell m\omega }^{{\rm{in,up}}}}$ of the Teukolsky equation, multiplied by the source term of this equation.
\begin{align}
& {\cal I}^{\infty,{\rm H}}_{\ell m\omega}=  \frac{1}{{W_{\rm C}}}\left[ {\left( {{A_{nn0}} + {A_{\bar mn0}}+{A_{\bar m\bar m0}}}\right)R_{\ell m}^{{\rm{in,up}}}} \right. \nonumber \\
 &- \left( {{A_{\bar mn1}} + {A_{\bar m\bar m1}}} \right){\left( {R_{\ell m}^{{\rm{in,up}}}} \right)^\prime }{\left. + {A_{\bar m\bar m2}}{{\left( {R_{\ell m}^{{\rm{in,up}}}} \right)}^{\prime \prime }} \right]},
\end{align}
where the prime denotes $\partial/\partial r$. The angular solution \eqref{eq:ATE_sol_phi} is encoded in the coefficients $A$, whose explicit expressions are provided in Refs.~\cite{Teukolsky:1973ha,Drasco:2005kz,OSullivan:2014ywd}.
The normalization is fixed by the conserved Wronskian
\begin{equation}
   {W_{\rm C}} = R_{\ell m}^{{\rm{up}}}\frac{\rm d}{{{\rm d}{r^*}}}R_{\ell m}^{{\rm{in}}} - R_{\ell m}^{{\rm{in}}}\frac{\rm d}{{{\rm d}{r^*}}}R_{\ell m}^{{\rm{up}}}= 2i\omega B_{\ell m}^{{\rm{inc}}},
\end{equation}
which ensures the correct flux normalization and is independent of the choice of radial coordinate.

Employing the Fourier identity
\begin{equation}
\int_{-\infty}^\infty e^{ix\lambda} d\lambda = 2\pi \delta(x),
\end{equation}
 the integration over $\lambda$ enforces energy conservation and collapses the continuous frequency spectrum into a discrete line spectrum:
\begin{equation}\label{eq:Zlmw_Integ}
  Z_{\ell m \omega}^{\infty ,{\rm{H}}} = \sum\limits_{k =  - \infty }^\infty  \sum\limits_{n =  - \infty }^\infty  Z_{\ell mnk}^{\infty ,{\rm{H}}} \delta (\omega  - {\omega _{mnk}}),
\end{equation}
where $Z_{\ell mnk}^{\infty ,{\rm{H}}} = 2\pi J_{\ell mnk}^{\infty ,{\rm{H}}}/{\Upsilon _t}$.

With the discrete mode amplitudes $Z_{\ell mnk}^{\infty,\rm H}$ fully determined, the total energy flux carried by gravitational radiation is evaluated by summing the time-averaged contributions over all harmonic modes.
The radiation is partitioned into two distinct channels: waves escaping to spatial infinity and waves entering the event horizon.
Their respective energy loss rates are given by
\begin{subequations}\label{eq:Flux_Inf_Hor}
  \begin{align}
      \dot {\cal E}=& \dot {\cal E}^\infty+\dot {\cal E}^{\rm H},\\
   \dot {\cal E}^\infty=& {\left\langle {\frac{{{\rm{d}}E}}{{{\rm{d}}t}}} \right\rangle ^\infty } = \sum\limits_{\ell mnk} {\frac{{|Z_{\ell mnk}^\infty {|^2}}}{{4\pi \omega _{mnk}^2}} }, \\
    \dot {\cal E}^{\rm H}=&{\left\langle {\frac{{{\rm{d}}E}}{{{\rm{d}}t}}} \right\rangle ^{\rm{H}}} = \sum\limits_{\ell mnk} {\frac{{{\alpha _{\ell m}}|Z_{\ell mnk}^{\rm{H}}{|^2}}}{{4\pi \omega _{mnk}^2}}}.
\end{align}
\end{subequations}
where the summation runs over the standard multipole and orbital harmonic indices,
\begin{align}
\sum\limits_{\ell mnk} {}  \to \sum\limits_{\ell  = 2}^\infty  {\sum\limits_{m = 1}^\ell  {\sum\limits_{n =  - \infty }^\infty  {\sum\limits_{k =  - \infty }^\infty  {} } } }.
\end{align}

The horizon energy flux is modulated by the transmission factor $\alpha_{\ell m\omega}$, which encapsulates the graybody filtering and superradiant scattering properties of the Kerr geometry. Its explicit form reads
\begin{align}
 &  \alpha_{\ell m\omega}=
 \frac{256(2Mr_+)^5 P (P ^2+4\varepsilon^2)(P ^2+16\varepsilon^2)\omega^3}
{|{\bf{C}}|^2},\\
 &  |{\bf{C}}|^2= (2\,\lambda+3)\,(96\,a^2\,\omega^2-48\,a\,\omega\,m)+144\,\omega^2\,(M^2-a^2) \nonumber\\
  &     + \left[(\lambda+2)^2+4\,a\,\omega\,m-4\,a^2\,\omega^2\right]\left[\lambda^2+36\,a\,\omega\,m-36\,a^2\,\omega^2\right],\nonumber\\
\end{align}
with $\varepsilon=\sqrt{M^2-a^2}/(4Mr_{+})$. This factor reduces to the Schwarzschild graybody factor in the limit $a \to 0$, and correctly encodes the superradiant amplification for modes satisfying $\omega < m\Omega_{\rm H}$, thereby ensuring global energy conservation between the asymptotic fluxes and the rate of orbital energy loss.

\begin{figure*}[htbp]
	\centering
\subfloat[SWSH functions for $2\leq \ell \leq 10$.\label{figa:SWSHs_fun_eig}]{\includegraphics[width=3.4in]{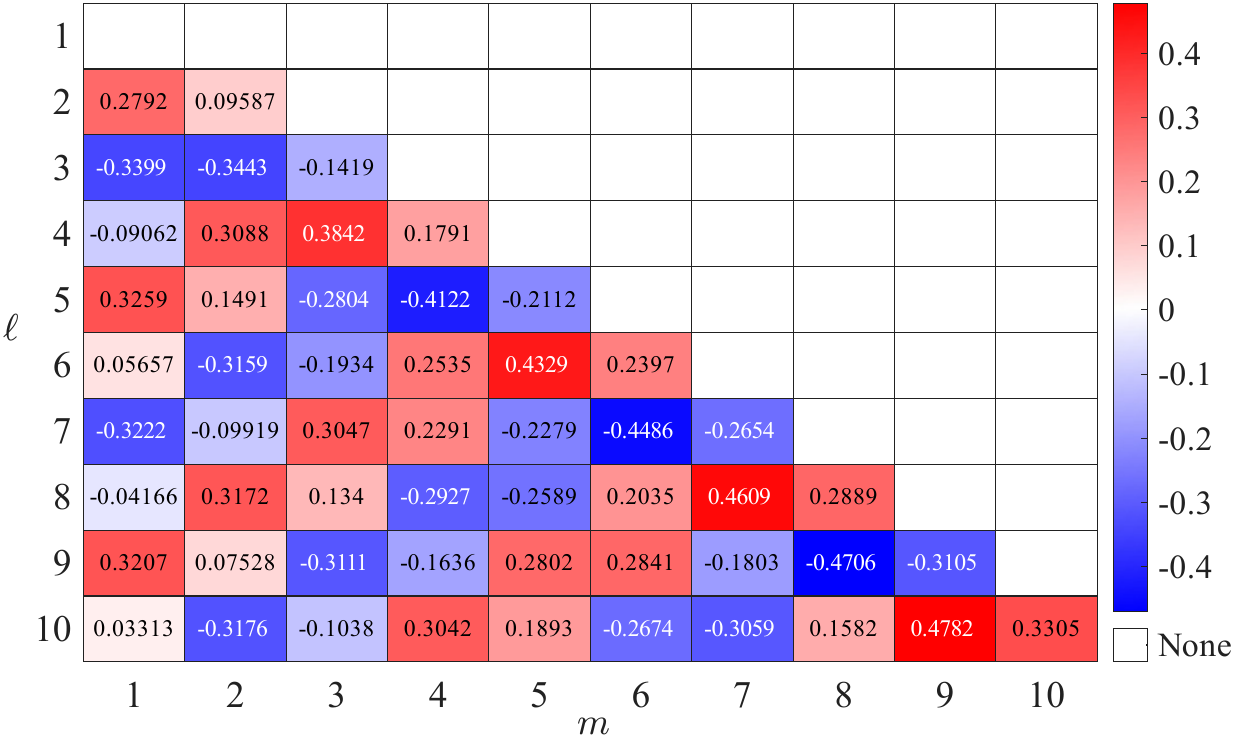}}\,\,
\subfloat[SWSH eigenvalues for $2\leq \ell \leq 10$.\label{figb:SWSHs_fun_eig}]{\includegraphics[width=3.4in]{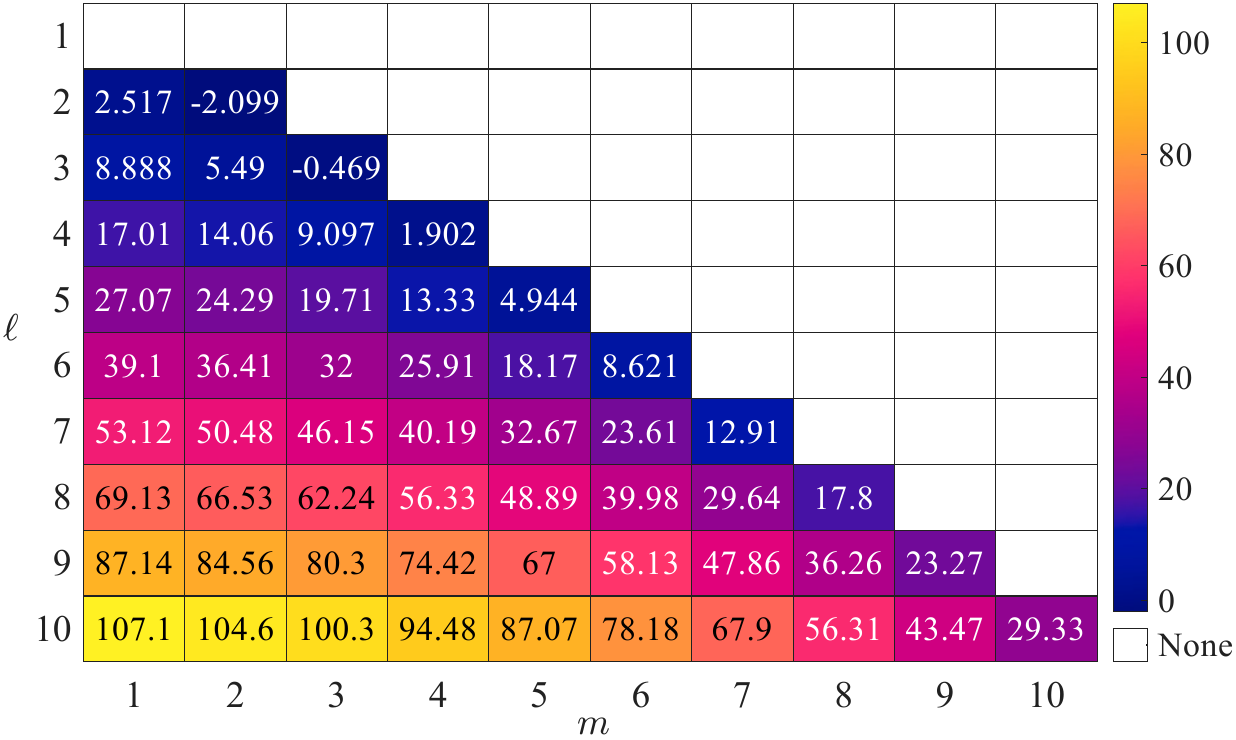}}\\
\caption{ SWSH functions and eigenvalues with the mechanical precision ($10^{-16}$) and the near-extremal spin $a=0.9999$}\label{fig:SWSHs_fun_eig}
\end{figure*}

\begin{figure*}[htbp]
	\centering
\subfloat[HeunC method via Wronskian derivative \eqref{eq:Norm_DWdq}.\label{figa:SWSHs_err}]{\includegraphics[width=3.4in]{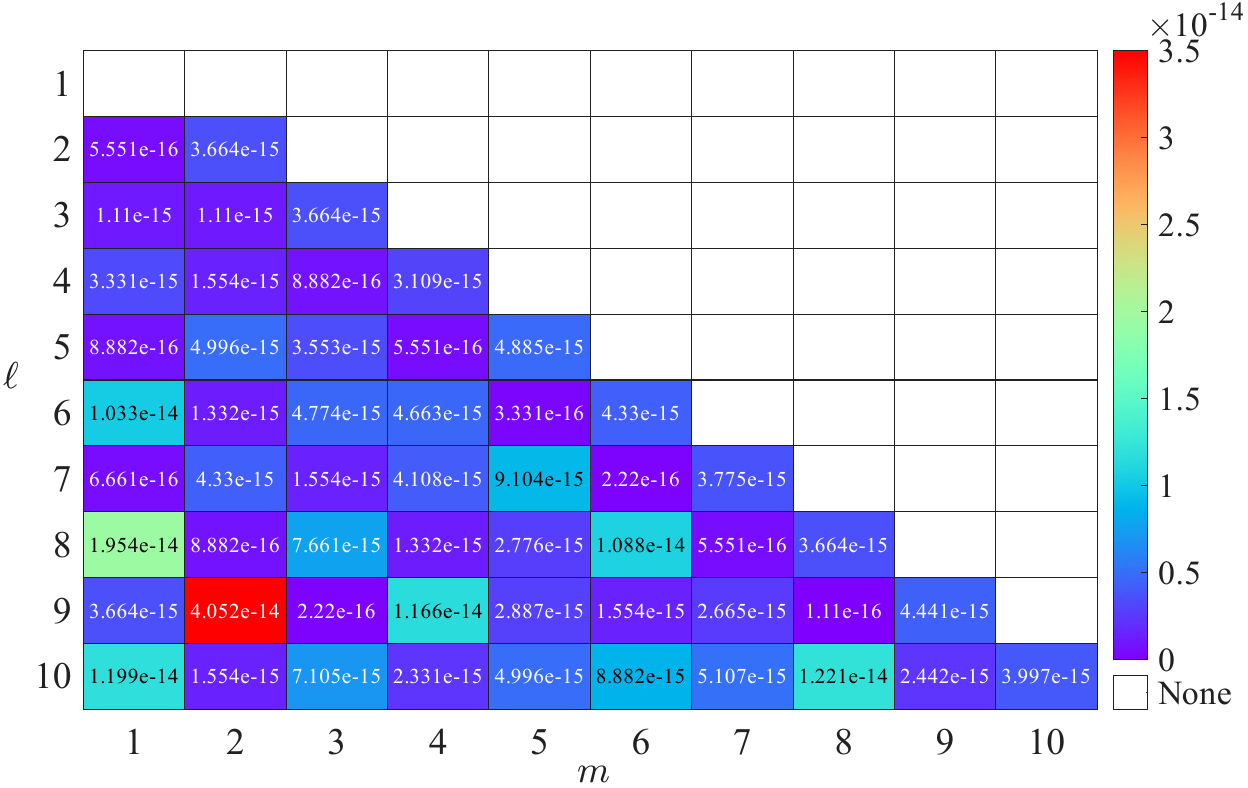}}\,\,
\subfloat[15-term spectral expansion method.\label{figb:SWSHs_err}]{\includegraphics[width=3.4in]{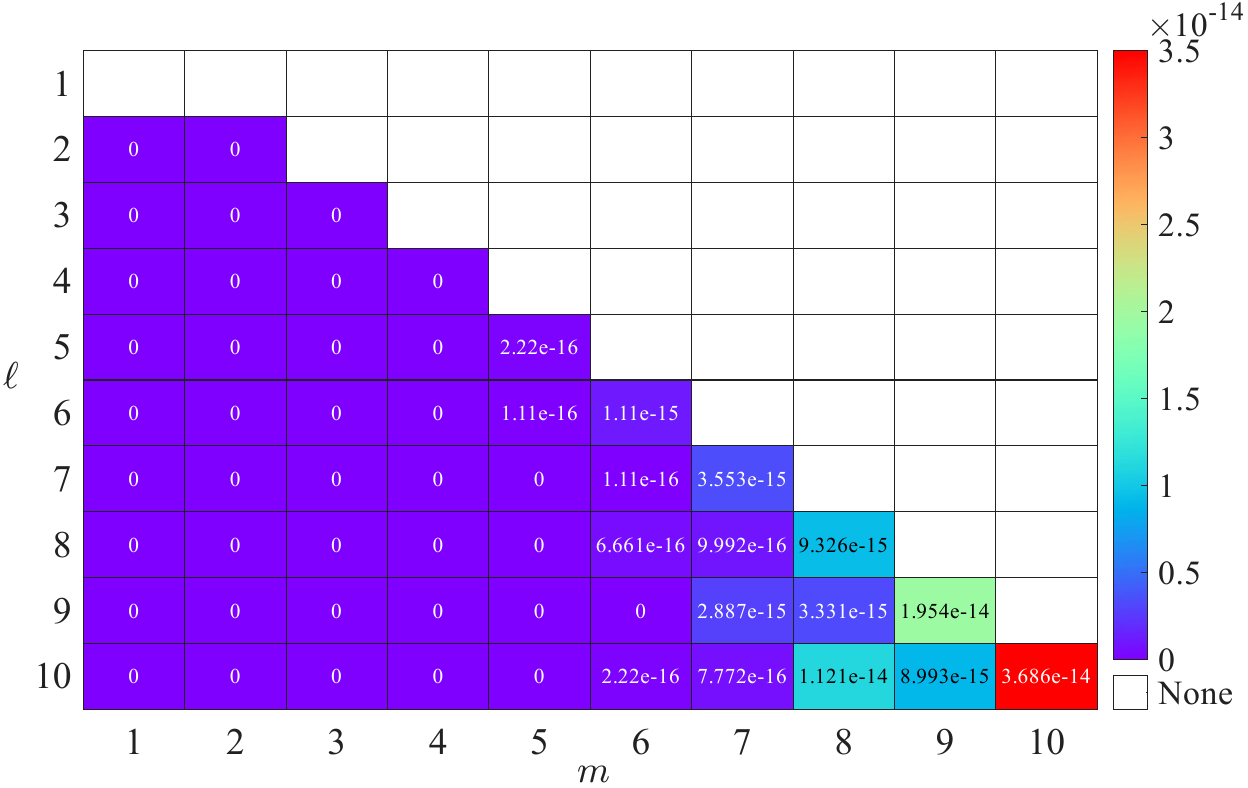}}\\
\subfloat[HeunC method via power-series summation \eqref{eq:Norm_Fac_2}.\label{figc:SWSHs_err}]{\includegraphics[width=3.4in]{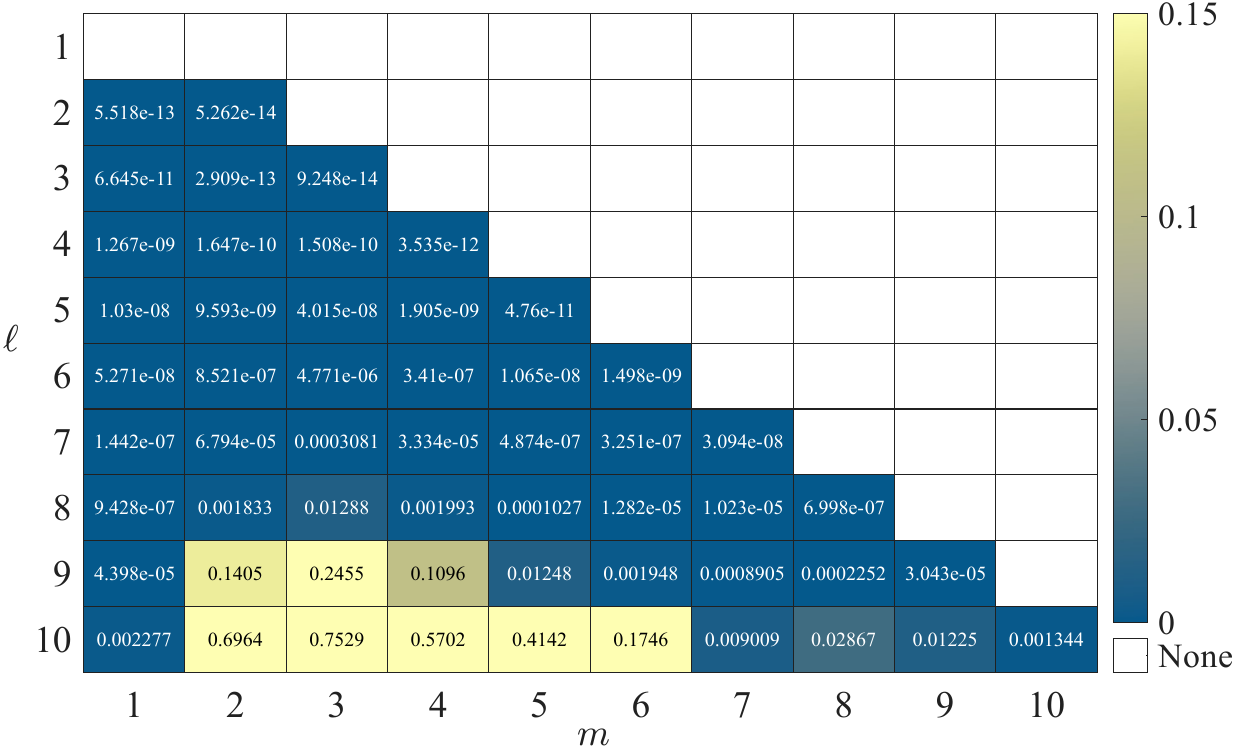}}\,\,
\subfloat[Leaver's continued-fraction method.\label{figd:SWSHs_err}]{\includegraphics[width=3.4in]{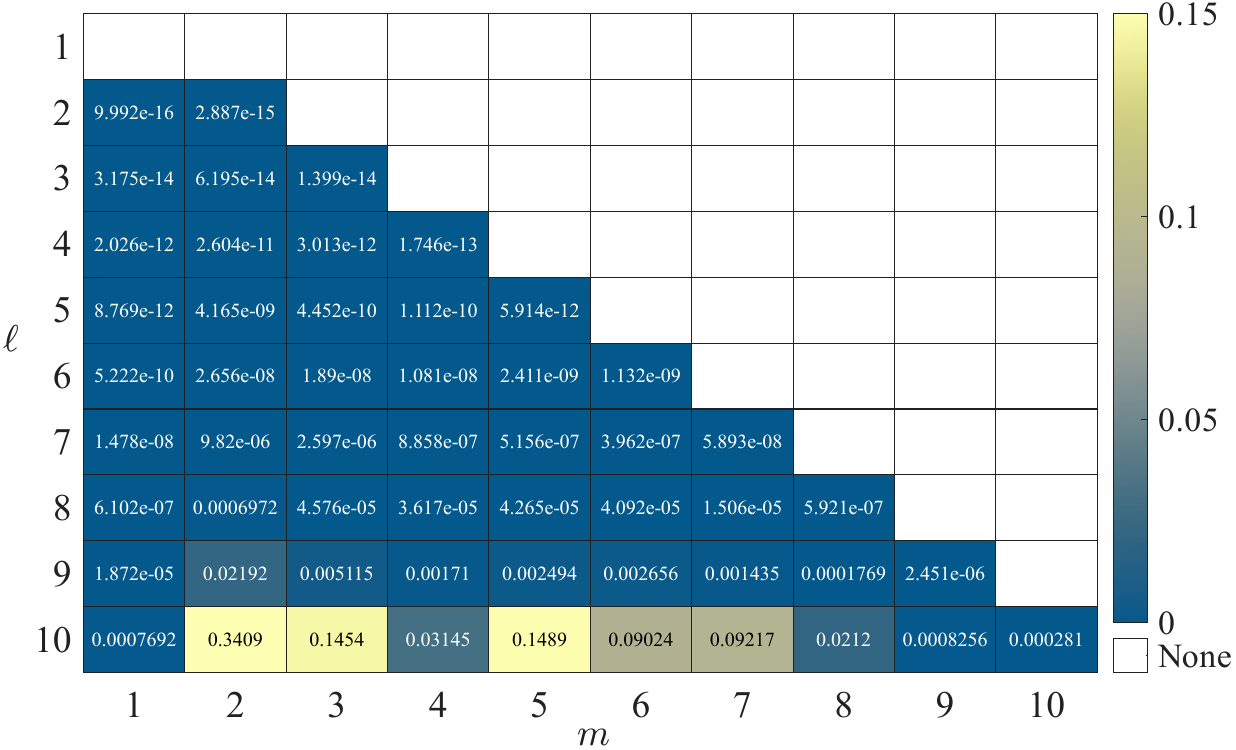}}
\caption{Relative errors of SWSH functions computed with machine precision ($10^{-16}$) and near-extremal spin $a=0.9999$.}\label{fig:SWSHs_err}
\end{figure*}

\begin{table*}[htbp]
  \centering
  \caption{Comparison of computational performance for computing SWSH functions with $2 \le \ell \le 10$. We compare the HeunC method (via Wronskian derivative), Leaver's method, and the spectral decomposition method. $L_\infty$ and $L_2$ errors are computed against a high-precision reference solution. Time is reported in milliseconds (ms).}
    \begin{tabular}{cccccccccccc}
    \toprule
     & \multicolumn{3}{c}{HeunC Method} &      & \multicolumn{3}{c}{Leaver's Method~\cite{BHPToolkit}} &      & \multicolumn{3}{c}{Spectral Decomposition Method~\cite{BHPToolkit}} \\
    \cmidrule(lr){2-4} \cmidrule(lr){6-8} \cmidrule(lr){10-12}
    $a$ & $L_\infty$-Error & $L_2$-Error & Time (ms) &      & $L_\infty$-Error & $L_2$-Error & Time (ms) &      & $L_\infty$-Error & $L_2$-Error & Time (ms) \\
    \midrule
    0.1   & $4.563 \times 10^{-14}$ & $5.064 \times 10^{-14}$ & 121    &      & $9.608 \times 10^{-15}$ & $1.583 \times 10^{-14}$ & 1178   &      & $1.519 \times 10^{-14}$ & $1.542 \times 10^{-14}$ & 540    \\
    0.3   & $5.618 \times 10^{-14}$ & $6.049 \times 10^{-14}$ & 118    &      & $5.536 \times 10^{-14}$ & $7.091 \times 10^{-14}$ & 2005   &      & $8.801 \times 10^{-15}$ & $1.051 \times 10^{-14}$ & 532    \\
    0.5   & $6.617 \times 10^{-14}$ & $7.122 \times 10^{-14}$ & 109    &      & $1.177 \times 10^{-14}$ & $1.269 \times 10^{-14}$ & 1809   &      & $1.238 \times 10^{-14}$ & $1.296 \times 10^{-14}$ & 467    \\
    0.7   & $5.362 \times 10^{-14}$ & $6.775 \times 10^{-14}$ & 115    &      & $8.799 \times 10^{-14}$ & $1.176 \times 10^{-13}$ & 1506   &      & $5.930 \times 10^{-14}$ & $6.402 \times 10^{-14}$ & 596    \\
    0.9   & $3.986 \times 10^{-14}$ & $5.223 \times 10^{-14}$ & 129    &      & $3.835 \times 10^{-14}$ & $4.106 \times 10^{-14}$ & 849    &      & $5.048 \times 10^{-14}$ & $5.495 \times 10^{-14}$ & 590    \\
    0.9999& $4.052 \times 10^{-14}$ & $5.778 \times 10^{-14}$ & 117    &      & $1.065 \times 10^{-14}$ & $1.495 \times 10^{-14}$ & 2267   &      & $3.488 \times 10^{-14}$ & $5.406 \times 10^{-14}$ & 727    \\
    \bottomrule
    \end{tabular}%
  \label{tab:Time_SWSHs}%
\end{table*}
\section{Comparisons with other methods}\label{sec:comparisons}
In this section, we comprehensively evaluate the numerical performance of the HeunC framework under double-precision arithmetic.
This choice is motivated by practical efficiency considerations: while arbitrary-precision arithmetic can in principle achieve higher accuracy, the computational cost typically grows superlinearly with the number of digits, often leading to prohibitive runtimes for the large number of mode evaluations required in generic-orbit flux calculations.
Our goal is therefore to demonstrate that the HeunC framework can deliver machine-precision accuracy using only double-precision arithmetic, thereby maximizing computational throughput without sacrificing reliability.

To establish a reliable benchmark for our comparisons, we generate reference solutions using the BHPToolkit~\cite{BHPToolkit}, a robust suite of black hole perturbation theory codes developed over many decades.
However, in our numerical simulations of generic orbits, we observed that the default implementation in the BHPToolkit relies on adaptive grids for the integration of \cref{eq:Jlmnk_def}, which can occasionally yield inaccurate results due to insufficient sampling (i.e., grid spacing that is too coarse).
This issue is more pronounced for high-order modes with large harmonic indices $(l,m,n,k)$, but can also occur for certain low-order configurations, such as $(l,m,n,k)=(4,4,3,4)$, depending on the orbital parameters.
To ensure the reliability of our benchmark data, we modified the integration module within the BHPToolkit to employ a sufficiently dense uniform grid for evaluating the radiative fluxes via the trapezoidal method, with a strict accuracy requirement of $10^{-60}$ or better.
While this straightforward quadrature method incurs a higher computational cost than adaptive schemes, it guarantees accurate integration results by adequately resolving the rapid oscillations of the integrand ${\bf G}_{\ell m \omega}^{\infty,{\rm H}}$ across the entire domain.
All numerical computations were performed on a laptop equipped with an Intel Core i7-12700H processor (base frequency 2.70 GHz).
In the following tables and figures, the relative error of a quantity $\mathcal{X}$ is defined as
\begin{equation}
    \blacktriangle \mathcal{X} = \left| \frac{\mathcal{X}_{\mathrm{Ref}} - \mathcal{X}_0}{\mathcal{X}_{\mathrm{Ref}}} \right|,
\end{equation}
where $\mathcal{X}_{\mathrm{Ref}}$ denotes the reference solution and $\mathcal{X}_0$ represents the numerical result obtained by a specific method. Additionally, we employ two standard error metrics: the $L_\infty$-error, which measures the maximum absolute error across the entire grid (representing the worst-case deviation), and the $L_2$-error, which represents the root-mean-square error and provides a measure of the average accuracy over the domain.

\begin{table*}[htbp!]
  \centering
  \caption{Comparison of three methods for GW fluxes of Schwarzschild BHs in circular orbits. $N$ denotes the effective working precision (or truncation order) optimized for each method. }
    \begin{tabular}{ccccccccc}
    \toprule
         & \multicolumn{2}{c}{HeunC Method ($N=16$)} &      & \multicolumn{2}{c}{BHPToolkit~\cite{BHPToolkit} ($N=22$)} &      & \multicolumn{2}{c}{Nekrasov--Shatashvili Method~\cite{Cipriani:2025ikx} ($N=40$)} \\
    \cmidrule(lr){2-3} \cmidrule(lr){5-6} \cmidrule(lr){8-9}
    $r/M$  & Error & Time (ms) &      & Error & Time (ms) &      & Error & Time (ms) \\
    \midrule
    6    & $8.036 \times 10^{-12}$ & 70     &      & $1.449 \times 10^{-11}$ & 12619   &      & $1.550 \times 10^{-10}$ & 54584   \\
    7    & $4.447 \times 10^{-12}$ & 73     &      & $3.052 \times 10^{-11}$ & 11681   &      & $9.413 \times 10^{-12}$ & 54197   \\
    8    & $3.778 \times 10^{-12}$ & 72     &      & $1.061 \times 10^{-12}$ & 10121   &      & $6.088 \times 10^{-13}$ & 57496   \\
    9    & $8.081 \times 10^{-12}$ & 75     &      & $4.914 \times 10^{-14}$ & 10527   &      & $5.118 \times 10^{-14}$ & 53777   \\
    10   & $3.285 \times 10^{-12}$ & 75     &      & $3.551 \times 10^{-13}$ & 11399   &      & $5.474 \times 10^{-15}$ & 56890   \\
    \bottomrule
    \end{tabular}%
  \label{tab:Nekrasov--Shatashvili_Errors}%
\end{table*}
\subsection{SWSH functions and eigenvalues}\label{subsec:spheroidal}
We begin by comparing the computation of SWSHs, which constitutes a foundational component of the Teukolsky formalism.
Lo developed a Julia package for computing GW fluxes, which employs a spectral method for solving SWSHs~\cite{Lo:2023fvv}.
While this implementation is computationally efficient, its numerical accuracy is typically limited to approximately $10^{-12}$.
Consequently, for our benchmark comparisons, we utilize the spectral method implemented in the BHPToolkit, which achieves significantly higher precision and serves as a more reliable standard.

\cref{fig:SWSHs_fun_eig} presents the SWSH functions and corresponding eigenvalues across the mode space $2 \leq \ell \leq 10$ for a near-extremal spin $a=0.9999$.
These panels serve as a baseline reference, illustrating the mode structure and eigenvalue spectrum generated by our framework.
The regular patterns observed in \cref{figa:SWSHs_fun_eig} and the monotonic behavior of the eigenvalues in \cref{figb:SWSHs_fun_eig} are consistent with the known analytical properties~\cite{Hughes:1999bq} of the ATE \eqref{eq:GFoATE}.
The quantitative assessment of numerical accuracy, stability, and computational efficiency is provided below through direct error analysis.

The accuracy and numerical stability of the different methods are compared in \cref{fig:SWSHs_err}.
The HeunC method based on the Wronskian derivative \eqref{eq:Norm_DWdq}, shown in \cref{figa:SWSHs_err}, demonstrates excellent precision, with relative errors remaining below $\sim 10^{-14}$ across the entire mode space ($2 \le \ell \le 10$).
This accuracy is comparable to that of the 15-term spectral expansion method depicted in \cref{figb:SWSHs_err}.
However, as detailed in \cref{tab:Time_SWSHs}, the Wronskian-based HeunC method is significantly more efficient.
In contrast, the HeunC method based on direct power-series summation \eqref{eq:Norm_Fac_2} (\cref{figc:SWSHs_err}) and Leaver's continued-fraction method (\cref{figd:SWSHs_err}) exhibit significant numerical instability in the near-extremal regime ($a=0.9999$).
As seen in the heatmaps, their errors grow rapidly for higher $\ell$ and $m$, reaching values as large as $10^{-1}$.
This degradation highlights the superiority of the Wronskian-based method for high-precision calculations involving near-extremal black holes.
In addition to accuracy, computational efficiency is critical. \cref{tab:Time_SWSHs} summarizes the performance across a range of black hole spin parameters ($0.1 \leq a \leq 0.9999$).
The HeunC method (via the Wronskian derivative \eqref{eq:Norm_DWdq}) consistently outperforms both Leaver's method and the spectral decomposition method by approximately one order of magnitude, especially near the extremal limit.
These results confirm that the HeunC-based algorithm offers a superior balance of efficiency and robust accuracy, making it highly suitable for extensive numerical parameter studies involving near-extremal Kerr black holes.

\subsection{Inspiral fluxes}\label{subsec:inspirals}
\subsubsection{Circular orbits for schwarzschild geometry}\label{subsubsec:circular_schwarz}
We first validate our framework against the simplest case: a test particle in an equatorial circular orbit around a Schwarzschild black hole.
This configuration serves as a critical benchmark because high-precision reference values are readily available through established methods.
In addition to the BHPToolkit implementation of the MST method~\cite{BHPToolkit}, we compare our results against the recently proposed Nekrasov--Shatashvili method~\cite{Bonelli:2021uvf,Bautista:2023sdf,Cipriani:2025ikx}.
The Nekrasov--Shatashvili method utilizes a profound connection between the CHE \eqref{eq:HeunC-Eq} and classical conformal blocks in two-dimensional conformal field theory~\cite{Bonelli:2021uvf}.
By expressing the connection formulas of the Heun class functions in terms of Nekrasov--Shatashvili functions, which are special functions originating from $\mathcal{N}=2$ supersymmetric gauge theories, this method offers a novel physical picture: a clean near-far factorization in which far-zone scattering and near-zone horizon absorption are naturally decoupled~\cite{Bautista:2023sdf}.
While this insight is valuable for understanding the analytic structure of black-hole perturbations,
the practical evaluation of Nekrasov--Shatashvili functions involves intricate series expansions and connection formulas that are computationally demanding~\cite{Bautista:2023sdf}.

\cref{tab:Nekrasov--Shatashvili_Errors} presents a quantitative comparison of the three methods for computing GW fluxes.
Here, the parameter $N$ denotes the effective working precision (or truncation order) optimized for each algorithm: $N=16$ corresponds to standard double-precision arithmetic ($\sim 10^{-16}$) for our HeunC method, whereas the MST and Nekrasov--Shatashvili methods require higher working precision ($N=22$ and $N=40$, respectively) to stabilize their numerical procedures.
The results reveal a striking efficiency advantage for the HeunC method.
Operating at standard machine precision ($N=16$), our method consistently achieves relative errors of order $10^{-12}$ with a computational cost of merely $\sim 70$ ms across all orbital radii.
In contrast, both the MST method in the BHPToolkit and the Nekrasov--Shatashvili method require substantially higher working precision and correspondingly longer evaluation times ($\sim 10$--$57$ s) to reach comparable accuracy.
This represents an efficiency gain of approximately two orders of magnitude over existing methods for double-precision applications.
It is worth noting that the MST and Nekrasov--Shatashvili methods can, in principle, achieve higher accuracy ($<10^{-14}$) when run at even higher precision ($N > 40$), but at a prohibitive computational cost that scales super-linearly with $N$.
For most astrophysical applications, including GW template generation and parameter estimation, double-precision accuracy ($\sim 10^{-12}$) is entirely sufficient.
In this regime, the HeunC method provides an optimal efficiency-accuracy trade-off, enabling rapid, high-throughput computations without sacrificing reliability.
This makes it particularly well-suited for large-scale parameter space explorations, in which millions of flux evaluations may be required.

\begin{figure*}[htbp]
	\centering
\subfloat[Real part of ${\bf G}_{\ell m \omega}^{\infty}$.]{\includegraphics[width=3.4in]{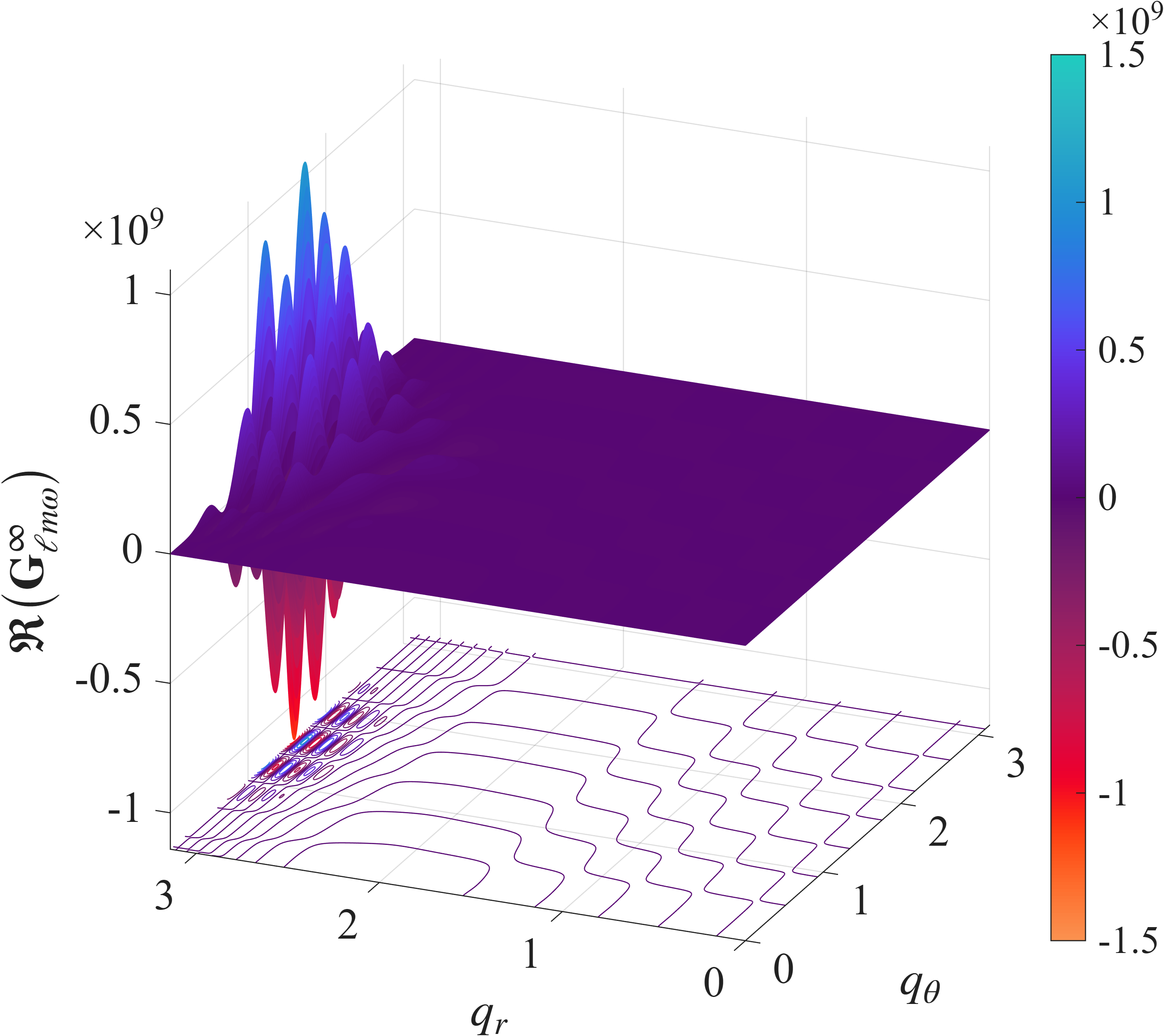}}\,\,
\subfloat[Imaginary part of ${\bf G}_{\ell m \omega}^{\infty}$.]{\includegraphics[width=3.4in]{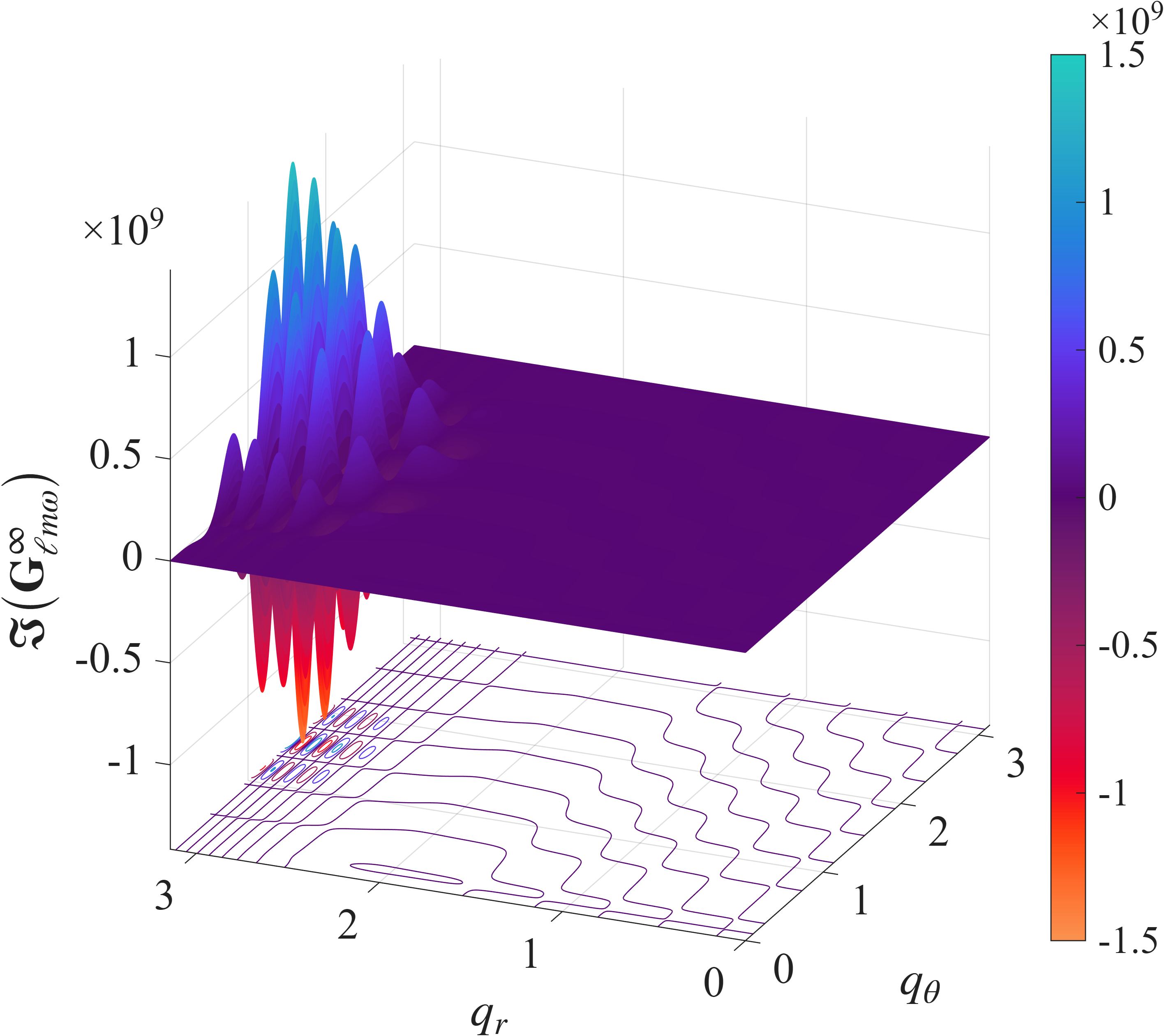}}\\
\subfloat[Real part of ${\bf G}_{\ell m \omega}^{{\rm{H}}}$.]{\includegraphics[width=3.4in]{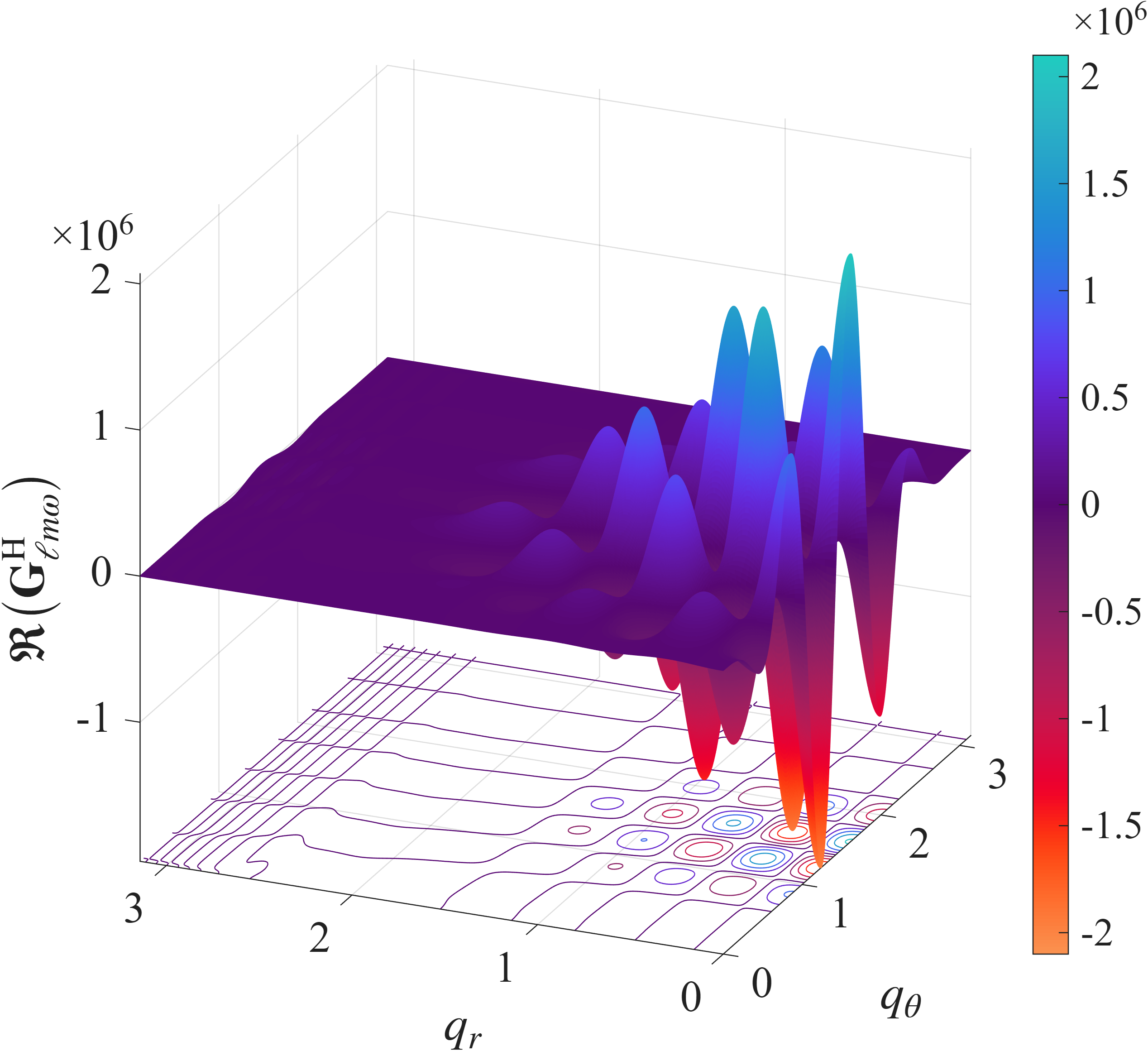}}\,\,
\subfloat[Imaginary part of ${\bf G}_{\ell m \omega}^{{\rm{H}}}$.]{\includegraphics[width=3.4in]{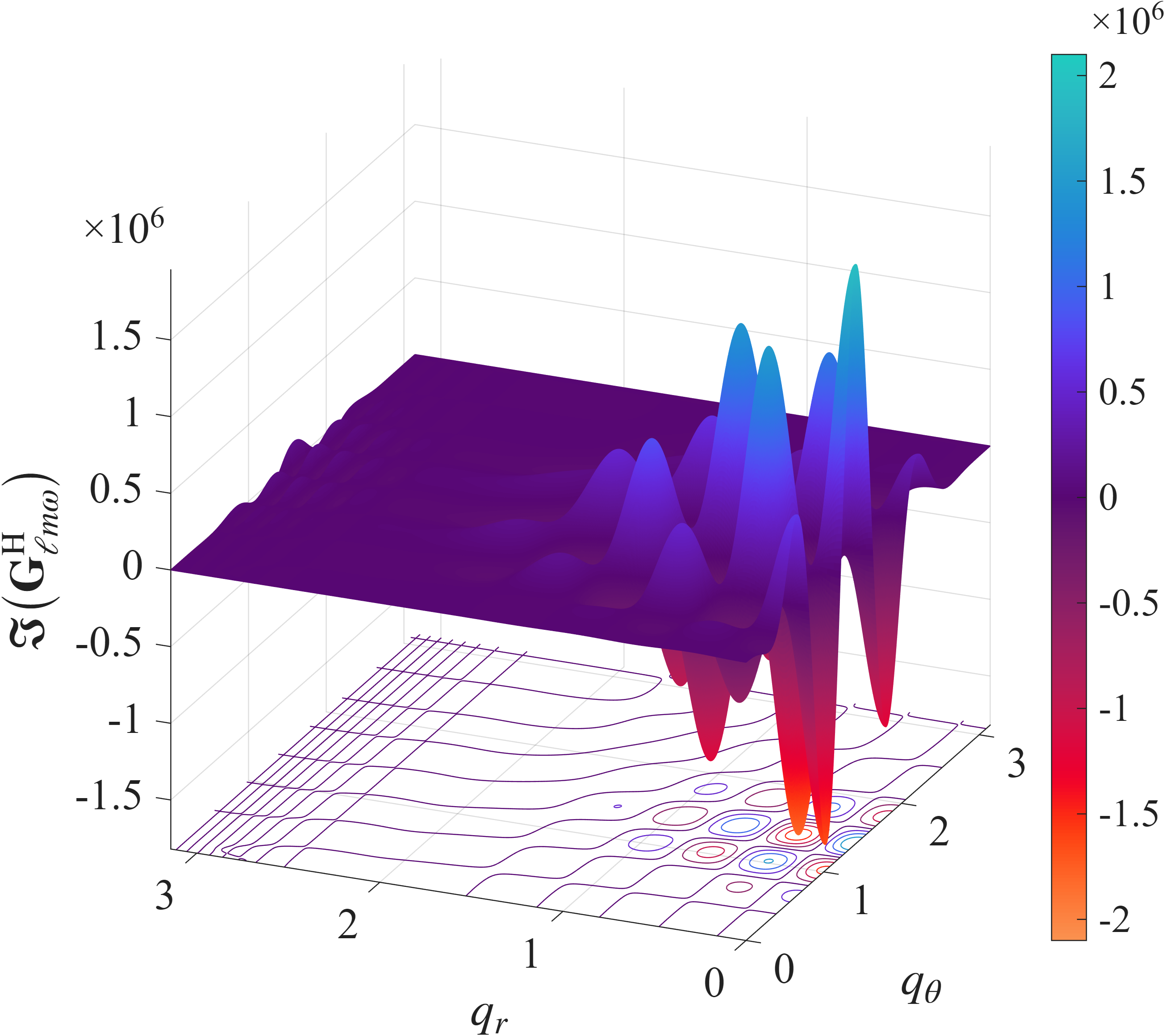}}
\caption{ Integrand ${\bf G}_{\ell m \omega}^{\infty, {\rm{H}}}$ with $(\ell,m,n,k)=(4,4,3,4)$ and $(a,p,e,x_I)=(0.9, 10, 0.7, 0.1)$.}\label{fig:Integrand4443}
\end{figure*}
\begin{figure*}[htbp]
	\centering
\subfloat[Real part of ${\bf G}_{\ell m \omega}^{\infty}$.]{\includegraphics[width=3.4in]{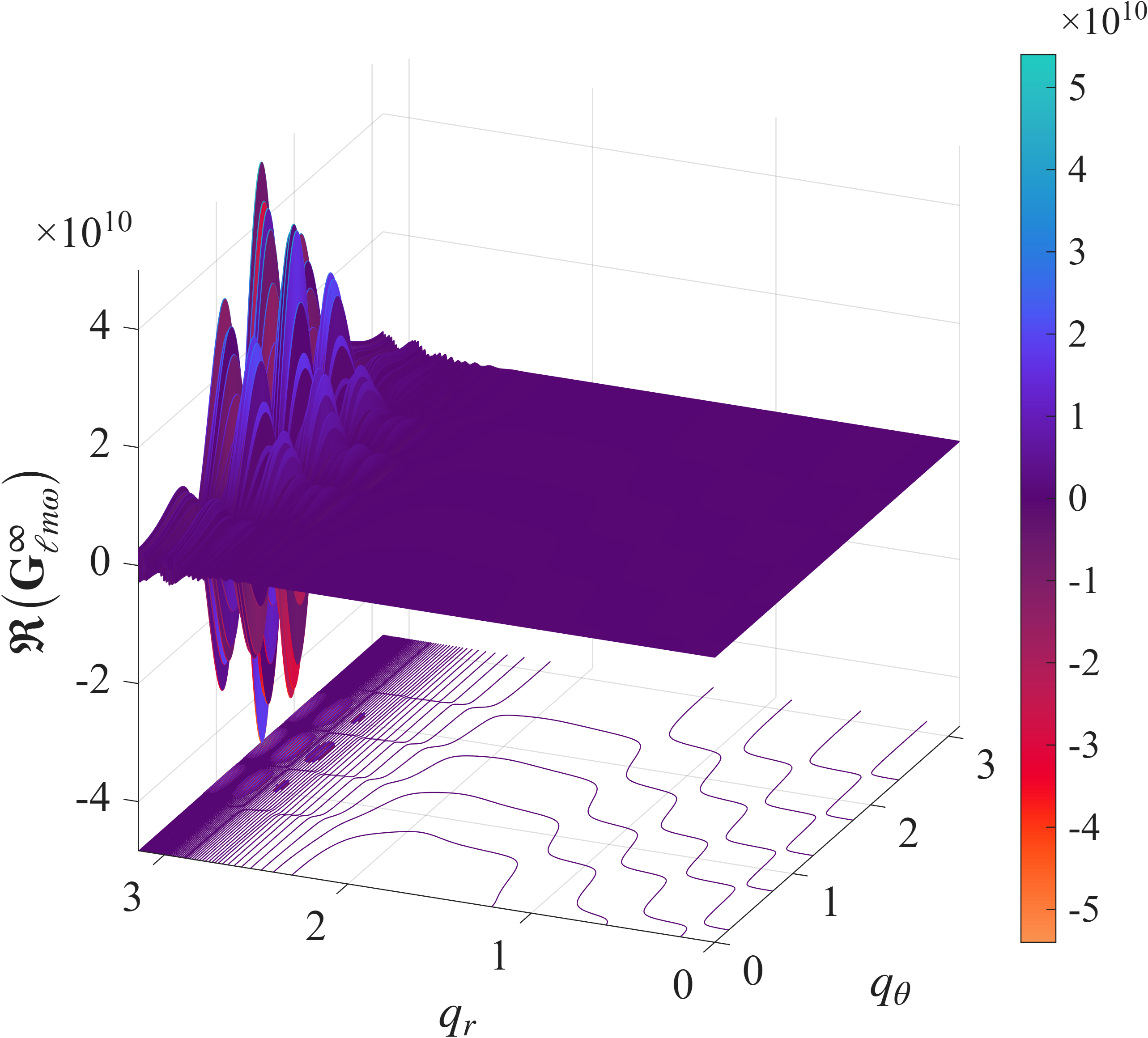}}\,\,
\subfloat[Imaginary part of ${\bf G}_{\ell m \omega}^{\infty}$.]{\includegraphics[width=3.4in]{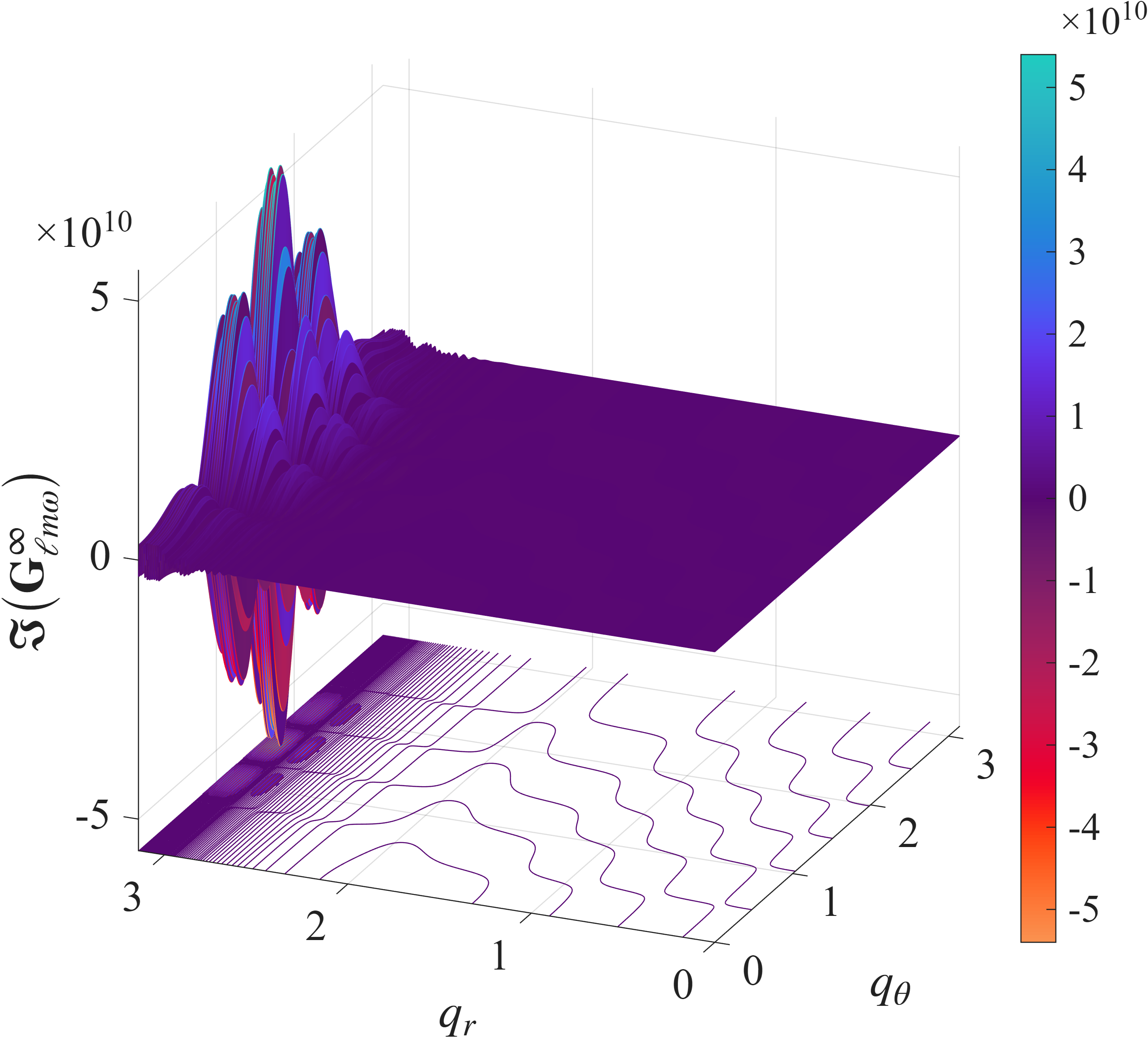}}\\
\subfloat[Real part of ${\bf G}_{\ell m \omega}^{{\rm{H}}}$.]{\includegraphics[width=3.4in]{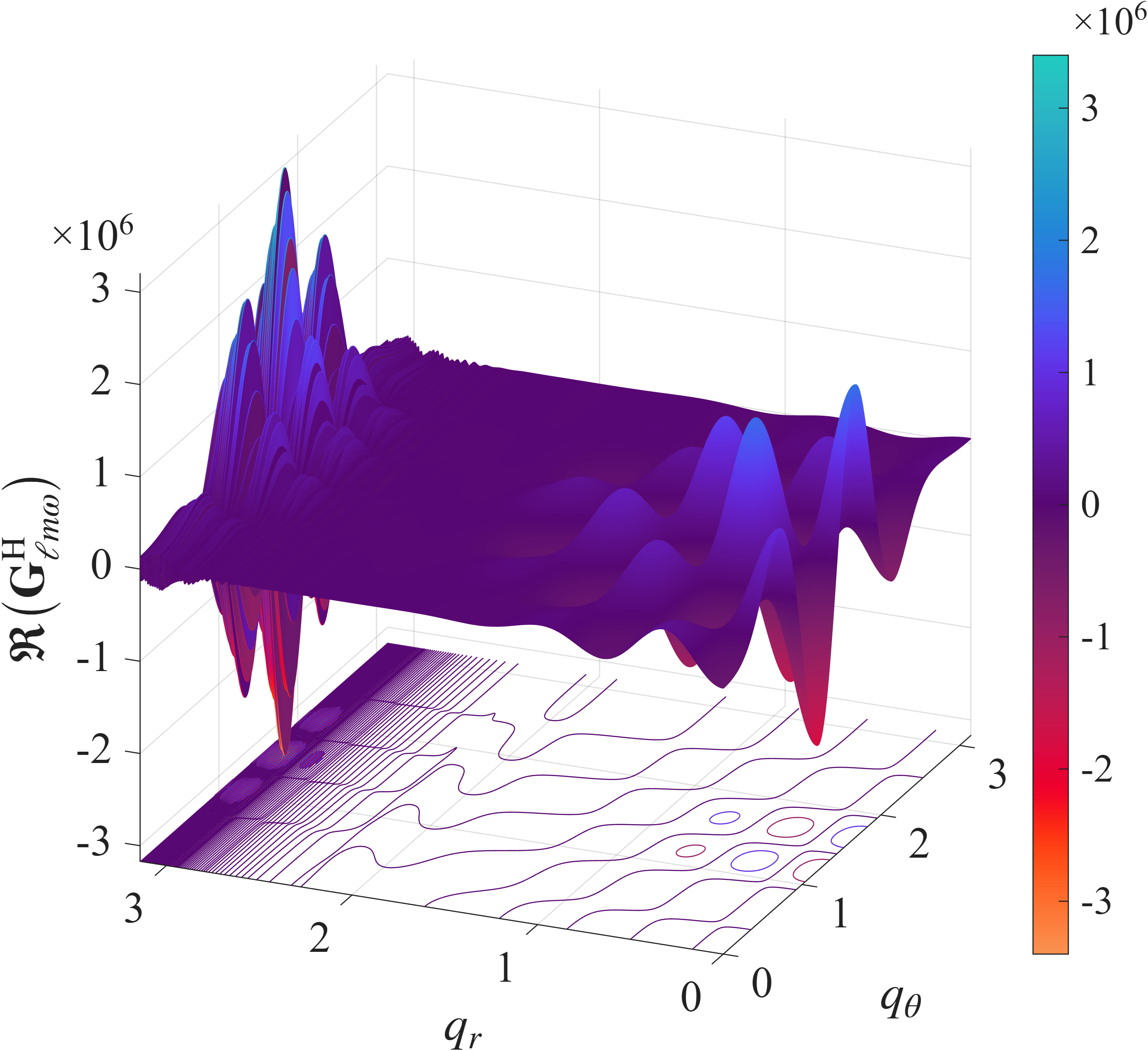}}\,\,
\subfloat[Imaginary part of ${\bf G}_{\ell m \omega}^{{\rm{H}}}$.]{\includegraphics[width=3.4in]{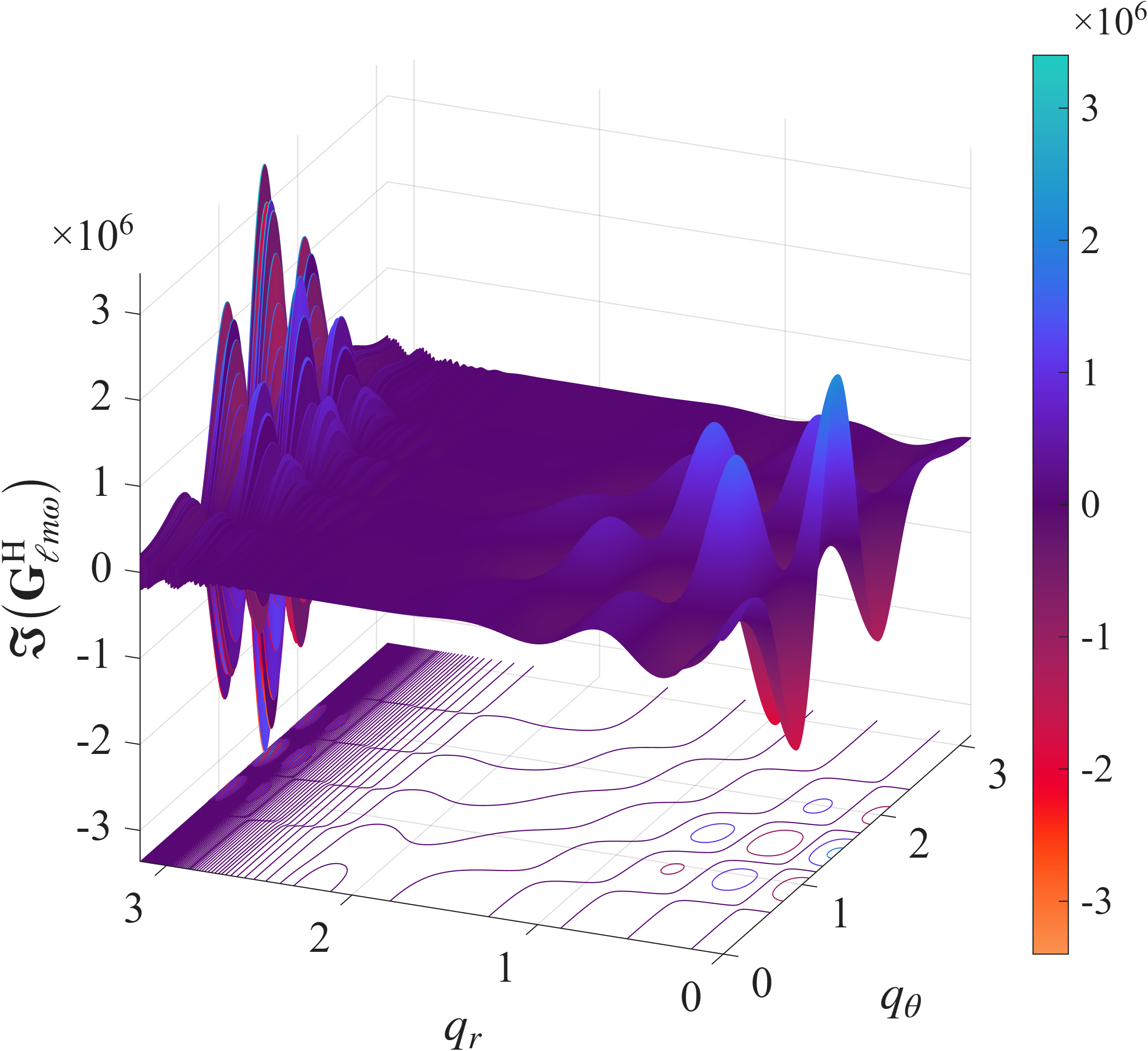}}
\caption{ Integrand ${\bf G}_{\ell m \omega}^{\infty, {\rm{H}}}$ with $(\ell,m,n,k)=(4,4,50,4)$ and $(a,p,e,x_I)=(0.9, 10, 0.9, 0.5)$.}\label{fig:Integrand44450}
\end{figure*}
\begin{figure*}[htbp]
	\centering
\subfloat[HeunC method.]{\includegraphics[width=3.4in]{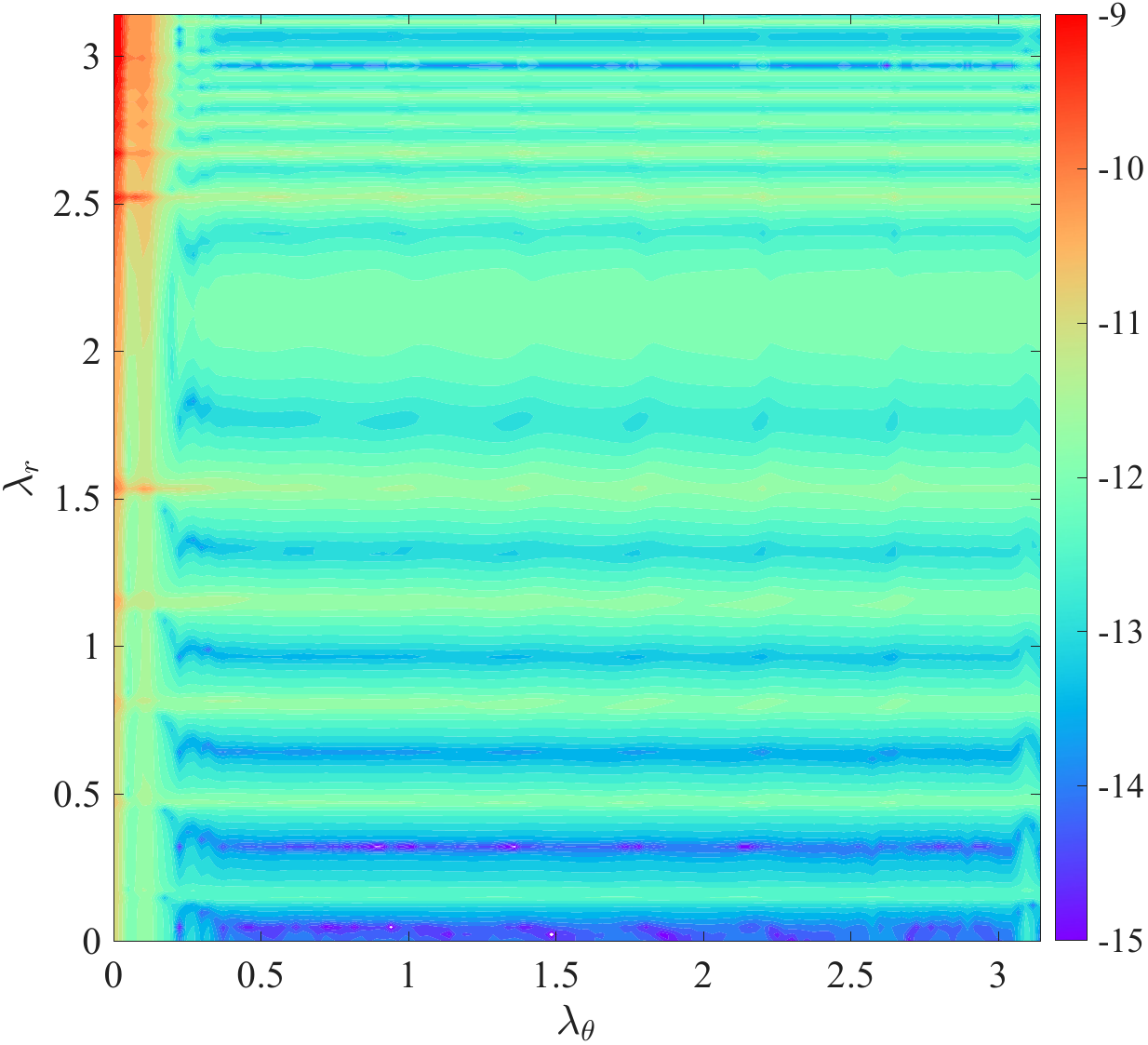}}\,\,
\subfloat[GSN method~\cite{GeneralizedSasakiNakamura}.]{\includegraphics[width=3.4in]{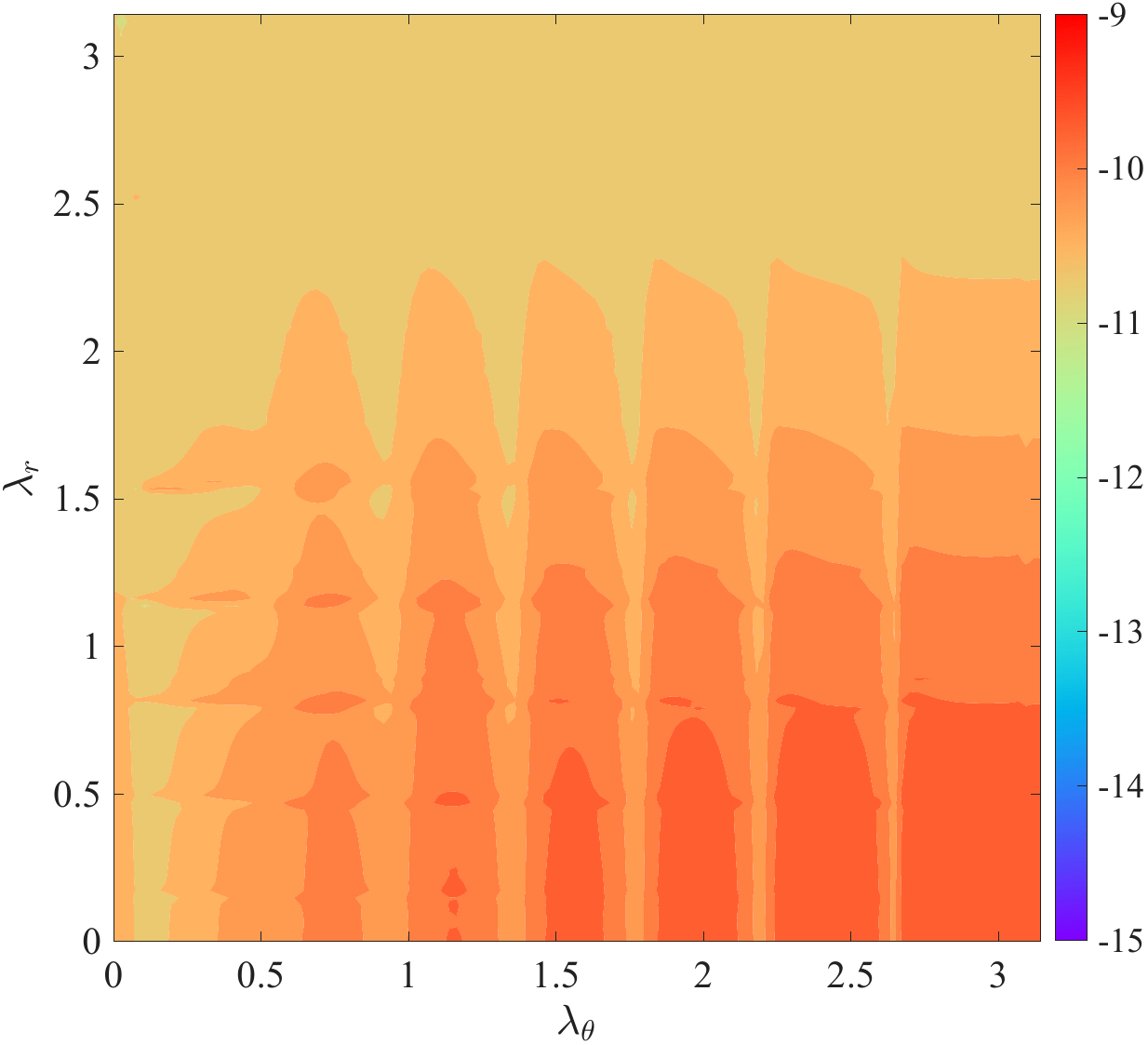}}
\caption{Logarithmic error $\log_{10}\blacktriangle \left| {{\bf{G}}_{\ell m \omega}^\infty } \right|$ for $(\ell,m,n,k)=(4,4,3,4)$ and $(a,p,e,x_I)=(0.1, 10, 0.7, 0.1)$. The integrand is evaluated on a $1000 \times 1000$ uniform grid in the orbital phase coordinates $(\lambda_r, \lambda_\theta)$.}
\label{fig:integrand_error}
\end{figure*}
\begin{table*}[htbp]
  \centering
  \caption{ Comparison of GW flux errors and computational cost. Scale $(\dot{\cal E}^\infty, \dot{\cal E}^{\rm H})$ denotes the order of magnitude of the absolute fluxes, respectively. $N_r$ and $N_\theta$ are the grid numbers for $\lambda_r$ and $\lambda_\theta$, respectively.}\label{tab:flux_comparison}
  \begin{tabular}{cccccccc}
    \toprule
    & $(l,m,n,k)$ & \multicolumn{2}{c}{ $(4,4,3,4)$} & & \multicolumn{2}{c}{$ (4,4,50,4)$} \\
    \cmidrule(r){3-4} \cmidrule(l){6-7}
    & $(a,p,e,x_I)$& $(0.1, 10, 0.7, 0.1)$ & $(0.9, 10, 0.7, 0.1)$ & & $(0.1, 10, 0.9, 0.5)$ & $(0.9, 10, 0.9, 0.5)$ \\ \cmidrule(){2-7}
    & Scale $(\dot {\cal E}^\infty, \dot {\cal E}^{\rm H})$ & $\sim(10^{-26}, 10^{-35})$ & $\sim( 10^{-19}, 10^{-25})$ & & $\sim (10^{-27}, 10^{-35})$ & $\sim (10^{-19}, 10^{-25})$ \\

    \midrule
    \multirow{4}{*}{HeunC Method}
    & $\blacktriangle\dot {\cal E}^\infty$ & $1.485 \times 10^{-7}$ & $4.058 \times 10^{-9}$ & & $9.501 \times 10^{-7}$ & $1.196 \times 10^{-10}$ \\
    & $\blacktriangle\dot {\cal E}^{\rm H}$ & $1.493 \times 10^{-7}$ & $6.625 \times 10^{-10}$ & & $3.650 \times 10^{-7}$ & $2.900 \times 10^{-9}$ \\
    & ${N_r} \times {N_\theta }$ & $60 \times 60$ & $64 \times 60$ & & $500 \times 200$ & $500 \times 128$ \\
    & Time(ms) & 59 & 98 & & 247 & 275 \\
    \midrule
    \multirow{4}{*}{GSN-Trapz~\cite{GeneralizedSasakiNakamura}}
    & $\blacktriangle\dot {\cal E}^\infty$ & $1.428 \times 10^{-5}$ & $3.229 \times 10^{-9}$ & & $1.274 \times 10^{-5}$ & $7.216 \times 10^{-10}$ \\
    & $\blacktriangle\dot {\cal E}^{\rm H}$ & $3.973 \times 10^{-7}$ & $8.900 \times 10^{-10}$ & & $1.978 \times 10^{-6}$ & $2.477 \times 10^{-10}$ \\
    & ${N_r} \times {N_\theta }$ & $64 \times 64$ & $64 \times 64$ & & $1024 \times 256$ & $512 \times 128$ \\
    & Time(ms) & 163 & 201 & & 764 & 292 \\
    \midrule
    \multirow{4}{*}{GSN-Levin~\cite{GeneralizedSasakiNakamura}}
    & $\blacktriangle\dot {\cal E}^\infty$ & $2.581 \times 10^{-6}$ & $8.485 \times 10^{-10}$ & & $2.081 \times 10^{-5}$ & $1.121 \times 10^{-8}$ \\
    & $\blacktriangle\dot {\cal E}^{\rm H}$ & $3.460 \times 10^{-7}$ & $3.643 \times 10^{-10}$ & & $3.683 \times 10^{-7}$ & $1.934 \times 10^{-9}$ \\
    & ${N_r} \times {N_\theta }$ & $256 \times 64$ & $256 \times 64$ & & $512 \times 256$ & $256 \times 32$ \\
    & Time(ms) & 1388 & 1240 & & 14975 & 449 \\
    \midrule
    \multirow{4}{*}{\texttt{pybhpt}~\cite{pybhpt}}
    & $\blacktriangle\dot {\cal E}^\infty$ & $2.954 \times 10^{-5}$ & $2.510 \times 10^{-9}$ & & $6.345 \times 10^{-5}$ & $1.616 \times 10^{-9}$ \\
    & $\blacktriangle\dot {\cal E}^{\rm H}$ & $3.034 \times 10^{-6}$ & $9.365 \times 10^{-10}$ & & $2.380 \times 10^{-6}$ & $4.736 \times 10^{-10}$ \\
    & ${N_r} \times {N_\theta }$ & $256 \times 256$ & $256 \times 256$ & & $2048 \times 2048$ & $2048 \times 2048$ \\
    & Time(ms) & 52 & 68 & & 790 & 770 \\
    \bottomrule
  \end{tabular}
\end{table*}

\begin{table*}[htbp]
  \centering
  \caption{{Comparison of GW flux errors and computational cost for the high-order $(l,m,n,k)=(8,8,50,1)$ with $e = 0.9$ and $x_I=0.005$.}}
  \label{tab:GWFflux-high_mode}
  \begin{tabular}{ccccccc}
    \toprule
         & $a$    & \multicolumn{2}{c}{$0.9$} &      & \multicolumn{2}{c}{$0.99$} \\
    \cmidrule{3-4}\cmidrule{6-7}
         & $p$    & $p_s{+}0.1$ & $p_s{+}0.05$ &      & $p_s{+}0.1$ & $p_s{+}0.05$ \\
    \cmidrule{2-7}
         & $(\dot{\mathcal{E}}^\infty, \dot{\mathcal{E}}^{\rm H})$ & $(10^{-22}, 10^{-27})$ & $(10^{-23}, 10^{-28})$ &      & $(10^{-21}, 10^{-26})$ & $(10^{-21}, 10^{-26})$ \\
    \midrule
    \multirow{4}[2]{*}{HeunC Method}
         & $\blacktriangle\dot {\cal E}^\infty$ & $3.894 \times 10^{-7}$ & $6.838 \times 10^{-6}$ &      & $8.191 \times 10^{-7}$ & $2.516 \times 10^{-6}$ \\
         & $\blacktriangle\dot {\cal E}^{\rm H}$ & $2.197 \times 10^{-7}$ & $7.774 \times 10^{-7}$ &      & $8.575 \times 10^{-7}$ & $3.938 \times 10^{-8}$ \\
         & ${N_r} \times {N_\theta }$ & $500 \times 60$ & $500 \times 60$ &      & $550 \times 60$ & $600 \times 60$ \\
         & Time(ms) & 225    & 280    &      & 578   & 364  \\
    \midrule
    \multirow{4}[2]{*}{GSN-Trapz~\cite{GeneralizedSasakiNakamura}}
         & $\blacktriangle\dot {\cal E}^\infty$ & $2.396 \times 10^{-5}$ & $2.396 \times 10^{-5}$ &      & $1.017 \times 10^{-6}$ & $7.560 \times 10^{-6}$ \\
         & $\blacktriangle\dot {\cal E}^{\rm H}$ & $2.682 \times 10^{-5}$ & $2.682 \times 10^{-5}$ &      & $3.607 \times 10^{-5}$ & $1.198 \times 10^{-5}$ \\
         & ${N_r} \times {N_\theta }$ & $\begin{array}{l}
2048 \times 64{\kern 1pt} {\kern 1pt} {\kern 1pt} (\dot{\mathcal{E}}^\infty)\\
512 \times 64{\kern 1pt} {\kern 1pt} {\kern 5pt} ( {\cal E}^{\rm H})
\end{array}$& $\begin{array}{l}
1024 \times 64{\kern 1pt} {\kern 1pt} {\kern 1pt} (\dot{\mathcal{E}}^\infty)\\
512 \times 64{\kern 1pt} {\kern 1pt} {\kern 5pt} ( {\cal E}^{\rm H})
\end{array}$ &      & $\begin{array}{l}
2048 \times 64{\kern 1pt} {\kern 1pt} {\kern 1pt} (\dot{\mathcal{E}}^\infty)\\
512 \times 64{\kern 1pt} {\kern 1pt} {\kern 5pt} ( {\cal E}^{\rm H})
\end{array}$ & $\begin{array}{l}
1024 \times 64{\kern 1pt} {\kern 1pt} {\kern 1pt} (\dot{\mathcal{E}}^\infty)\\
512 \times 64{\kern 1pt} {\kern 1pt} {\kern 5pt} ( {\cal E}^{\rm H})
\end{array}$ \\
         & Time(ms) & 396  & 364  &      & 867  & 784  \\
    \midrule
    \multirow{4}[2]{*}{GSN-Levin~\cite{GeneralizedSasakiNakamura}}
         & $\blacktriangle\dot {\cal E}^\infty$ & $1.269 \times 10^{-4}$ & $6.793 \times 10^{-1}$ &      & $1.263 \times 10^{-4}$ & $1.263 \times 10^{-4}$ \\
         & $\blacktriangle\dot {\cal E}^{\rm H}$ & $1.266 \times 10^{-4}$ & $6.793 \times 10^{-1}$ &      & $1.264 \times 10^{-4}$ & $1.264 \times 10^{-4}$ \\
         & ${N_r} \times {N_\theta }$ & $1024 \times 256$ & $1024 \times 256$ &      & $1024 \times 256$ & $1024 \times 256$ \\
         & Time(ms) & 98813  & 139233  &      & 98813  & 99753  \\
    \midrule
    \multirow{4}[2]{*}{\texttt{pybhpt}~\cite{pybhpt}}
         & $\blacktriangle\dot {\cal E}^\infty$ & $1.822 \times 10^{-7}$ & $2.954 \times 10^{-5}$ &      & $1.093 \times 10^{-8}$ & $3.982 \times 10^{-8}$ \\
         & $\blacktriangle\dot {\cal E}^{\rm H}$ & $2.111 \times 10^{-7}$ & $3.034 \times 10^{-6}$ &      & $9.964 \times 10^{-8}$ & $7.165 \times 10^{-7}$ \\
         & ${N_r} \times {N_\theta }$ & $2048 \times 2048$ & $2048 \times 2048$ &      & $2048 \times 2048$ & $2048 \times 2048$ \\
         & Time(ms) & 90000  & 26300  &      & 90000  & 93000  \\
    \bottomrule
  \end{tabular}
\end{table*}

\begin{table*}[htbp]
  \centering
  \caption{Comparison of error and computational time for different methods of the total radiative flux $\dot {\cal E}$.
  The parameters of the generic orbit at $p=10M$, $e=0.7$ and $x_I = 0.005$ are $\ell_{\max} = 5$, $1 \le n \le 3$, and $1 \le k \le 4$.}
  \label{tab:comparison}
  \begin{tabular}{ccccccccc}
    \toprule
    & \multicolumn{2}{c}{HeunC Method} & & \multicolumn{2}{c}{GSN Method~\cite{GeneralizedSasakiNakamura}} & & \multicolumn{2}{c}{\texttt{pybhpt}~\cite{pybhpt}} \\
    \cmidrule(r){2-3} \cmidrule(lr){5-6} \cmidrule(l){8-9}
    $a/M$ & Error & Time(s) & & Error & Time(s) & & Error & Time(s) \\
    \midrule
    0.1 & $2.610 \times 10^{-11}$ & 9.3  & & $3.911 \times 10^{-11}$ & 28.0 & & $1.169 \times 10^{-10}$ & 96.0 \\
    0.3 & $2.750 \times 10^{-11}$ & 9.5  & & $4.208 \times 10^{-11}$ & 35.4 & & $1.179 \times 10^{-10}$ & 126.0 \\
    0.5 & $3.487 \times 10^{-11}$ & 9.5  & & $4.127 \times 10^{-11}$ & 25.9 & & $1.172 \times 10^{-10}$ & 114.0 \\
    0.7 & $1.837 \times 10^{-11}$ & 10.5 & & $3.926 \times 10^{-11}$ & 28.9 & & $1.138 \times 10^{-10}$ & 39.2 \\
    0.9 & $2.987 \times 10^{-11}$ & 11.0 & & $4.135 \times 10^{-11}$ & 30.3 & & $1.084 \times 10^{-10}$ & 38.7 \\
    \bottomrule
  \end{tabular}
\end{table*}
\subsubsection{General orbits for Kerr geometry}\label{subsubsec:General_Kerr}
The most stringent test for any flux calculation method lies in generic (inclined and eccentric) orbits around a Kerr black hole. In this regime, the integrand ${\bf G}_{\ell m \omega}^{\infty ,{\rm{H}}}$ exhibits complex oscillatory behavior that poses significant challenges for numerical quadrature.
To illustrate these features, \Cref{fig:Integrand4443,fig:Integrand44450} display the real and imaginary parts of the integrand for the modes $(l,m,n,k)=(4,4,3,4)$ and $(4,4,50,4)$, respectively.
Several distinct characteristics emerge from the discretized profiles.
Here, $\lambda_r$ and $\lambda_\theta$ denote the decoupled Mino-time phases for the radial and polar motions, respectively, with $\lambda_r = \Upsilon_r \lambda$ ranging from $0$ to $2\pi$ over one radial period.
The periastron passage corresponds to $\lambda_r = 0$ (or equivalently $2\pi$), while apastron occurs at $\lambda_r = \pi$.
For the flux at infinity ${\bf G}_{\ell m \omega}^{\infty}$, oscillations are predominantly concentrated near $\lambda_r \to \pi$ (apastron) even for low-order indices, while the region near $\lambda_r \to 0$ remains relatively smooth.
As the harmonic indices increase (e.g., $n$ from 3 to 50), the oscillation frequency near $\lambda_r \to \pi$ intensifies dramatically, requiring dense sampling to resolve the rapid phase variations.
The flux at the event horizon ${\bf G}_{\ell m \omega}^{{\rm{H}}}$ exhibits markedly different behavior.
For low-order indices, it shows slight oscillation across the domain from $\lambda_r=0$ to the intermediate region ($\lambda_r \approx \pi/2$), with only negligible undulations near $\lambda_r \to \pi$.
As the indices increase significantly, these undulations near $\lambda_r \to \pi$ become more pronounced, whereas the $0 \to \pi/2$ region remains smooth and gently undulating.

To quantify the numerical advantages of the proposed framework, we systematically compare our results against established implementations, including the GSN method~\cite{GeneralizedSasakiNakamura} and the \texttt{pybhpt} package~\cite{pybhpt}.
All benchmarks are performed in standard double precision.
We first examine the pointwise accuracy of the integrand.
\cref{fig:integrand_error} displays the logarithmic error distribution $\log_{10} \blacktriangle |{\bf G}_{\ell m \omega}^\infty|$ across the phase space.
The HeunC method achieves excellent accuracy across the integration domain, with errors typically between $10^{-15}$ and $10^{-10}$, except near the polar boundary $\lambda_\theta \to 0$ where coordinate singularities naturally increase the error.
In stark contrast, the GSN method exhibits errors that are systematically larger by several orders of magnitude throughout the domain, typically ranging from $10^{-11}$ to $10^{-9}$.
For clarity, this comparison is restricted to the asymptotic integrand ${\bf G}_{\ell m \omega}^\infty$.
This is because the GSN code computes the horizon contribution ${\bf G}_{\ell m \omega}^{\rm H}$ using a spin weight of $s=+2$ and distinct normalization conventions, which would introduce additional transformation steps and complicate a direct pointwise correspondence.
Focusing on ${\bf G}_{\ell m \omega}^\infty$ (where both methods use $s=-2$) allows for a rigorous and unambiguous assessment of the underlying quadrature performance.

The integrated flux accuracy and computational efficiency are summarized in \cref{tab:flux_comparison} for representative the modes $(4,4,3,4)$ and $(4,4,50,4)$. In constructing this comparison, we aim to provide a fair assessment by configuring each method to achieve comparable accuracy with similar grid resolutions where possible.
However, it should be noted that the GSN and \texttt{pybhpt} implementations constrain their grid sizes to powers of two ($2^n$), limiting the flexibility in grid selection.
Our HeunC method employs the trapezoidal method for numerical integration: for low-order modes, we use a uniform grid, whereas for high-order modes, we adopt an adaptive grid strategy as detailed in Appendix~\ref{app:bi_power_mapping}.
The reported runtimes reflect these different quadrature strategies: our implementation employs the bi-power mapping with the trapezoidal rule over the reduced domain $[0,\pi]$~\footnote{Due to the periodicity of the integrand ${\bf G}_{\ell m}^{\infty ,{\rm{H}}}$ in \cref{eq:Jlmnk_def}, the full integral can be obtained by integrating over $[0,\pi]$ and multiplying the result by 4.}, GSN-Trapz uses a uniform trapezoidal rule over $[0,\pi]$, GSN-Levin utilizes a Levin method~\cite{Lo:2023fvv} specialized for oscillatory integrals over $[0,\pi]$, and \texttt{pybhpt} applies a uniform trapezoidal rule over the full period $[0,2\pi]$.

{
To satisfy the requirement for a rigorous stress-test under extreme conditions, we evaluate the high-order mode $(l,m,n,k)=(8,8,50,1)$ for highly eccentric ($e=0.9$) near-polar orbits ($x_I=0.005$) approaching the separatrix~\cite{Stein:2019buj}. The separatrix $p_s$, defining the limit of stable bound orbits, is computed via the \texttt{KerrGeoSeparatrix} function in the BHPtoolkit~\cite{BHPToolkit}.The results, summarized in \cref{tab:GWFflux-high_mode}, reveal the limitations of existing numerical schemes in the ``whirl-and-zoom'' regime. While \texttt{pybhpt} provides high precision, its computational cost escalates to approximately $90$~s per mode due to arbitrary-precision overhead. In contrast, our HeunC framework maintains sub-second runtimes (200--600 ms), achieving a speedup of over $250$ times compared to \texttt{pybhpt}. Furthermore, our method consistently outperforms GSN-Trapz in both speed and accuracy; for the $a=0.9, p=p_s+0.05$ case, HeunC is not only faster but also two orders of magnitude more precise ($\mathcal{O}(10^{-7})$ vs $\mathcal{O}(10^{-5})$).Crucially, we find that the HeunC framework with adaptive bi-power mapping remains robust where other methods struggle. For instance, the GSN-Levin method becomes numerically unstable at $e=0.9$, yielding relative errors as high as $10^{-1}$. Finally, we focus on the $k=1$ mode for this near-polar benchmark, as high-$k$ modes currently exhibit divergent results across all tested packages, likely due to unresolved polar angular structures.
}

Despite these methodological differences, the HeunC framework demonstrates robust performance across all tested configurations.
As illustrated in \cref{fig:integrand_error}, the HeunC method maintains superior pointwise accuracy of the integrand across the entire phase space, which directly translates into more accurate flux calculations.
For the mode $(4,4,3,4)$ with $a=0.1$, the HeunC method not only achieves a $\blacktriangle\dot{\cal E}^\infty$ approximately two orders of magnitude smaller than that of the GSN-Trapz method, but also demonstrates superior computational efficiency, requiring only 59~ms compared to 163~ms for GSN-Trapz.
Notably, as the radial harmonic index increases to $n=50$, the HeunC method's computational cost scales favorably, maintaining competitive runtimes while achieving higher accuracy than both GSN variants. This demonstrates the effectiveness of our adaptive integration strategy in resolving highly oscillatory high-order modes where traditional uniform grids struggle.

A key finding from \cref{tab:flux_comparison} is the exceptional efficiency of the HeunC framework in the high-frequency regime, which warrants a brief comment. For the highly oscillatory mode $(4,4,50,4)$, the GSN-Levin method requires $14975$~ms to converge. This is because it converts oscillatory quadrature into a Chebyshev spectral collocation problem, incurring an $\mathcal{O}(N^3)$ overhead that becomes prohibitive as the grid resolution increases to resolve high harmonic indices. In contrast, our bi-power mapping method relies on an analytic coordinate transformation coupled with a uniform trapezoidal rule. This formulation yields linear scaling with grid resolution and negligible algebraic overhead, enabling rapid convergence in only $247$~ms. This approximately 60-fold speedup, combined with the superior precision shown in \cref{tab:flux_comparison}, highlights the robustness of our adaptive strategy. Even compared to the basic GSN-Trapz and \texttt{pybhpt} implementations, the HeunC framework consistently delivers better precision with significantly lower computational overhead, making it ideal for large-scale data generation where high-order modes predominate.

The cumulative performance is further validated in \cref{tab:comparison}, which reports the total radiative flux $\dot {\cal E}$ summed over low-order modes $(\ell,m,n,k)$ for a near-polar orbit ($\iota \approx 89.71^\circ$, corresponding to $x_I = 0.005$).
We restrict the comparison to this subset because generating the high-precision reference solutions with the BHPToolkit becomes increasingly costly as the harmonic indices grow, making a full high-order benchmark computationally prohibitive.
The HeunC method maintains consistent precision across the full spin range, achieving relative errors of order $10^{-11}$---consistently lower than the GSN method and approximately an order of magnitude smaller than \texttt{pybhpt}.
More importantly, the HeunC framework demonstrates superior computational efficiency: it is approximately 3--4 times faster than the GSN method and 3--13 times faster than \texttt{pybhpt}, depending on the spin parameter.
Crucially, the HeunC method exhibits exceptional stability: its runtime varies by less than 20\% across the entire spin range $0.1 \leq a/M \leq 0.9$, whereas \texttt{pybhpt} shows fluctuations exceeding 200\% (from $\sim$126\,s at $a=0.3$ to $\sim$39\,s at $a=0.9$), reflecting its sensitivity to the underlying arbitrary-precision arithmetic overhead.
This robustness, combined with its favorable scaling with mode number, makes the HeunC framework particularly well-suited for assembling complete energy fluxes that require summation over thousands of high-order modes.
Taken together, these benchmarks confirm that the proposed framework achieves a favorable balance of numerical stability and computational efficiency, providing a reliable tool for strong-field flux calculations across generic Kerr orbits.

\begin{figure*}[htbp]
	\centering
\subfloat[$\ell=m=n=k=2$]{\includegraphics[width=3.4in]{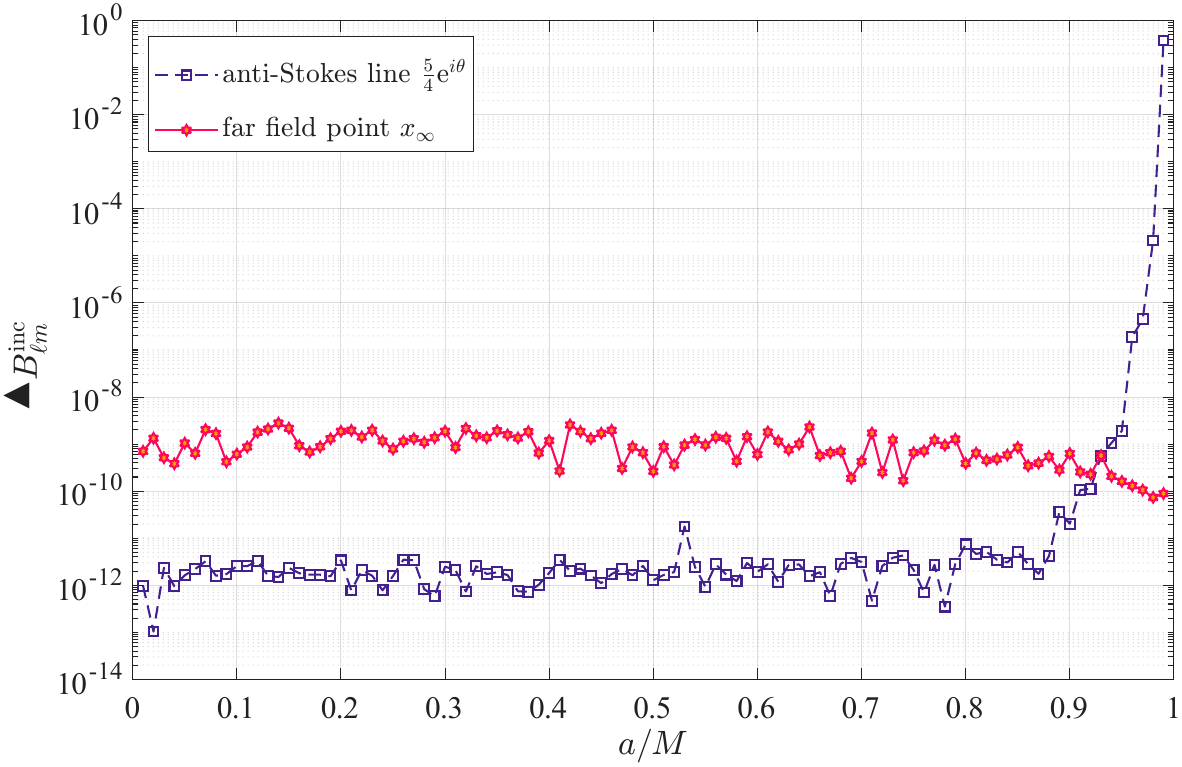}}\,\,
\subfloat[$\ell=m=n=k=4$.]{\includegraphics[width=3.4in]{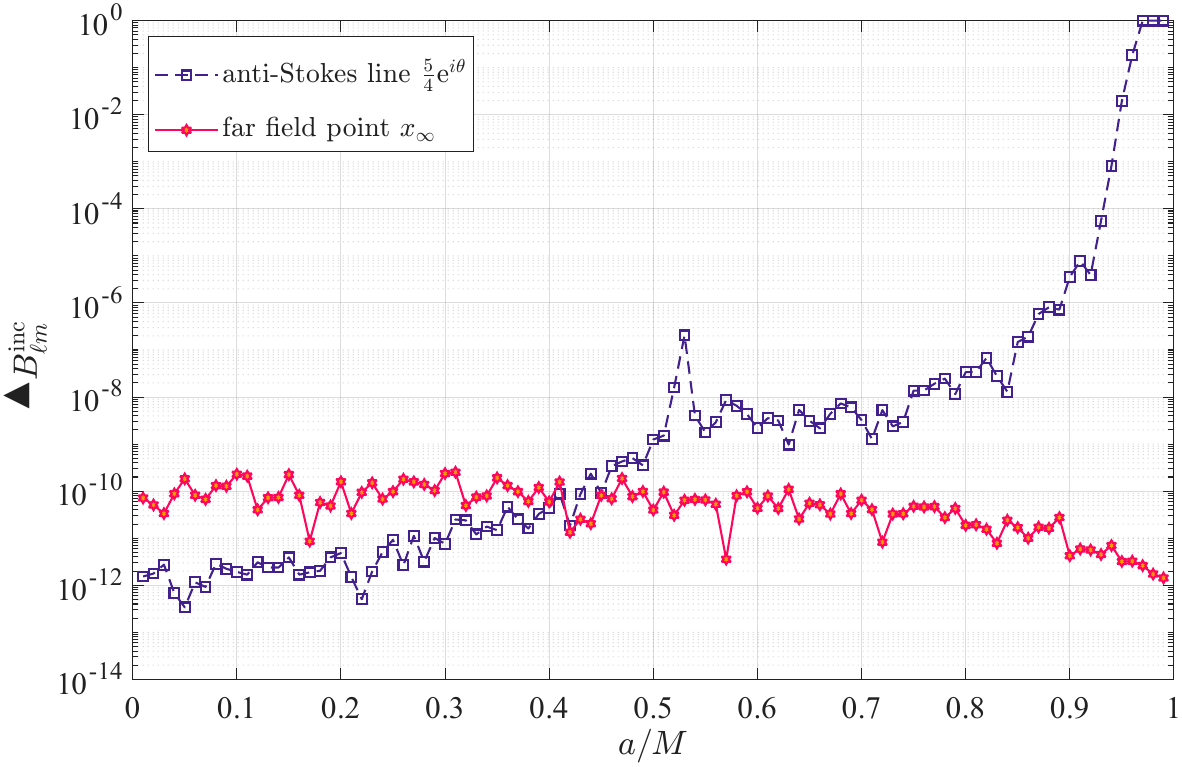}}\\
\subfloat[$\ell=m=n=k=6$.]{\includegraphics[width=3.4in]{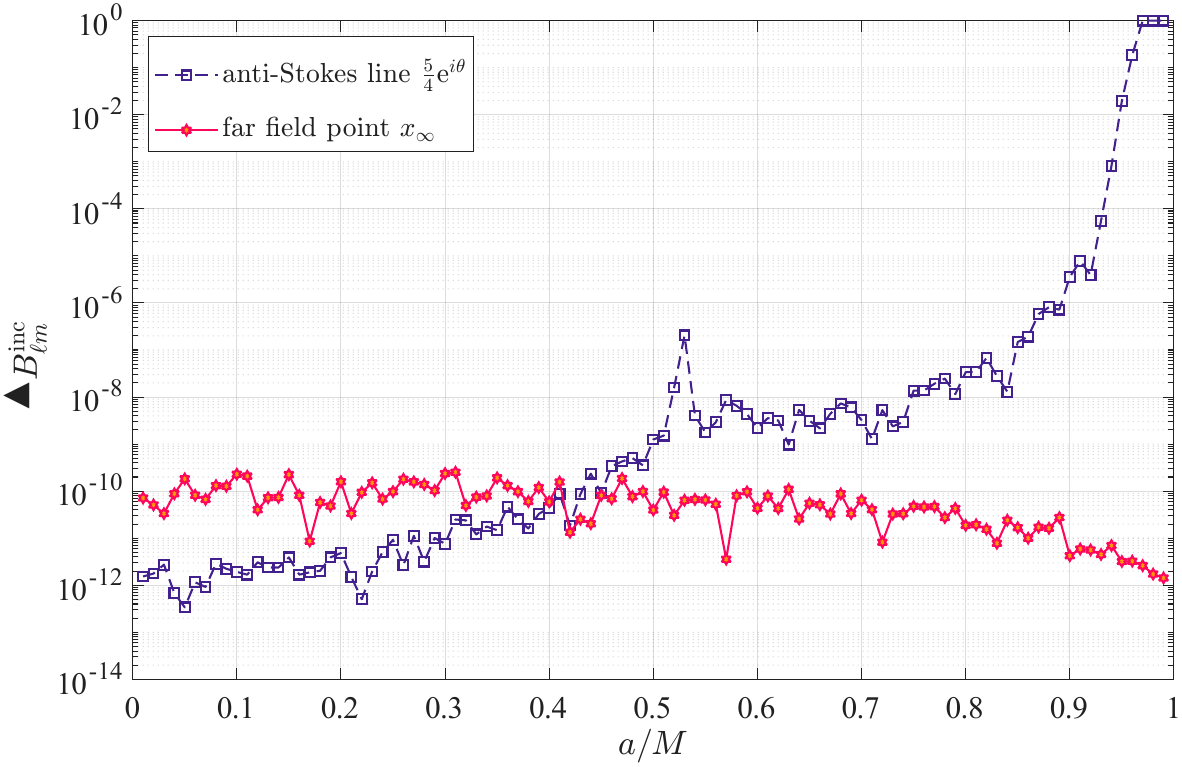}}\,\,
\subfloat[$\ell=m=n=k=8$.]{\includegraphics[width=3.4in]{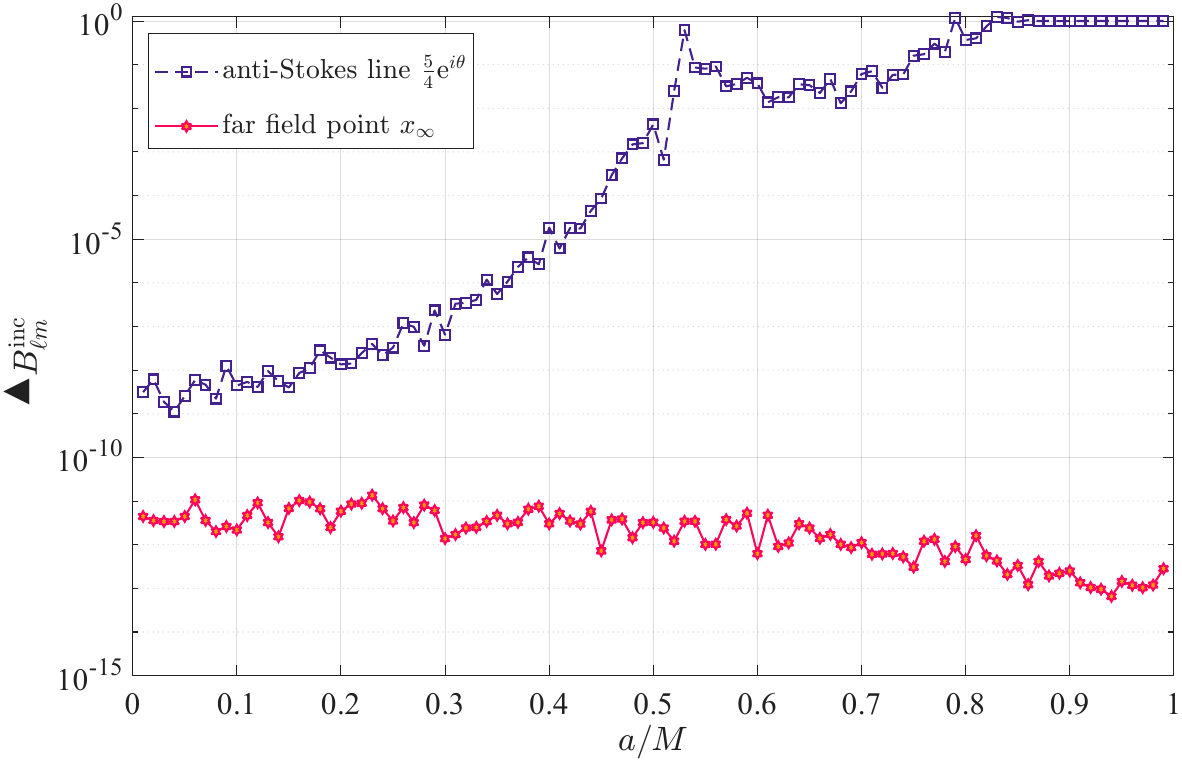}}
\caption{{ Logarithmic errors of the amplitude ${B_{\ell m}^{\rm inc}(\omega)}$ for the modes $\ell=\{2,4,6,8\}$ with $a=e\in[0,0.99]$ and $(p,x_I)=(p_s+0.1, 0.5)$. The adaptive far-field strategy (red solid) remains stable across all modes, whereas fixed matching at the anti-Stokes line (blue dashed) fails for high-order harmonics (panel d).}
}\label{fig:LogErrAmp}
\end{figure*}

\begin{figure*}[htbp]
	\centering
\subfloat[QNM eigenvalues.\label{fig:qnm_eigenvalues}]{\includegraphics[width=3.48in]{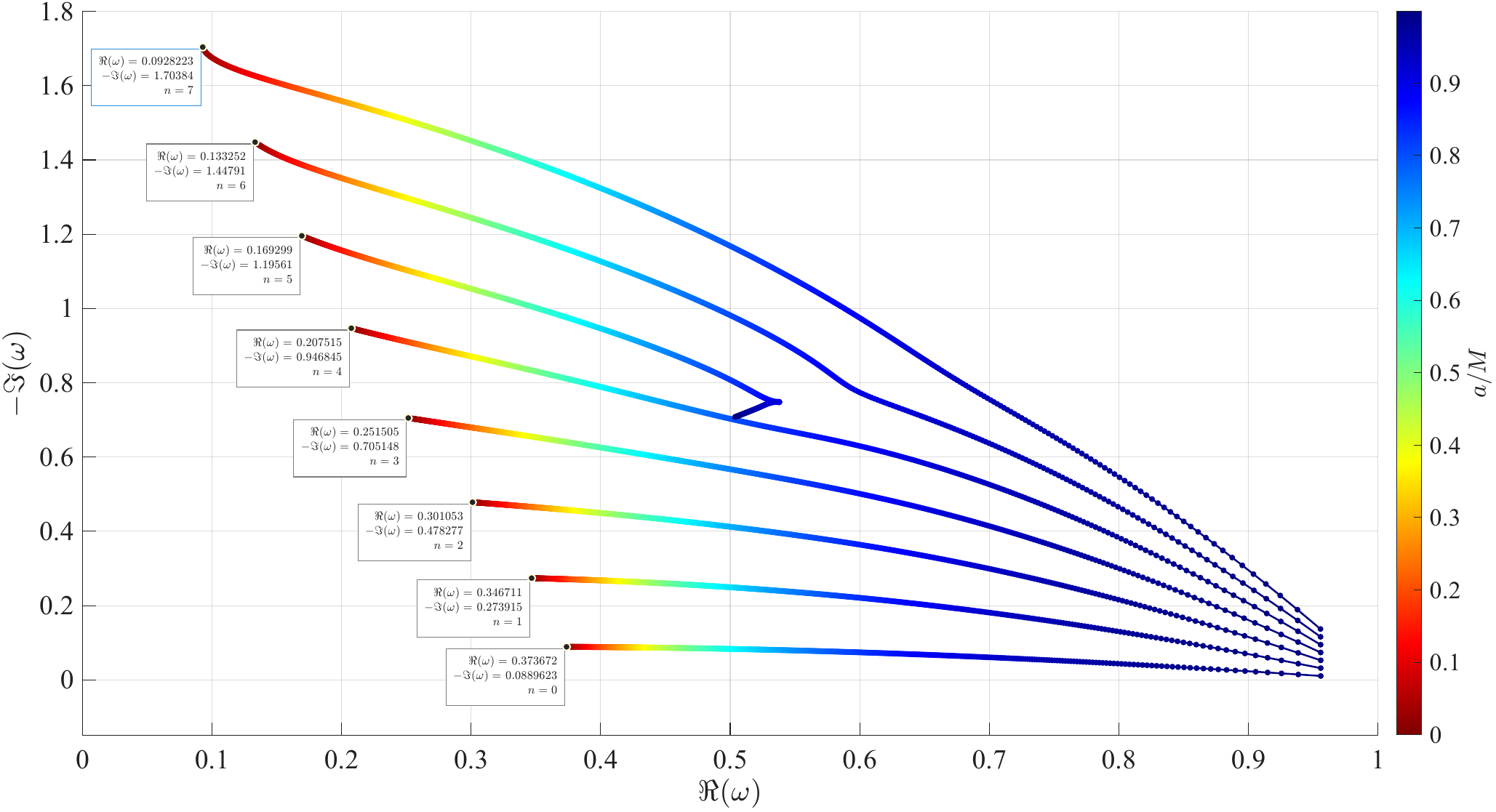}}\,\,
\subfloat[Errors.\label{fig:qnm_errors}]{\includegraphics[width=3.4in]{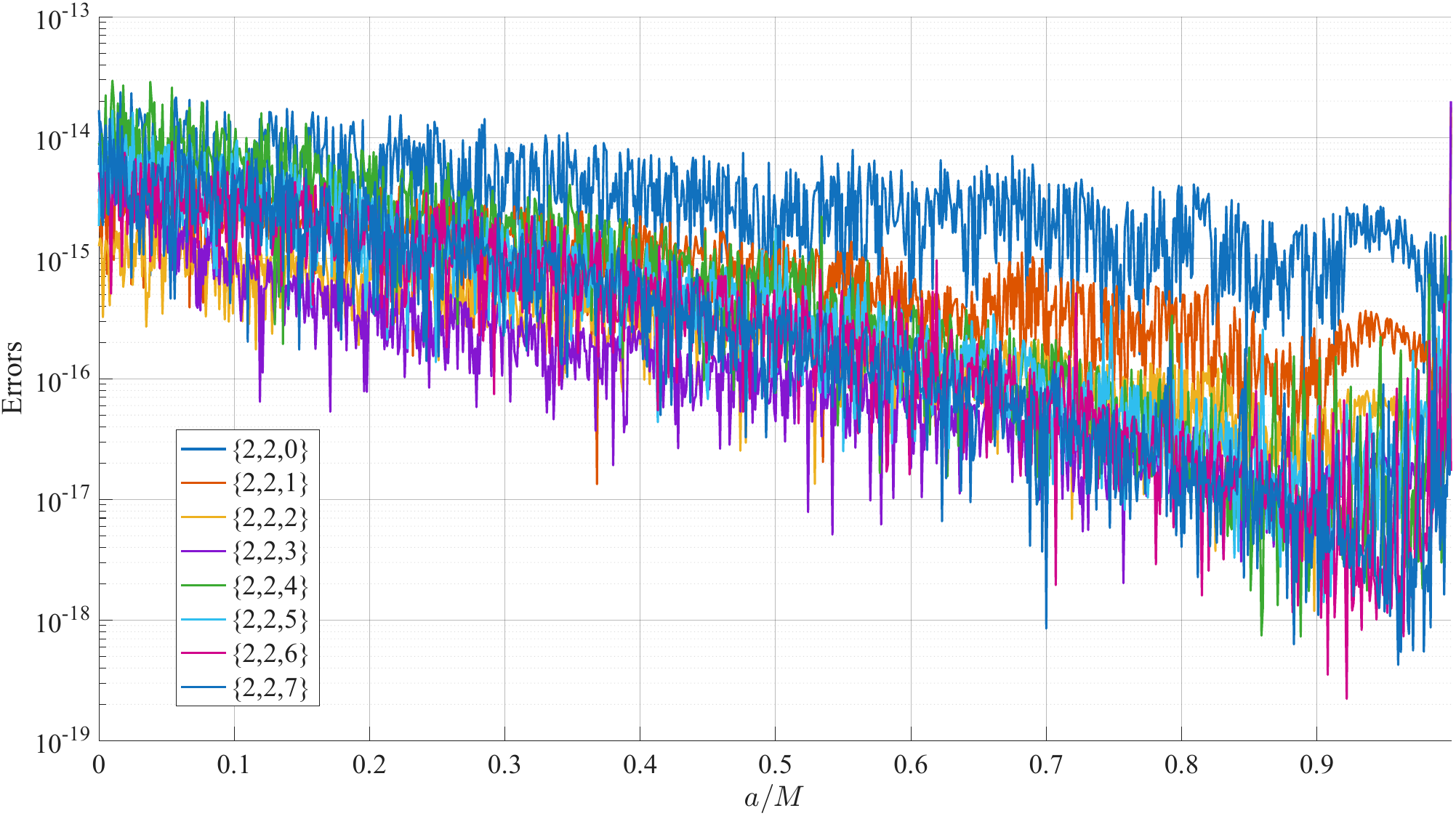}}\\
\caption{QNM eigenvalues and errors with $\ell = m=2$, $0\leq n \leq 7$ and $a \in [0,0.999]$.}\label{fig:QNM_Err}
\end{figure*}
\begin{table}[htbp]
  \centering
  \caption{Comparison of the normalized amplitude residuals $\mathcal{R}_{\rm amp}$ for the QNM sequence $\ell = m=2$, $0 \leq \hat{n} \leq 7$. The $L_2$ and $L_\infty$ norms are computed by sampling the spin range $a \in [0, 0.999]$.}
    \begin{tabular}{ccccc}
    \toprule
    & \multicolumn{2}{c}{BHPToolkit (MST)} & \multicolumn{2}{c}{HeunC Method} \\
    \cmidrule(lr){2-3} \cmidrule(lr){4-5}
    $\hat{n}$ & $L_2$-Residual & $L_\infty$-Residual & $L_2$-Residual & $L_\infty$-Residual \\
    \midrule
    0 & $1.155 \times 10^{-14}$ & $1.116 \times 10^{-14}$ & $1.720 \times 10^{-13}$ & $2.380 \times 10^{-14}$ \\
    1 & $1.576 \times 10^{-14}$ & $1.574 \times 10^{-14}$ & $5.390 \times 10^{-14}$ & $1.486 \times 10^{-14}$ \\
    2 & $1.908 \times 10^{-14}$ & $1.908 \times 10^{-14}$ & $2.438 \times 10^{-14}$ & $1.782 \times 10^{-14}$ \\
    3 & $2.175 \times 10^{-14}$ & $2.171 \times 10^{-14}$ & $3.090 \times 10^{-14}$ & $1.998 \times 10^{-14}$ \\
    4 & $5.882 \times 10^{-13}$ & $1.016 \times 10^{-13}$ & $1.381 \times 10^{-13}$ & $2.957 \times 10^{-14}$ \\
    5 & $7.270 \times 10^{-14}$ & $1.673 \times 10^{-14}$ & $7.290 \times 10^{-14}$ & $1.675 \times 10^{-14}$ \\
    6 & $5.035 \times 10^{-14}$ & $8.800 \times 10^{-15}$ & $5.076 \times 10^{-14}$ & $9.203 \times 10^{-15}$ \\
    7 & $2.988 \times 10^{-14}$ & $8.672 \times 10^{-15}$ & $6.699 \times 10^{-14}$ & $1.685 \times 10^{-14}$ \\
    \bottomrule
    \end{tabular}
  \label{tab:QNM_error_comparison}
\end{table}

\subsection{QNM tests}\label{subsec:ringdown}
QNMs characterize the ringdown phase of perturbed black holes and provide a stringent test for the numerical accuracy of the scattering amplitude framework. We validate our HeunC framework by evaluating the asymptotic amplitudes at the discrete QNM frequencies obtained from Ref.~\cite{Chen:2025sbz}.
At a QNM frequency $\omega_{\rm QNM}$, the homogeneous RTE solution must satisfy purely ingoing boundary conditions at the event horizon and purely outgoing conditions at infinity. This physical requirement implies that the incident amplitude should vanish, i.e., $B_{\ell m}^{\rm inc}(\omega_{\rm QNM}) = 0$. To quantify the numerical violation of this condition, we define the normalized incident amplitude residual as
\begin{equation}
  \mathcal{R}_{\rm amp} \equiv \left| \frac{B_{\ell m}^{\rm inc}(\omega)}{B_{\ell m}^{\rm ref}(\omega)} \right|.
\end{equation}
{ Since the accurate evaluation of $B_{\ell m}^{\text{inc}}$ is highly sensitive to the numerical stability of the matching procedure described in \cref{subsec:teukolsky_heun}, the QNM test serves as an ideal benchmark for validating our adaptive matching strategy.
To demonstrate the necessity of this approach, we compare the relative errors of the incident amplitude in \cref{fig:LogErrAmp},
contrasting a fixed matching point strategy—specifically $x_\pm^\star = -1.25$, which corresponds to the anti-Stokes line on the negative real axis for real orbital frequencies—with our dynamic far-field point strategy ($x_\infty = {\mathfrak{R}}\mathrm{e}^{i\theta}/{|\tilde{\epsilon}|}$).
As illustrated across the four panels of \cref{fig:LogErrAmp}, while the fixed-point strategy maintains acceptable precision for low-order modes such as $\ell=2$, it begins to degrade significantly in the high-spin regime.
For higher-order harmonics such as the $(\ell, m, n, k) = (8, 8, 8, 8)$ mode, the fixed-point strategy exhibits a catastrophic failure as the black hole spin increases, with errors escalating to $\mathcal{O}(1)$.
This numerical instability arises because the increased centrifugal potential barrier renders the near-horizon Frobenius series ill-conditioned when matched at small radii.
By contrast, matching at the adaptive far-field point $x_\infty$ yields a stable relative error of order $10^{-12}$ regardless of the spin or mode order.  }

The numerical performance across the full spin range $a \in [0, 0.999]$ for the dominant $\ell=m=2$ modes (with overtones $0 \leq \hat{n} \leq 7$) is summarized in \cref{fig:QNM_Err} and \cref{tab:QNM_error_comparison}. Here, the $L_2$ and $L_\infty$ metrics characterize the average and maximum residuals, respectively, sampled over the spin parameter $a$.
As shown in \cref{tab:QNM_error_comparison}, the residuals remain bounded between $10^{-18}$ and $10^{-14}$ across the entire parameter space, demonstrating exceptional numerical stability even in the near-extremal regime. A direct comparison with the MST method (as implemented in the BHPToolkit) in \cref{tab:QNM_error_comparison} confirms that the HeunC framework achieves comparable machine-precision accuracy, maintaining residuals well below the requirements for high-precision GW data analysis.

The robust performance of the HeunC framework arises from its direct treatment of the RTE.
By expressing the homogeneous RTE solutions in terms of HeunC functions, we enforce the correct asymptotic boundary conditions without relying on recurrence relations or continued-fraction expansions that often suffer from numerical stiffness. Furthermore, the solver directly yields the asymptotic amplitudes required to compute QNM excitation factors, which quantify the coupling strength between a perturbation source and each QNM~\cite{Berti:2006wq,Zhang:2013ksa,DellaRocca:2025zbe}.
Unlike traditional approaches that require evaluating two-sided infinite series and solving for the renormalized angular momentum $\nu$, the HeunC connection formulas provide a closed analytical pathway to the scattering amplitudes, eliminating redundant computational overhead and avoiding branch-cut instabilities associated with $\nu$-search algorithms~\cite{Berti:2006wq,Zhang:2013ksa,DellaRocca:2025zbe}.

Combined with the complete QNM spectra and angular separation constants established in our previous work~\cite{Chen:2025sbz}, the present method supplies the missing radial scattering amplitudes \eqref{eq:Rin_Amp} required to construct full ringdown waveforms.
While the former study efficiently resolved the eigenvalue problem across branch cuts and highly damped regimes, it did not yield the non-vanishing asymptotic coefficients that determine mode excitation strengths. The current framework directly evaluates these amplitudes at the discrete QNM frequencies, ensuring uniform machine-precision accuracy across the entire spin parameter space.
Collectively, these two components enable efficient extraction of excitation factors for generic overtone sequences, providing a reliable computational tool for interpreting ringdown signals and testing strong-field gravity with current and next-generation GW observatories.

\section{Conclusion}\label{sec:conclusion}
This work presents a highly efficient theoretical framework for computing GW energy fluxes from generic bound orbits in Kerr spacetime.
By formulating both the angular and radial Teukolsky equations exactly in terms of HeunC functions, we circumvent the primary bottleneck of our previous work~\cite{Chen:2023lsa} and the traditional MST formalism: the costly computation of the renormalized angular momentum $\nu$.
The framework rests on two complementary pillars:
(i) Motygin's hybrid analytic continuation algorithm, which computes the connection coefficients required to construct globally convergent radial homogeneous solutions and directly extract the scattering amplitudes, thereby enabling efficient flux evaluation without recourse to the auxiliary parameter $\nu$;
and (ii) an adaptive bi-power mapping quadrature that efficiently resolves the highly oscillatory source integrals encountered in generic orbits.

The physical robustness of the framework has been rigorously validated through multiple independent applications.
These include the computation of SWSHs and their eigenvalues across the full spin range $0 \leq a/M \leq 0.9999$, the evaluation of GW energy fluxes for generic orbits, and the precise verification of the asymptotic scattering amplitudes \eqref{eq:Rin_Amp} at known QNM frequencies, where the solver accurately reproduces the vanishing incidence amplitude condition to machine precision.
Extensive numerical benchmarks summarized in \cref{tab:comparison} quantify the performance of the proposed framework under standard double-precision arithmetic. For the total radiative flux summed over 168 low-order modes, the HeunC method achieves relative errors of order $10^{-11}$ across the full spin range $0.1 \leq a/M \leq 0.9$, consistently outperforming existing tools by up to an order of magnitude in precision, while reducing computational times by factors of 3--4 and 3--13 compared to \texttt{GeneralizedSasakiNakamura.jl} and \texttt{pybhpt}, respectively.
The runtime of the HeunC method also exhibits modest variation ($<20\%$) across this spin range, in contrast to the stronger spin-dependence observed in some alternative implementations.
These results suggest that the unified HeunC formulation, combined with the adaptive bi-power quadrature, offers a favorable trade-off between numerical stability and computational efficiency for strong-field flux calculations.
The efficiency gain becomes more pronounced when assembling complete energy fluxes that require summation over thousands of high-order modes, as the method's cost scales gently with mode number while avoiding the overhead of arbitrary-precision arithmetic.
In challenging regimes such as high-order harmonics or orbits with large inclinations or eccentricities, the bi-power mapping strategy effectively suppresses quadrature errors near periastron passage, addressing a known limitation of uniform or symmetric integration grids.
In summary, these comprehensive validations and benchmarks confirm that the HeunC framework provides a reliable, efficient, and versatile backend for strong-field black hole perturbation theory.

Looking forward, the efficiency and precision of this framework make it particularly well-suited for next-generation GW astronomy, providing a natural foundation for generating accurate frequency-domain waveform templates for EMRIs, computing near-horizon tidal response and absorption effects, and modeling ringdown signals in the strong-field regime.
A defining feature of our approach is its unified treatment: both the angular and radial Teukolsky equations are solved exactly within the same HeunC basis. Unlike the hybrid algorithms employed in \texttt{pybhpt}, \texttt{GeneralizedSasakiNakamura.jl}, the BHPToolkit, or our earlier work~\cite{Chen:2023lsa}, which rely on mixed solution strategies, this consistent formulation eliminates basis-matching overhead and streamlines the construction of Green's functions.
Such uniformity is especially advantageous for second-order self-force, where repeated, high-precision evaluations of homogeneous and inhomogeneous mode functions are computationally demanding.
While recent analytical studies on metric reconstruction have successfully utilized HeunC solutions~\cite{Berens:2024czo}, their practical numerical evaluation has historically remained a bottleneck, akin to the challenges faced by other analytically elegant but numerically intensive formalisms such as the Nekrasov--Shatashvili method~\cite{Cipriani:2025ikx}.
By delivering a robust and high-throughput numerical implementation, our framework directly bridges this gap.
Furthermore, the unified Heun formalism is inherently portable: for any type-D black hole or perturbation field, one only needs to map the governing differential equation to the corresponding Heun parameters to immediately compute scattering amplitudes, fluxes, or response functions.
Future work will therefore focus on extending the bi-power mapping strategy to efficiently compute inhomogeneous solutions for extended sources, integrating the solver into adiabatic EMRI evolution codes, and applying it to high-order self-force computations.
In summary, the HeunC method represents a competitive alternative for precision and computational efficiency in Kerr perturbation theory, providing a practical, scalable, and broadly applicable numerical backend to support both theoretical investigations and data analysis in the high-precision era of GW astronomy.

\section*{Data Availability}

The data that support the findings of this article are openly available~\cite{ChenGWFlux}, embargo periods may apply.

\section*{Acknowledgement}

This work is supported by the National Key Research and Development Program of China (Nos. 2021YFC2203001), the National Natural Science Foundation of China (No.~12475049).
This work makes use of the Black Hole Perturbation Toolkit~\cite{BHPToolkit} and Motygin's HeunC Code~\cite{motygin_confluent_Heun_functions}.

\appendix
%
%
%
%
%
\section{SYMMETRIES OF HEUNC SOLUTIONS}\label{app:Symmetries_HC_TE}

\begin{enumerate}
  \item The symmetry of the parameter $\alpha$ in CHE \eqref{eq:HeunC-Eq} can be given as follows,
\begin{align}\label{eq:symmetry_HeunC_alpha}
  {{\rm HeunC}} (\alpha ,\beta ,\gamma ,\delta ,\eta;x) = {\mathrm{e}^{ - \alpha x}}  {{\rm HeunC}}( - \alpha ,\beta ,\gamma ,\delta,&\eta; \,x), \nonumber \\
   \beta \notin \mathbb{Z}.&
\end{align}
where $\mathbb{Z}$ is the set of integers.

  \item  The symmetry of the parameter $\gamma$ in CHE \eqref{eq:HeunC-Eq} can be given as follows,
  \begin{align}\label{eq:symmetry_HeunC_gamma}
  {\rm{HeunC}}( {\alpha ,\beta ,\gamma ,\delta ,\eta;x}) ={( {1 - x})^{ - \gamma }}{\rm{HeunC}}( \alpha ,\beta , - &\gamma ,\delta,\eta;x ), \nonumber \\
  \beta \notin \mathbb{Z}.&
\end{align}
   \item The symmetry of the parameter $\beta$ is only displayed in the Teukolsky equations. It can be seen that the parameters $(\alpha,\beta,\gamma)$ in the general solution \eqref{eq:GSol_ATE} of the ATE \eqref{eq:GFoATE} and the general solution \eqref{eq:GSol_RTE} of the RTE \eqref{eq:GFoRTE} are interchangeable.
       The symmetry properties of the general solution ${S} _{\ell m}$ and ${{R} _{\ell m}}$ are as follows,
\begin{subequations}\label{eq:symetry_ATE}
  \begin{align}
&{{S} _{\ell m}}(\alpha ,\beta ,\gamma ;x) = {{S} _{\ell m}}( - \alpha ,\beta ,\gamma ;x),\\
&{{S} _{\ell m}}(\alpha ,\beta ,\gamma ;x) = {{S} _{\ell m}}(\alpha ,\beta , - \gamma ;x),\\
&{{S} _{\ell m}}(\alpha ,\beta ,\gamma ;x) = {{S} _{\ell m}}(\alpha , - \beta ,\gamma ;x),\,\,\,{\rm if} \,\,\,{D_1} = {D_2}.
  \end{align}
\end{subequations}
and
\begin{subequations}\label{eq:symetry_RTE}
  \begin{align}
&{{R} _{\ell m}}(\alpha ,\beta ,\gamma ;x) = {{R} _{\ell m}}( - \alpha ,\beta ,\gamma ;x),\\
&{{R} _{\ell m}}(\alpha ,\beta ,\gamma ;x) = {{R} _{\ell m}}(\alpha ,\beta , - \gamma ;x),\\
&{{R} _{\ell m}}(\alpha ,\beta ,\gamma ;x) = {{R} _{\ell m}}(\alpha , - \beta ,\gamma ;x),\,\,\,{\rm if} \,\,\, {C_1} = {C_2}.
 \end{align}
\end{subequations}
where $\{D_1,D_2\}$ and $\{C_1,C_2\}$ are the constants in the general solutions \eqref{eq:GFoATE} and \eqref{eq:GSol_RTE}, respectively.

\end{enumerate}

\section{NORMALIZATION FACTOR FOR ATES}\label{app:normalization}
The normalization factor $\mathcal{N}$ is defined by the standard orthonormality condition for spin-weighted spheroidal harmonics,
\begin{equation}\label{eq:Norm_Fac_0}
  \mathcal{N}^2 = \int_0^{2\pi}\!\!\int_0^\pi \left|S_\theta(x)\,\HCa(x) \right|^2 \sin\theta\,{\rm d}\theta\,{\rm d}\phi,
\end{equation}
where $S_\theta(x)$ is the S-homotopic prefactor defined in \cref{eq:S-homotopic-theta-d2} and $\HCa(x)$ denotes the HeunC solution \eqref{eq:HCSol-x0}.

\subsection{Power-series summation}\label{subapp:power_series}
The regular local solution $\HCa(x)$ admits a power-series expansion~\cite{Motygin2018}:
\begin{equation}\label{eq:HeunC_series}
  \HCa(x) = \sum_{n = 0}^\infty b_n x^n,
\end{equation}
where the coefficients $b_n$ satisfy the three-term recurrence relation associated with the first parametrization:
\begin{equation}\label{eq:recur_relation}
P_n b_n = Q_n b_{n-1} + R_n b_{n-2},
\end{equation}
with
\begin{subequations}\label{eq:recur_coeffs}
\begin{align}
P_n &= n(n + \beta), \label{eq:Pn}\\
Q_n &= (n - 1)(n + \beta + \gamma - \alpha) \nonumber \\
    &\quad + \eta + \tfrac{1}{2}\big[(\beta + 1)(\gamma + 1 - \alpha) - 1\big], \label{eq:Qn}\\
R_n &= \alpha\left(n - 1 + \tfrac{\beta + \gamma}{2}\right) + \delta. \label{eq:Rn}
\end{align}
\end{subequations}
The initial conditions are $b_{-1}=0$ and $b_0=1$.

Substituting \cref{eq:HeunC_series} into \cref{eq:Norm_Fac_0} and using the measure transformation ${\rm d}\Omega = 4\pi\,{\rm d}x$, the normalization factor becomes
\begin{equation}\label{eq:Norm_Fac_1}
  \mathcal{N}^2 = 4\pi \int_0^1 \left| x^{\beta+i}(1 - x)^\gamma e^{\alpha x} \right| \sum_{i = 0}^{n_{\max}} \sum_{j = 0}^i b_j b_{i - j} \, {\rm d}x,
\end{equation}
where the double sum represents the discrete convolution of the series coefficients. For moderate precision requirements, fast convolution techniques (e.g., FFT-based methods) may be employed. However, for high-order spin-weighted harmonics, the coefficients often exhibit strong oscillations or stem from ill-conditioned recurrences \eqref{eq:recur_relation}, necessitating high-precision arithmetic. In such cases, direct explicit summation is preferred to avoid numerical instability.

To enhance computational efficiency, \cref{eq:Norm_Fac_1} can be reorganized as
\begin{equation}\label{eq:Norm_Fac_2_app}
\mathcal{N} = \sqrt{4\pi \sum_{i=0}^{n_{\max}} \sum_{j=0}^i b_j b_{i-j} \, |I(i)|},
\end{equation}
where the overlap integral kernel $I(i)$ is
\begin{equation}\label{eq:Integ_core}
I(i) = \int_0^1 \mathrm{e}^{\alpha x} x^{i + \beta} (1 - x)^\gamma \, {\rm d}x.
\end{equation}
Using the standard integral formula (Eq.~3.383.1 of Ref.~\cite{book2015249}),
\begin{align}\label{eq:hyperp_integ}
\int_0^u x^{\nu - 1}(u - x)^{\mu - 1} \mathrm{e}^{\kappa x} \, {\rm d}x = &\, \mathrm{B}(\mu,\nu) u^{\mu + \nu - 1} {}_1F_1(\nu; \mu + \nu; \kappa u), \nonumber \\
  &\, \mathrm{Re}(\mu) > 0,\ \mathrm{Re}(\nu) > 0,
\end{align}
where $\mathrm{B}(\mu,\nu) = \Gamma(\mu)\Gamma(\nu)/\Gamma(\mu+\nu)$ is the Beta function.
Applying this to \cref{eq:Integ_core} yields
\begin{equation}\label{eq:Integ_core_simp}
  I(i) = \frac{\Gamma(\gamma+1)\Gamma(i+\beta+1)}{\Gamma(i+\beta+\gamma+2)} \, {}_1F_1(i+\beta+1, i+\beta+\gamma+2; \alpha).
\end{equation}

\subsection{Alternative form via Wronskian derivative}\label{subapp:wronskian}
For higher-order modes, the power-series summation often requires precision significantly exceeding standard double-precision arithmetic, leading to substantial computational overhead. To circumvent this, we adopt an alternative method based on the Sturm--Liouville structure of the CHE. Berens \textit{et al.} demonstrated that this structure allows the normalization integral to be expressed via the derivative of the Wronskian with respect to the eigenparameter $\tilde{q}$~\cite{Becker1997,Berens:2024czo}, thereby bypassing explicit quadrature. The normalization factor \eqref{eq:Norm_Fac_0} then takes the form
\begin{equation}\label{eq:Norm_New}
  \mathcal{N} = \sqrt{2\pi I_n},
\end{equation}
where $I_n$ is the weighted self-overlap integral
\begin{equation}\label{eq:HeunC_Integral}
  I_n = 2\int_0^1 w(x) \left[ \HCa(x) \right]^2 {\rm d}x,
\end{equation}
with the weight function $w(x)=| x^{\tilde \gamma - 1} (x - 1)^{\tilde \delta - 1} e^{\tilde\epsilon x} |$.

Let $f_0(\tilde{q}, x)=\HCa(x)$ denote the solution to the CHE \eqref{eq:HeunC-Eq} for an arbitrary parameter $\tilde{q}$, and $f_0(\tilde{q}_n, x)$ the eigenfunction corresponding to the discrete eigenvalue\footnote{According to \cref{eq:HeunC_Para_Relat}, $\tilde{q}_n$ implicitly incorporates the eigenvalue $A_{\ell m}$.} $\tilde{q}_n$. Both satisfy the self-adjoint CHE. Taking the difference of the two equations yields the Lagrange identity:
\begin{align}\label{eq:identity1}
\frac{\rm d}{{\rm d}x}\Bigg[ p(x)\Big( f_0(\tilde{q}_n,& x)\partial_x f_0(\tilde{q}, x) - f_0(\tilde{q}, x) \partial_x f_0(\tilde{q}_n, x) \Big) \Bigg] \nonumber \\
&= (\tilde{q} - \tilde{q}_n) w(x) f_0(\tilde{q}, x) f_0(\tilde{q}_n, x),
\end{align}
where $p(x) = x^{\tilde \gamma} (x - 1)^{\tilde \delta} e^{\tilde \epsilon x}$ is the leading coefficient in the self-adjoint form. Integrating \cref{eq:identity1} from $0$ to $x$ and noting that the lower boundary term vanishes due to regularity at $x=0$, we obtain
\begin{align}
&\int_0^x w(\xi) f_0(\tilde{q}, \xi) f_0(\tilde{q}_n, \xi) \, {\rm d}\xi \nonumber \\
&= \frac{p(x)}{\tilde{q} - \tilde{q}_n} \left[ f_0(\tilde{q}_n, x) \frac{\partial f_0(\tilde{q}, x)}{\partial x} - f_0(\tilde{q}, x) \frac{\partial f_0(\tilde{q}_n, x)}{\partial x} \right].
\end{align}

To recover the self-overlap integral \eqref{eq:HeunC_Integral}, we take the limit $\tilde{q} \to \tilde{q}_n$. Applying L'Hôpital's rule to the right-hand side yields
\begin{align}\label{eq:Norm_integral}
 \frac{1}{2}I_n &= \int_0^x w(\zeta) [f_0(\tilde{q}_n, \zeta)]^2 {\rm d}\zeta \nonumber\\
 &= p(x)\left[ f_0 \frac{\partial^2 f_0}{\partial \tilde{q} \partial x} - \frac{\partial f_0}{\partial \tilde{q}} \frac{\partial f_0}{\partial x} \right]_{\tilde{q}=\tilde{q}_n}.
\end{align}

To evaluate the full integral, we take the limit $x \to 1$. Near $x=1$, the general solution decomposes into two linearly independent local branches:
\begin{equation}\label{eq:f0_dec}
  f_0(\tilde{q}, x) = A(\tilde{q}) f_1(\tilde{q}, x) + B(\tilde{q}) \tilde{f}_1(\tilde{q}, x),
\end{equation}
where
\begin{align}
  f_1(\tilde q,x) &= \mathrm{HeunC}(\tilde q - \tilde \alpha, -\tilde \alpha, \tilde \delta, \tilde \gamma, -\tilde \epsilon; 1 - x), \label{eq:f1_def}\\
  \tilde{f}_1(\tilde q,x) &= (1 - x)^{1 - \tilde \delta} \mathrm{HeunC}(\tilde q - \tilde \alpha - (\tilde \epsilon + \tilde \gamma)(1 - \tilde \delta), \nonumber\\
 &\quad -\tilde \alpha - \tilde \epsilon(1 - \tilde \delta), 2 - \tilde \delta, \tilde \gamma, -\tilde \epsilon; 1 - x). \label{eq:f1t_def}
\end{align}

Regularity at the south pole $x=1$ for an eigenvalue $\tilde{q}=\tilde{q}_n$ forces $B(\tilde{q}_n)=0$, which implies
\begin{equation}\label{eq:Aq}
  A(\tilde{q}_n) = \left. \frac{f_0(\tilde{q}, x)}{f_1(\tilde{q}, x)} \right|_{\tilde{q}=\tilde{q}_n}.
\end{equation}
Substituting \Cref{eq:Asy-HCA-x0,eq:f0_dec} into \cref{eq:Norm_integral} and taking $x \to 1$, we find
\begin{equation}\label{eq:Norm_integralv2}
   \frac{1}{2}I_n = \lim_{x \to 1} \left[ p(x) A(\tilde{q}) \frac{{\rm d}B}{{\rm d}\tilde{q}} f_1(\tilde{q}, x) \partial_x \tilde{f}_1(\tilde{q}, x) \right]_{\tilde{q} = \tilde{q}_n}.
\end{equation}

Differentiating \cref{eq:f0_dec} with respect to $x$ and $\tilde{q}$ allows us to express the derivative of $B$ as
\begin{equation}\label{eq:dBdq}
  \frac{{\rm d}B}{{\rm d}\tilde{q}}(\tilde{q}_n) = \left. \frac{\partial W}{\partial \tilde{q}} \left( \tilde{f}_1 \partial_x f_1 - f_1 \partial_x \tilde{f}_1 \right)^{-1} \right|_{\tilde{q}=\tilde{q}_n},
\end{equation}
where $W(\tilde{q},x) = f_0 \partial_x f_1 - f_1 \partial_x f_0$ is the Wronskian of $f_0$ and $f_1$. Inserting \Cref{eq:Asy-HCA-x0,eq:Aq,eq:dBdq} into \cref{eq:Norm_integralv2} yields
\begin{equation}
   I_n = -2p(x) \frac{\partial W(\tilde{q}_n, x)}{\partial \tilde{q}} \frac{f_0(\tilde{q}_n, x)}{f_1(\tilde{q}_n, x)}.
\end{equation}

Finally, the normalization factor \eqref{eq:Norm_New} can be compactly written as
\begin{equation}\label{eq:Norm_DWdq_app}
{\cal N} = \sqrt {\left|( - 1)^{\tilde\gamma }4\pi p(x)\frac{\partial W}{\partial \tilde{q}}\frac{{{f_0}}}{{{f_1}}} \right|}.
\end{equation}
The derivative $\partial W / \partial \tilde{q}$ in \cref{eq:Norm_DWdq} is most efficiently evaluated using the complex-step differentiation technique. For analytic functions, this method extracts the derivative from a single function evaluation with machine-precision accuracy, completely avoiding truncation errors and subtractive cancellation.

\section{ADAPTIVE RADIAL QUADRATURE VIA BI-POWER MAPPING}\label{app:bi_power_mapping}
The integrand profiles displayed in \Cref{fig:Integrand4443,fig:Integrand44450} reveal a critical challenge for numerical quadrature: the oscillatory structure is highly localized near $\lambda_r \to \pi$ (apastron passage), while the region $\lambda_r \in [0, \pi/2]$ remains relatively smooth.
A uniform grid or symmetric mapping would therefore allocate excessive points to the smooth region while risking severe undersampling near $\lambda_r \to \pi$, where the integrand exhibits rapid phase variations.
To resolve this inefficiency, we implement a bi-power mapping method for the radial integration in \cref{eq:Jlmnk_def}.
Instead of integrating over a uniform grid, we perform a coordinate transformation $\lambda_r = \lambda_r(\xi)$ with $\xi \in [0,1]$.
The radial component of the double integral is transformed as
\begin{equation}
\int_0^\pi {\bf G}_{\ell m}^{\infty ,{\rm{H}}}(\lambda_r) \, d\lambda_r = \int_0^1 {\bf G}_{\ell m}^{\infty ,{\rm{H}}}\bigl(\lambda_r(\xi)\bigr) \frac{d\lambda_r}{d\xi} \, d\xi.
\end{equation}
The mapping function is defined as
\begin{equation}
\lambda_r(\xi) = \pi \cdot \frac{\xi^{p_l}}{\xi^{p_l} + (1-\xi)^{p_h}},
\end{equation}
where $p_l$ and $p_h$ are positive exponents controlling the sampling density near $\lambda_r=0$ and $\lambda_r=\pi$, respectively.
In practice, $p_l$ governs the node clustering near periapsis ($\lambda_r \to 0$), with typical values in the range $1.5 \lesssim p_l \lesssim 4.0$; larger values yield denser sampling where the integrand varies rapidly due to strong-field radial motion. The parameter $p_h$ controls clustering near apastron ($\lambda_r \to \pi$), where horizon-absorbed contributions and high-frequency oscillations often require enhanced resolution; we recommend $5.0 \lesssim p_h \lesssim 8.0$ for orbits with large eccentricity or strong-field penetration. For the representative case discussed in \cref{tab:flux_comparison} with $(a,p,e,x_I) = (0.1, 10, 0.9, 0.5)$ and mode $(l,m,k,n) = (4,4,4,50)$, we adopt $p_l = 3.8$ and $p_h = 5.0$, which yield exponential convergence of the bi-power mapped quadrature to high precision.

The analytical Jacobian factor is
\begin{equation}
\frac{d\lambda_r}{d\xi} = \pi \cdot \frac{p_l \xi^{p_l-1}(1-\xi)^{p_h} + p_h \xi^{p_l}(1-\xi)^{p_h-1}}{\bigl[\xi^{p_l} + (1-\xi)^{p_h}\bigr]^2}.
\end{equation}
{
In our implementation, we consistently adopt fixed empirical exponents $p_l = 3.8$ and $p_h = 5.0$. While the optimal exponents could theoretically be tuned for specific orbital configurations and harmonic indices $(\ell, m, n, k)$, our sensitivity analysis indicates that the quadrature precision is remarkably robust to minor variations in these parameters. Specifically, the relative error in the integrated flux typically fluctuates by less than one order of magnitude across a wide range of $p_l$ and $p_h$ values. This static configuration effectively resolves the highly localized oscillations near apastron across the entire parameter space while avoiding the significant computational overhead required for per-mode parameter optimization.
}

This method offers three critical advantages over standard quadrature schemes.
First, independent tuning of $p_l$ and $p_h$ allows precise concentration of grid points where oscillations occur ($\lambda_r \to \pi$), while keeping the grid sparse in the smooth region.
For example, setting $p_h \gg p_l$ creates a dense cluster near apastron passage without over-sampling the region near $\lambda_r \to 0$.
Second, the ratio $p_l/p_h$ shifts the point of minimum sampling density to align with the stable intermediate zone ($\lambda_r \approx 1.8\text{--}2.6$), maximizing quadrature efficiency by avoiding redundant evaluations where the integrand varies slowly.
Third, the mapping function and its Jacobian are analytical, introducing negligible computational overhead while ensuring perfect boundary coverage and numerical stability.
By coupling this optimized integration method with the HeunC framework, we achieve rapid, high-precision flux evaluations for generic Kerr orbits, effectively overcoming the numerical bottlenecks associated with highly oscillatory integrands.

\bibliography{GWFbib}

\begin{thebibliography}{113}%
\makeatletter
\providecommand \@ifxundefined [1]{%
 \@ifx{#1\undefined}
}%
\providecommand \@ifnum [1]{%
 \ifnum #1\expandafter \@firstoftwo
 \else \expandafter \@secondoftwo
 \fi
}%
\providecommand \@ifx [1]{%
 \ifx #1\expandafter \@firstoftwo
 \else \expandafter \@secondoftwo
 \fi
}%
\providecommand \natexlab [1]{#1}%
\providecommand \enquote  [1]{``#1''}%
\providecommand \bibnamefont  [1]{#1}%
\providecommand \bibfnamefont [1]{#1}%
\providecommand \citenamefont [1]{#1}%
\providecommand \href@noop [0]{\@secondoftwo}%
\providecommand \href [0]{\begingroup \@sanitize@url \@href}%
\providecommand \@href[1]{\@@startlink{#1}\@@href}%
\providecommand \@@href[1]{\endgroup#1\@@endlink}%
\providecommand \@sanitize@url [0]{\catcode `\\12\catcode `\$12\catcode
  `\&12\catcode `\#12\catcode `\^12\catcode `\_12\catcode `\%12\relax}%
\providecommand \@@startlink[1]{}%
\providecommand \@@endlink[0]{}%
\providecommand \url  [0]{\begingroup\@sanitize@url \@url }%
\providecommand \@url [1]{\endgroup\@href {#1}{\urlprefix }}%
\providecommand \urlprefix  [0]{URL }%
\providecommand \Eprint [0]{\href }%
\providecommand \doibase [0]{https://doi.org/}%
\providecommand \selectlanguage [0]{\@gobble}%
\providecommand \bibinfo  [0]{\@secondoftwo}%
\providecommand \bibfield  [0]{\@secondoftwo}%
\providecommand \translation [1]{[#1]}%
\providecommand \BibitemOpen [0]{}%
\providecommand \bibitemStop [0]{}%
\providecommand \bibitemNoStop [0]{.\EOS\space}%
\providecommand \EOS [0]{\spacefactor3000\relax}%
\providecommand \BibitemShut  [1]{\csname bibitem#1\endcsname}%
\let\auto@bib@innerbib\@empty
\bibitem [{\citenamefont {Abbott}\ \emph
  {et~al.}(2016{\natexlab{a}})\citenamefont {Abbott} \emph
  {et~al.}}]{LIGOScientific:2016wkq}%
  \BibitemOpen
  \bibfield  {author} {\bibinfo {author} {\bibfnamefont {T.~D.}\ \bibnamefont
  {Abbott}} \emph {et~al.} (\bibinfo {collaboration} {LIGO Scientific,
  Virgo}),\ }\bibfield  {title} {\bibinfo {title} {{Improved analysis of
  GW150914 using a fully spin-precessing waveform Model}},\ }\href
  {https://doi.org/10.1103/PhysRevX.6.041014} {\bibfield  {journal} {\bibinfo
  {journal} {Phys. Rev. X}\ }\textbf {\bibinfo {volume} {6}},\ \bibinfo {pages}
  {041014} (\bibinfo {year} {2016}{\natexlab{a}})},\ \Eprint
  {https://arxiv.org/abs/1606.01210} {arXiv:1606.01210 [gr-qc]} \BibitemShut
  {NoStop}%
\bibitem [{\citenamefont {Abbott}\ \emph
  {et~al.}(2016{\natexlab{b}})\citenamefont {Abbott} \emph
  {et~al.}}]{LIGOScientific:2016dsl}%
  \BibitemOpen
  \bibfield  {author} {\bibinfo {author} {\bibfnamefont {B.~P.}\ \bibnamefont
  {Abbott}} \emph {et~al.} (\bibinfo {collaboration} {LIGO Scientific,
  Virgo}),\ }\bibfield  {title} {\bibinfo {title} {{Binary black hole mergers
  in the first advanced LIGO observing run}},\ }\href
  {https://doi.org/10.1103/PhysRevX.6.041015} {\bibfield  {journal} {\bibinfo
  {journal} {Phys. Rev. X}\ }\textbf {\bibinfo {volume} {6}},\ \bibinfo {pages}
  {041015} (\bibinfo {year} {2016}{\natexlab{b}})},\ \bibinfo {note} {[Erratum:
  Phys.Rev.X 8, 039903 (2018)]},\ \Eprint {https://arxiv.org/abs/1606.04856}
  {arXiv:1606.04856 [gr-qc]} \BibitemShut {NoStop}%
\bibitem [{\citenamefont {Abbott}\ \emph {et~al.}(2019)\citenamefont {Abbott}
  \emph {et~al.}}]{LIGOScientific:2018mvr}%
  \BibitemOpen
  \bibfield  {author} {\bibinfo {author} {\bibfnamefont {B.~P.}\ \bibnamefont
  {Abbott}} \emph {et~al.} (\bibinfo {collaboration} {LIGO Scientific,
  Virgo}),\ }\bibfield  {title} {\bibinfo {title} {{GWTC-1: A
  gravitational-wave transient catalog of compact binary mergers observed by
  LIGO and Virgo during the first and second observing runs}},\ }\href
  {https://doi.org/10.1103/PhysRevX.9.031040} {\bibfield  {journal} {\bibinfo
  {journal} {Phys. Rev. X}\ }\textbf {\bibinfo {volume} {9}},\ \bibinfo {pages}
  {031040} (\bibinfo {year} {2019})},\ \Eprint
  {https://arxiv.org/abs/1811.12907} {arXiv:1811.12907 [astro-ph.HE]}
  \BibitemShut {NoStop}%
\bibitem [{\citenamefont {Abbott}\ \emph {et~al.}(2021)\citenamefont {Abbott}
  \emph {et~al.}}]{LIGOScientific:2020ibl}%
  \BibitemOpen
  \bibfield  {author} {\bibinfo {author} {\bibfnamefont {R.}~\bibnamefont
  {Abbott}} \emph {et~al.} (\bibinfo {collaboration} {LIGO Scientific,
  Virgo}),\ }\bibfield  {title} {\bibinfo {title} {{GWTC-2: Compact binary
  coalescences observed by LIGO and Virgo during the first half of the third
  observing run}},\ }\href {https://doi.org/10.1103/PhysRevX.11.021053}
  {\bibfield  {journal} {\bibinfo  {journal} {Phys. Rev. X}\ }\textbf {\bibinfo
  {volume} {11}},\ \bibinfo {pages} {021053} (\bibinfo {year} {2021})},\
  \Eprint {https://arxiv.org/abs/2010.14527} {arXiv:2010.14527 [gr-qc]}
  \BibitemShut {NoStop}%
\bibitem [{\citenamefont {Abbott}\ \emph {et~al.}(2023)\citenamefont {Abbott}
  \emph {et~al.}}]{KAGRA:2021duu}%
  \BibitemOpen
  \bibfield  {author} {\bibinfo {author} {\bibfnamefont {R.}~\bibnamefont
  {Abbott}} \emph {et~al.} (\bibinfo {collaboration} {KAGRA, VIRGO, LIGO
  Scientific}),\ }\bibfield  {title} {\bibinfo {title} {{Population of merging
  compact binaries inferred using gravitational waves through GWTC-3}},\ }\href
  {https://doi.org/10.1103/PhysRevX.13.011048} {\bibfield  {journal} {\bibinfo
  {journal} {Phys. Rev. X}\ }\textbf {\bibinfo {volume} {13}},\ \bibinfo
  {pages} {011048} (\bibinfo {year} {2023})},\ \Eprint
  {https://arxiv.org/abs/2111.03634} {arXiv:2111.03634 [astro-ph.HE]}
  \BibitemShut {NoStop}%
\bibitem [{\citenamefont {Amaro-Seoane}\ \emph {et~al.}(2017)\citenamefont
  {Amaro-Seoane} \emph {et~al.}}]{LISA:2017pwj}%
  \BibitemOpen
  \bibfield  {author} {\bibinfo {author} {\bibfnamefont {P.}~\bibnamefont
  {Amaro-Seoane}} \emph {et~al.} (\bibinfo {collaboration} {LISA}),\ }\bibfield
   {title} {\bibinfo {title} {{Laser Interferometer Space Antenna}},\
  }\href@noop {} {\  (\bibinfo {year} {2017})},\ \Eprint
  {https://arxiv.org/abs/1702.00786} {arXiv:1702.00786 [astro-ph.IM]}
  \BibitemShut {NoStop}%
\bibitem [{\citenamefont {Colpi}\ \emph {et~al.}(2024)\citenamefont {Colpi}
  \emph {et~al.}}]{LISA:2024hlh}%
  \BibitemOpen
  \bibfield  {author} {\bibinfo {author} {\bibfnamefont {M.}~\bibnamefont
  {Colpi}} \emph {et~al.} (\bibinfo {collaboration} {LISA}),\ }\bibfield
  {title} {\bibinfo {title} {{LISA Definition Study Report}},\ }\href@noop {}
  {\  (\bibinfo {year} {2024})},\ \Eprint {https://arxiv.org/abs/2402.07571}
  {arXiv:2402.07571 [astro-ph.CO]} \BibitemShut {NoStop}%
\bibitem [{\citenamefont {Luo}\ \emph {et~al.}(2016)\citenamefont {Luo} \emph
  {et~al.}}]{TianQin:2015yph}%
  \BibitemOpen
  \bibfield  {author} {\bibinfo {author} {\bibfnamefont {J.}~\bibnamefont
  {Luo}} \emph {et~al.} (\bibinfo {collaboration} {TianQin}),\ }\bibfield
  {title} {\bibinfo {title} {{TianQin: a space-borne gravitational wave
  detector}},\ }\href {https://doi.org/10.1088/0264-9381/33/3/035010}
  {\bibfield  {journal} {\bibinfo  {journal} {Classical Quantum Gravity}\
  }\textbf {\bibinfo {volume} {33}},\ \bibinfo {pages} {035010} (\bibinfo
  {year} {2016})},\ \Eprint {https://arxiv.org/abs/1512.02076}
  {arXiv:1512.02076 [astro-ph.IM]} \BibitemShut {NoStop}%
\bibitem [{\citenamefont {Mei}\ \emph {et~al.}(2021)\citenamefont {Mei} \emph
  {et~al.}}]{TianQin:2020hid}%
  \BibitemOpen
  \bibfield  {author} {\bibinfo {author} {\bibfnamefont {J.}~\bibnamefont
  {Mei}} \emph {et~al.} (\bibinfo {collaboration} {TianQin}),\ }\bibfield
  {title} {\bibinfo {title} {{The TianQin project: current progress on science
  and technology}},\ }\href {https://doi.org/10.1093/ptep/ptaa114} {\bibfield
  {journal} {\bibinfo  {journal} {Prog. Theor. Exp. Phys.}\ }\textbf {\bibinfo
  {volume} {2021}},\ \bibinfo {pages} {05A107} (\bibinfo {year} {2021})},\
  \Eprint {https://arxiv.org/abs/2008.10332} {arXiv:2008.10332 [gr-qc]}
  \BibitemShut {NoStop}%
\bibitem [{\citenamefont {Li}\ \emph {et~al.}(2025)\citenamefont {Li} \emph
  {et~al.}}]{Li:2024rnk}%
  \BibitemOpen
  \bibfield  {author} {\bibinfo {author} {\bibfnamefont {E.-K.}\ \bibnamefont
  {Li}} \emph {et~al.},\ }\bibfield  {title} {\bibinfo {title} {{Gravitational
  wave astronomy with TianQin}},\ }\href
  {https://doi.org/10.1088/1361-6633/adc9be} {\bibfield  {journal} {\bibinfo
  {journal} {Rept. Prog. Phys.}\ }\textbf {\bibinfo {volume} {88}},\ \bibinfo
  {pages} {056901} (\bibinfo {year} {2025})},\ \Eprint
  {https://arxiv.org/abs/2409.19665} {arXiv:2409.19665 [astro-ph.GA]}
  \BibitemShut {NoStop}%
\bibitem [{\citenamefont {Ruan}\ \emph {et~al.}(2020)\citenamefont {Ruan},
  \citenamefont {Guo}, \citenamefont {Cai},\ and\ \citenamefont
  {Zhang}}]{Ruan:2018tsw}%
  \BibitemOpen
  \bibfield  {author} {\bibinfo {author} {\bibfnamefont {W.-H.}\ \bibnamefont
  {Ruan}}, \bibinfo {author} {\bibfnamefont {Z.-K.}\ \bibnamefont {Guo}},
  \bibinfo {author} {\bibfnamefont {R.-G.}\ \bibnamefont {Cai}},\ and\ \bibinfo
  {author} {\bibfnamefont {Y.-Z.}\ \bibnamefont {Zhang}},\ }\bibfield  {title}
  {\bibinfo {title} {{Taiji program: Gravitational-wave sources}},\ }\href
  {https://doi.org/10.1142/S0217751X2050075X} {\bibfield  {journal} {\bibinfo
  {journal} {Int. J. Mod. Phys. A}\ }\textbf {\bibinfo {volume} {35}},\
  \bibinfo {pages} {2050075} (\bibinfo {year} {2020})},\ \Eprint
  {https://arxiv.org/abs/1807.09495} {arXiv:1807.09495 [gr-qc]} \BibitemShut
  {NoStop}%
\bibitem [{\citenamefont {Gong}\ \emph {et~al.}(2021)\citenamefont {Gong},
  \citenamefont {Luo},\ and\ \citenamefont {Wang}}]{Gong:2021gvw}%
  \BibitemOpen
  \bibfield  {author} {\bibinfo {author} {\bibfnamefont {Y.}~\bibnamefont
  {Gong}}, \bibinfo {author} {\bibfnamefont {J.}~\bibnamefont {Luo}},\ and\
  \bibinfo {author} {\bibfnamefont {B.}~\bibnamefont {Wang}},\ }\bibfield
  {title} {\bibinfo {title} {{Concepts and status of Chinese space
  gravitational wave detection projects}},\ }\href
  {https://doi.org/10.1038/s41550-021-01480-3} {\bibfield  {journal} {\bibinfo
  {journal} {Nature Astron.}\ }\textbf {\bibinfo {volume} {5}},\ \bibinfo
  {pages} {881} (\bibinfo {year} {2021})},\ \Eprint
  {https://arxiv.org/abs/2109.07442} {arXiv:2109.07442 [astro-ph.IM]}
  \BibitemShut {NoStop}%
\bibitem [{\citenamefont {Amaro-Seoane}(2018)}]{Amaro-Seoane:2012lgq}%
  \BibitemOpen
  \bibfield  {author} {\bibinfo {author} {\bibfnamefont {P.}~\bibnamefont
  {Amaro-Seoane}},\ }\bibfield  {title} {\bibinfo {title} {{Relativistic
  dynamics and extreme mass ratio inspirals}},\ }\href
  {https://doi.org/10.1007/s41114-018-0013-8} {\bibfield  {journal} {\bibinfo
  {journal} {Living Rev. Rel.}\ }\textbf {\bibinfo {volume} {21}},\ \bibinfo
  {pages} {4} (\bibinfo {year} {2018})},\ \Eprint
  {https://arxiv.org/abs/1205.5240} {arXiv:1205.5240 [astro-ph.CO]}
  \BibitemShut {NoStop}%
\bibitem [{\citenamefont {Seoane}\ \emph {et~al.}(2023)\citenamefont {Seoane}
  \emph {et~al.}}]{LISA:2022yao}%
  \BibitemOpen
  \bibfield  {author} {\bibinfo {author} {\bibfnamefont {P.~A.}\ \bibnamefont
  {Seoane}} \emph {et~al.} (\bibinfo {collaboration} {LISA}),\ }\bibfield
  {title} {\bibinfo {title} {{Astrophysics with the Laser Interferometer Space
  Antenna}},\ }\href {https://doi.org/10.1007/s41114-022-00041-y} {\bibfield
  {journal} {\bibinfo  {journal} {Living Rev. Rel.}\ }\textbf {\bibinfo
  {volume} {26}},\ \bibinfo {pages} {2} (\bibinfo {year} {2023})},\ \Eprint
  {https://arxiv.org/abs/2203.06016} {arXiv:2203.06016 [gr-qc]} \BibitemShut
  {NoStop}%
\bibitem [{\citenamefont {Kawamura}\ \emph {et~al.}(2021)\citenamefont
  {Kawamura} \emph {et~al.}}]{Kawamura:2020pcg}%
  \BibitemOpen
  \bibfield  {author} {\bibinfo {author} {\bibfnamefont {S.}~\bibnamefont
  {Kawamura}} \emph {et~al.},\ }\bibfield  {title} {\bibinfo {title} {{Current
  status of space gravitational wave antenna DECIGO and B-DECIGO}},\ }\href
  {https://doi.org/10.1093/ptep/ptab019} {\bibfield  {journal} {\bibinfo
  {journal} {Prog. Theor. Exp. Phys.}\ }\textbf {\bibinfo {volume} {2021}},\
  \bibinfo {pages} {05A105} (\bibinfo {year} {2021})},\ \Eprint
  {https://arxiv.org/abs/2006.13545} {arXiv:2006.13545 [gr-qc]} \BibitemShut
  {NoStop}%
\bibitem [{\citenamefont {Abac}\ \emph {et~al.}(2025)\citenamefont {Abac} \emph
  {et~al.}}]{ET:2025xjr}%
  \BibitemOpen
  \bibfield  {author} {\bibinfo {author} {\bibfnamefont {A.}~\bibnamefont
  {Abac}} \emph {et~al.} (\bibinfo {collaboration} {ET}),\ }\bibfield  {title}
  {\bibinfo {title} {{The Science of the Einstein Telescope}},\ }\href@noop {}
  {\  (\bibinfo {year} {2025})},\ \Eprint {https://arxiv.org/abs/2503.12263}
  {arXiv:2503.12263 [gr-qc]} \BibitemShut {NoStop}%
\bibitem [{\citenamefont {Reitze}\ \emph {et~al.}(2019)\citenamefont {Reitze}
  \emph {et~al.}}]{Reitze:2019iox}%
  \BibitemOpen
  \bibfield  {author} {\bibinfo {author} {\bibfnamefont {D.}~\bibnamefont
  {Reitze}} \emph {et~al.},\ }\bibfield  {title} {\bibinfo {title} {{Cosmic
  Explorer: The U.S. contribution to gravitational-wave astronomy beyond
  LIGO}},\ }\href@noop {} {\bibfield  {journal} {\bibinfo  {journal} {Bull. Am.
  Astron. Soc.}\ }\textbf {\bibinfo {volume} {51}},\ \bibinfo {pages} {035}
  (\bibinfo {year} {2019})},\ \Eprint {https://arxiv.org/abs/1907.04833}
  {arXiv:1907.04833 [astro-ph.IM]} \BibitemShut {NoStop}%
\bibitem [{\citenamefont {Punturo}\ \emph {et~al.}(2010)\citenamefont {Punturo}
  \emph {et~al.}}]{Punturo:2010zz}%
  \BibitemOpen
  \bibfield  {author} {\bibinfo {author} {\bibfnamefont {M.}~\bibnamefont
  {Punturo}} \emph {et~al.},\ }\bibfield  {title} {\bibinfo {title} {{The
  Einstein Telescope: A third-generation gravitational wave observatory}},\
  }\href {https://doi.org/10.1088/0264-9381/27/19/194002} {\bibfield  {journal}
  {\bibinfo  {journal} {Classical Quantum Gravity}\ }\textbf {\bibinfo {volume}
  {27}},\ \bibinfo {pages} {194002} (\bibinfo {year} {2010})}\BibitemShut
  {NoStop}%
\bibitem [{\citenamefont {Babak}\ \emph {et~al.}(2017)\citenamefont {Babak},
  \citenamefont {Gair}, \citenamefont {Sesana}, \citenamefont {Barausse},
  \citenamefont {Sopuerta}, \citenamefont {Berry}, \citenamefont {Berti},
  \citenamefont {Amaro-Seoane}, \citenamefont {Petiteau},\ and\ \citenamefont
  {Klein}}]{Babak:2017tow}%
  \BibitemOpen
  \bibfield  {author} {\bibinfo {author} {\bibfnamefont {S.}~\bibnamefont
  {Babak}}, \bibinfo {author} {\bibfnamefont {J.}~\bibnamefont {Gair}},
  \bibinfo {author} {\bibfnamefont {A.}~\bibnamefont {Sesana}}, \bibinfo
  {author} {\bibfnamefont {E.}~\bibnamefont {Barausse}}, \bibinfo {author}
  {\bibfnamefont {C.~F.}\ \bibnamefont {Sopuerta}}, \bibinfo {author}
  {\bibfnamefont {C.~P.~L.}\ \bibnamefont {Berry}}, \bibinfo {author}
  {\bibfnamefont {E.}~\bibnamefont {Berti}}, \bibinfo {author} {\bibfnamefont
  {P.}~\bibnamefont {Amaro-Seoane}}, \bibinfo {author} {\bibfnamefont
  {A.}~\bibnamefont {Petiteau}},\ and\ \bibinfo {author} {\bibfnamefont
  {A.}~\bibnamefont {Klein}},\ }\bibfield  {title} {\bibinfo {title} {{Science
  with the space-based interferometer LISA. V: Extreme mass-ratio inspirals}},\
  }\href {https://doi.org/10.1103/PhysRevD.95.103012} {\bibfield  {journal}
  {\bibinfo  {journal} {Phys. Rev. D}\ }\textbf {\bibinfo {volume} {95}},\
  \bibinfo {pages} {103012} (\bibinfo {year} {2017})},\ \Eprint
  {https://arxiv.org/abs/1703.09722} {arXiv:1703.09722 [gr-qc]} \BibitemShut
  {NoStop}%
\bibitem [{\citenamefont {Berry}\ \emph {et~al.}(2019)\citenamefont {Berry},
  \citenamefont {Hughes}, \citenamefont {Sopuerta}, \citenamefont {Chua},
  \citenamefont {Heffernan}, \citenamefont {Holley-Bockelmann}, \citenamefont
  {Mihaylov}, \citenamefont {Miller},\ and\ \citenamefont
  {Sesana}}]{Berry:2019wgg}%
  \BibitemOpen
  \bibfield  {author} {\bibinfo {author} {\bibfnamefont {C.~P.~L.}\
  \bibnamefont {Berry}}, \bibinfo {author} {\bibfnamefont {S.~A.}\ \bibnamefont
  {Hughes}}, \bibinfo {author} {\bibfnamefont {C.~F.}\ \bibnamefont
  {Sopuerta}}, \bibinfo {author} {\bibfnamefont {A.~J.~K.}\ \bibnamefont
  {Chua}}, \bibinfo {author} {\bibfnamefont {A.}~\bibnamefont {Heffernan}},
  \bibinfo {author} {\bibfnamefont {K.}~\bibnamefont {Holley-Bockelmann}},
  \bibinfo {author} {\bibfnamefont {D.~P.}\ \bibnamefont {Mihaylov}}, \bibinfo
  {author} {\bibfnamefont {M.~C.}\ \bibnamefont {Miller}},\ and\ \bibinfo
  {author} {\bibfnamefont {A.}~\bibnamefont {Sesana}},\ }\bibfield  {title}
  {\bibinfo {title} {{The unique potential of extreme mass-ratio inspirals for
  gravitational-wave astronomy}},\ }\href@noop {} {\bibfield  {journal}
  {\bibinfo  {journal} {Bull. Am. Astron. Soc.}\ }\textbf {\bibinfo {volume}
  {51}},\ \bibinfo {pages} {42} (\bibinfo {year} {2019})},\ \Eprint
  {https://arxiv.org/abs/1903.03686} {arXiv:1903.03686 [astro-ph.HE]}
  \BibitemShut {NoStop}%
\bibitem [{\citenamefont {Barack}\ and\ \citenamefont
  {Cutler}(2004)}]{Barack:2003fp}%
  \BibitemOpen
  \bibfield  {author} {\bibinfo {author} {\bibfnamefont {L.}~\bibnamefont
  {Barack}}\ and\ \bibinfo {author} {\bibfnamefont {C.}~\bibnamefont
  {Cutler}},\ }\bibfield  {title} {\bibinfo {title} {{LISA capture sources:
  Approximate waveforms, signal-to-noise ratios, and parameter estimation
  accuracy}},\ }\href {https://doi.org/10.1103/PhysRevD.69.082005} {\bibfield
  {journal} {\bibinfo  {journal} {Phys. Rev. D}\ }\textbf {\bibinfo {volume}
  {69}},\ \bibinfo {pages} {082005} (\bibinfo {year} {2004})},\ \Eprint
  {https://arxiv.org/abs/gr-qc/0310125} {arXiv:gr-qc/0310125} \BibitemShut
  {NoStop}%
\bibitem [{\citenamefont {Gair}\ \emph {et~al.}(2013)\citenamefont {Gair},
  \citenamefont {Vallisneri}, \citenamefont {Larson},\ and\ \citenamefont
  {Baker}}]{Gair:2012nm}%
  \BibitemOpen
  \bibfield  {author} {\bibinfo {author} {\bibfnamefont {J.~R.}\ \bibnamefont
  {Gair}}, \bibinfo {author} {\bibfnamefont {M.}~\bibnamefont {Vallisneri}},
  \bibinfo {author} {\bibfnamefont {S.~L.}\ \bibnamefont {Larson}},\ and\
  \bibinfo {author} {\bibfnamefont {J.~G.}\ \bibnamefont {Baker}},\ }\bibfield
  {title} {\bibinfo {title} {{Testing general relativity with low-frequency,
  space-based gravitational-wave detectors}},\ }\href
  {https://doi.org/10.12942/lrr-2013-7} {\bibfield  {journal} {\bibinfo
  {journal} {Living Rev. Rel.}\ }\textbf {\bibinfo {volume} {16}},\ \bibinfo
  {pages} {7} (\bibinfo {year} {2013})},\ \Eprint
  {https://arxiv.org/abs/1212.5575} {arXiv:1212.5575 [gr-qc]} \BibitemShut
  {NoStop}%
\bibitem [{\citenamefont {Maselli}\ \emph {et~al.}(2022)\citenamefont
  {Maselli}, \citenamefont {Franchini}, \citenamefont {Gualtieri},
  \citenamefont {Sotiriou}, \citenamefont {Barsanti},\ and\ \citenamefont
  {Pani}}]{Maselli:2021men}%
  \BibitemOpen
  \bibfield  {author} {\bibinfo {author} {\bibfnamefont {A.}~\bibnamefont
  {Maselli}}, \bibinfo {author} {\bibfnamefont {N.}~\bibnamefont {Franchini}},
  \bibinfo {author} {\bibfnamefont {L.}~\bibnamefont {Gualtieri}}, \bibinfo
  {author} {\bibfnamefont {T.~P.}\ \bibnamefont {Sotiriou}}, \bibinfo {author}
  {\bibfnamefont {S.}~\bibnamefont {Barsanti}},\ and\ \bibinfo {author}
  {\bibfnamefont {P.}~\bibnamefont {Pani}},\ }\bibfield  {title} {\bibinfo
  {title} {{Detecting fundamental fields with LISA observations of
  gravitational waves from extreme mass-ratio inspirals}},\ }\href
  {https://doi.org/10.1038/s41550-021-01589-5} {\bibfield  {journal} {\bibinfo
  {journal} {Nature Astron.}\ }\textbf {\bibinfo {volume} {6}},\ \bibinfo
  {pages} {464} (\bibinfo {year} {2022})},\ \Eprint
  {https://arxiv.org/abs/2106.11325} {arXiv:2106.11325 [gr-qc]} \BibitemShut
  {NoStop}%
\bibitem [{\citenamefont {Speri}\ \emph {et~al.}(2026)\citenamefont {Speri},
  \citenamefont {Barsanti}, \citenamefont {Maselli}, \citenamefont {Sotiriou},
  \citenamefont {Warburton}, \citenamefont {van~de Meent}, \citenamefont
  {Chua}, \citenamefont {Burke},\ and\ \citenamefont {Gair}}]{Speri:2024qak}%
  \BibitemOpen
  \bibfield  {author} {\bibinfo {author} {\bibfnamefont {L.}~\bibnamefont
  {Speri}}, \bibinfo {author} {\bibfnamefont {S.}~\bibnamefont {Barsanti}},
  \bibinfo {author} {\bibfnamefont {A.}~\bibnamefont {Maselli}}, \bibinfo
  {author} {\bibfnamefont {T.~P.}\ \bibnamefont {Sotiriou}}, \bibinfo {author}
  {\bibfnamefont {N.}~\bibnamefont {Warburton}}, \bibinfo {author}
  {\bibfnamefont {M.}~\bibnamefont {van~de Meent}}, \bibinfo {author}
  {\bibfnamefont {A.~J.~K.}\ \bibnamefont {Chua}}, \bibinfo {author}
  {\bibfnamefont {O.}~\bibnamefont {Burke}},\ and\ \bibinfo {author}
  {\bibfnamefont {J.}~\bibnamefont {Gair}},\ }\bibfield  {title} {\bibinfo
  {title} {{Probing fundamental physics with extreme mass ratio inspirals: Full
  Bayesian inference for scalar charge}},\ }\href
  {https://doi.org/10.1103/cnhz-6zlk} {\bibfield  {journal} {\bibinfo
  {journal} {Phys. Rev. D}\ }\textbf {\bibinfo {volume} {113}},\ \bibinfo
  {pages} {023036} (\bibinfo {year} {2026})},\ \Eprint
  {https://arxiv.org/abs/2406.07607} {arXiv:2406.07607 [gr-qc]} \BibitemShut
  {NoStop}%
\bibitem [{\citenamefont {Bonga}\ \emph {et~al.}(2019)\citenamefont {Bonga},
  \citenamefont {Yang},\ and\ \citenamefont {Hughes}}]{Bonga:2019ycj}%
  \BibitemOpen
  \bibfield  {author} {\bibinfo {author} {\bibfnamefont {B.}~\bibnamefont
  {Bonga}}, \bibinfo {author} {\bibfnamefont {H.}~\bibnamefont {Yang}},\ and\
  \bibinfo {author} {\bibfnamefont {S.~A.}\ \bibnamefont {Hughes}},\ }\bibfield
   {title} {\bibinfo {title} {{Tidal resonance in extreme mass-ratio
  inspirals}},\ }\href {https://doi.org/10.1103/PhysRevLett.123.101103}
  {\bibfield  {journal} {\bibinfo  {journal} {Phys. Rev. Lett.}\ }\textbf
  {\bibinfo {volume} {123}},\ \bibinfo {pages} {101103} (\bibinfo {year}
  {2019})},\ \Eprint {https://arxiv.org/abs/1905.00030} {arXiv:1905.00030
  [gr-qc]} \BibitemShut {NoStop}%
\bibitem [{\citenamefont {Speri}\ \emph {et~al.}(2023)\citenamefont {Speri},
  \citenamefont {Antonelli}, \citenamefont {Sberna}, \citenamefont {Babak},
  \citenamefont {Barausse}, \citenamefont {Gair},\ and\ \citenamefont
  {Katz}}]{Speri:2022upm}%
  \BibitemOpen
  \bibfield  {author} {\bibinfo {author} {\bibfnamefont {L.}~\bibnamefont
  {Speri}}, \bibinfo {author} {\bibfnamefont {A.}~\bibnamefont {Antonelli}},
  \bibinfo {author} {\bibfnamefont {L.}~\bibnamefont {Sberna}}, \bibinfo
  {author} {\bibfnamefont {S.}~\bibnamefont {Babak}}, \bibinfo {author}
  {\bibfnamefont {E.}~\bibnamefont {Barausse}}, \bibinfo {author}
  {\bibfnamefont {J.~R.}\ \bibnamefont {Gair}},\ and\ \bibinfo {author}
  {\bibfnamefont {M.~L.}\ \bibnamefont {Katz}},\ }\bibfield  {title} {\bibinfo
  {title} {{Probing accretion physics with gravitational waves}},\ }\href
  {https://doi.org/10.1103/PhysRevX.13.021035} {\bibfield  {journal} {\bibinfo
  {journal} {Phys. Rev. X}\ }\textbf {\bibinfo {volume} {13}},\ \bibinfo
  {pages} {021035} (\bibinfo {year} {2023})},\ \Eprint
  {https://arxiv.org/abs/2207.10086} {arXiv:2207.10086 [gr-qc]} \BibitemShut
  {NoStop}%
\bibitem [{\citenamefont {Khalvati}\ \emph {et~al.}(2025)\citenamefont
  {Khalvati}, \citenamefont {Santini}, \citenamefont {Duque}, \citenamefont
  {Speri}, \citenamefont {Gair}, \citenamefont {Yang},\ and\ \citenamefont
  {Brito}}]{Khalvati:2024tzz}%
  \BibitemOpen
  \bibfield  {author} {\bibinfo {author} {\bibfnamefont {H.}~\bibnamefont
  {Khalvati}}, \bibinfo {author} {\bibfnamefont {A.}~\bibnamefont {Santini}},
  \bibinfo {author} {\bibfnamefont {F.}~\bibnamefont {Duque}}, \bibinfo
  {author} {\bibfnamefont {L.}~\bibnamefont {Speri}}, \bibinfo {author}
  {\bibfnamefont {J.}~\bibnamefont {Gair}}, \bibinfo {author} {\bibfnamefont
  {H.}~\bibnamefont {Yang}},\ and\ \bibinfo {author} {\bibfnamefont
  {R.}~\bibnamefont {Brito}},\ }\bibfield  {title} {\bibinfo {title} {{Impact
  of relativistic waveforms in LISA{\textquoteright}s science objectives with
  extreme-mass-ratio inspirals}},\ }\href
  {https://doi.org/10.1103/PhysRevD.111.082010} {\bibfield  {journal} {\bibinfo
   {journal} {Phys. Rev. D}\ }\textbf {\bibinfo {volume} {111}},\ \bibinfo
  {pages} {082010} (\bibinfo {year} {2025})},\ \Eprint
  {https://arxiv.org/abs/2410.17310} {arXiv:2410.17310 [gr-qc]} \BibitemShut
  {NoStop}%
\bibitem [{\citenamefont {Gair}\ \emph {et~al.}(2010)\citenamefont {Gair},
  \citenamefont {Tang},\ and\ \citenamefont {Volonteri}}]{Gair:2010yu}%
  \BibitemOpen
  \bibfield  {author} {\bibinfo {author} {\bibfnamefont {J.~R.}\ \bibnamefont
  {Gair}}, \bibinfo {author} {\bibfnamefont {C.}~\bibnamefont {Tang}},\ and\
  \bibinfo {author} {\bibfnamefont {M.}~\bibnamefont {Volonteri}},\ }\bibfield
  {title} {\bibinfo {title} {{LISA extreme-mass-ratio inspiral events as probes
  of the black hole mass function}},\ }\href
  {https://doi.org/10.1103/PhysRevD.81.104014} {\bibfield  {journal} {\bibinfo
  {journal} {Phys. Rev. D}\ }\textbf {\bibinfo {volume} {81}},\ \bibinfo
  {pages} {104014} (\bibinfo {year} {2010})},\ \Eprint
  {https://arxiv.org/abs/1004.1921} {arXiv:1004.1921 [astro-ph.GA]}
  \BibitemShut {NoStop}%
\bibitem [{\citenamefont {Chapman-Bird}\ \emph {et~al.}(2023)\citenamefont
  {Chapman-Bird}, \citenamefont {Berry},\ and\ \citenamefont
  {Woan}}]{Chapman-Bird:2022tvu}%
  \BibitemOpen
  \bibfield  {author} {\bibinfo {author} {\bibfnamefont {C.~E.~A.}\
  \bibnamefont {Chapman-Bird}}, \bibinfo {author} {\bibfnamefont {C.~P.~L.}\
  \bibnamefont {Berry}},\ and\ \bibinfo {author} {\bibfnamefont
  {G.}~\bibnamefont {Woan}},\ }\bibfield  {title} {\bibinfo {title} {{Rapid
  determination of LISA sensitivity to extreme mass ratio inspirals with
  machine learning}},\ }\href {https://doi.org/10.1093/mnras/stad1397}
  {\bibfield  {journal} {\bibinfo  {journal} {Mon. Not. Roy. Astron. Soc.}\
  }\textbf {\bibinfo {volume} {522}},\ \bibinfo {pages} {6043} (\bibinfo {year}
  {2023})},\ \Eprint {https://arxiv.org/abs/2212.06166} {arXiv:2212.06166
  [astro-ph.HE]} \BibitemShut {NoStop}%
\bibitem [{\citenamefont {Amaro-Seoane}\ \emph {et~al.}(2007)\citenamefont
  {Amaro-Seoane}, \citenamefont {Gair}, \citenamefont {Freitag}, \citenamefont
  {Coleman~Miller}, \citenamefont {Mandel}, \citenamefont {Cutler},\ and\
  \citenamefont {Babak}}]{Amaro-Seoane:2007osp}%
  \BibitemOpen
  \bibfield  {author} {\bibinfo {author} {\bibfnamefont {P.}~\bibnamefont
  {Amaro-Seoane}}, \bibinfo {author} {\bibfnamefont {J.~R.}\ \bibnamefont
  {Gair}}, \bibinfo {author} {\bibfnamefont {M.}~\bibnamefont {Freitag}},
  \bibinfo {author} {\bibfnamefont {M.}~\bibnamefont {Coleman~Miller}},
  \bibinfo {author} {\bibfnamefont {I.}~\bibnamefont {Mandel}}, \bibinfo
  {author} {\bibfnamefont {C.~J.}\ \bibnamefont {Cutler}},\ and\ \bibinfo
  {author} {\bibfnamefont {S.}~\bibnamefont {Babak}},\ }\bibfield  {title}
  {\bibinfo {title} {{Astrophysics, detection and science applications of
  intermediate- and extreme mass-ratio inspirals}},\ }\href
  {https://doi.org/10.1088/0264-9381/24/17/R01} {\bibfield  {journal} {\bibinfo
   {journal} {Classical Quantum Gravity}\ }\textbf {\bibinfo {volume} {24}},\
  \bibinfo {pages} {R113} (\bibinfo {year} {2007})},\ \Eprint
  {https://arxiv.org/abs/astro-ph/0703495} {arXiv:astro-ph/0703495}
  \BibitemShut {NoStop}%
\bibitem [{\citenamefont {MacLeod}\ and\ \citenamefont
  {Hogan}(2008)}]{MacLeod:2007jd}%
  \BibitemOpen
  \bibfield  {author} {\bibinfo {author} {\bibfnamefont {C.~L.}\ \bibnamefont
  {MacLeod}}\ and\ \bibinfo {author} {\bibfnamefont {C.~J.}\ \bibnamefont
  {Hogan}},\ }\bibfield  {title} {\bibinfo {title} {{Precision of Hubble
  constant derived using black hole binary absolute distances and statistical
  redshift information}},\ }\href {https://doi.org/10.1103/PhysRevD.77.043512}
  {\bibfield  {journal} {\bibinfo  {journal} {Phys. Rev. D}\ }\textbf {\bibinfo
  {volume} {77}},\ \bibinfo {pages} {043512} (\bibinfo {year} {2008})},\
  \Eprint {https://arxiv.org/abs/0712.0618} {arXiv:0712.0618 [astro-ph]}
  \BibitemShut {NoStop}%
\bibitem [{\citenamefont {Laghi}\ \emph {et~al.}(2021)\citenamefont {Laghi},
  \citenamefont {Tamanini}, \citenamefont {Del~Pozzo}, \citenamefont {Sesana},
  \citenamefont {Gair}, \citenamefont {Babak},\ and\ \citenamefont
  {Izquierdo-Villalba}}]{Laghi:2021pqk}%
  \BibitemOpen
  \bibfield  {author} {\bibinfo {author} {\bibfnamefont {D.}~\bibnamefont
  {Laghi}}, \bibinfo {author} {\bibfnamefont {N.}~\bibnamefont {Tamanini}},
  \bibinfo {author} {\bibfnamefont {W.}~\bibnamefont {Del~Pozzo}}, \bibinfo
  {author} {\bibfnamefont {A.}~\bibnamefont {Sesana}}, \bibinfo {author}
  {\bibfnamefont {J.}~\bibnamefont {Gair}}, \bibinfo {author} {\bibfnamefont
  {S.}~\bibnamefont {Babak}},\ and\ \bibinfo {author} {\bibfnamefont
  {D.}~\bibnamefont {Izquierdo-Villalba}},\ }\bibfield  {title} {\bibinfo
  {title} {{Gravitational-wave cosmology with extreme mass-ratio inspirals}},\
  }\href {https://doi.org/10.1093/mnras/stab2741} {\bibfield  {journal}
  {\bibinfo  {journal} {Mon. Not. Roy. Astron. Soc.}\ }\textbf {\bibinfo
  {volume} {508}},\ \bibinfo {pages} {4512} (\bibinfo {year} {2021})},\ \Eprint
  {https://arxiv.org/abs/2102.01708} {arXiv:2102.01708 [astro-ph.CO]}
  \BibitemShut {NoStop}%
\bibitem [{\citenamefont {Auclair}\ \emph {et~al.}(2023)\citenamefont {Auclair}
  \emph {et~al.}}]{LISACosmologyWorkingGroup:2022jok}%
  \BibitemOpen
  \bibfield  {author} {\bibinfo {author} {\bibfnamefont {P.}~\bibnamefont
  {Auclair}} \emph {et~al.} (\bibinfo {collaboration} {LISA Cosmology Working
  Group}),\ }\bibfield  {title} {\bibinfo {title} {{Cosmology with the Laser
  Interferometer Space Antenna}},\ }\href
  {https://doi.org/10.1007/s41114-023-00045-2} {\bibfield  {journal} {\bibinfo
  {journal} {Living Rev. Rel.}\ }\textbf {\bibinfo {volume} {26}},\ \bibinfo
  {pages} {5} (\bibinfo {year} {2023})},\ \Eprint
  {https://arxiv.org/abs/2204.05434} {arXiv:2204.05434 [astro-ph.CO]}
  \BibitemShut {NoStop}%
\bibitem [{\citenamefont {Burke}\ \emph {et~al.}(2024)\citenamefont {Burke},
  \citenamefont {Piovano}, \citenamefont {Warburton}, \citenamefont {Lynch},
  \citenamefont {Speri}, \citenamefont {Kavanagh}, \citenamefont {Wardell},
  \citenamefont {Pound}, \citenamefont {Durkan},\ and\ \citenamefont
  {Miller}}]{Burke:2023lno}%
  \BibitemOpen
  \bibfield  {author} {\bibinfo {author} {\bibfnamefont {O.}~\bibnamefont
  {Burke}}, \bibinfo {author} {\bibfnamefont {G.~A.}\ \bibnamefont {Piovano}},
  \bibinfo {author} {\bibfnamefont {N.}~\bibnamefont {Warburton}}, \bibinfo
  {author} {\bibfnamefont {P.}~\bibnamefont {Lynch}}, \bibinfo {author}
  {\bibfnamefont {L.}~\bibnamefont {Speri}}, \bibinfo {author} {\bibfnamefont
  {C.}~\bibnamefont {Kavanagh}}, \bibinfo {author} {\bibfnamefont
  {B.}~\bibnamefont {Wardell}}, \bibinfo {author} {\bibfnamefont
  {A.}~\bibnamefont {Pound}}, \bibinfo {author} {\bibfnamefont
  {L.}~\bibnamefont {Durkan}},\ and\ \bibinfo {author} {\bibfnamefont
  {J.}~\bibnamefont {Miller}},\ }\bibfield  {title} {\bibinfo {title}
  {{Assessing the importance of first postadiabatic terms for small-mass-ratio
  binaries}},\ }\href {https://doi.org/10.1103/PhysRevD.109.124048} {\bibfield
  {journal} {\bibinfo  {journal} {Phys. Rev. D}\ }\textbf {\bibinfo {volume}
  {109}},\ \bibinfo {pages} {124048} (\bibinfo {year} {2024})},\ \Eprint
  {https://arxiv.org/abs/2310.08927} {arXiv:2310.08927 [gr-qc]} \BibitemShut
  {NoStop}%
\bibitem [{\citenamefont {Liang}\ and\ \citenamefont {Wang}(2026)}]{wanghe}%
  \BibitemOpen
  \bibfield  {author} {\bibinfo {author} {\bibfnamefont {B.}~\bibnamefont
  {Liang}}\ and\ \bibinfo {author} {\bibfnamefont {H.}~\bibnamefont {Wang}},\
  }\bibfield  {title} {\bibinfo {title} {Recent advances in simulation-based
  inference for gravitational wave data analysis},\ }\href
  {https://doi.org/10.61977/ati2025020} {\bibfield  {journal} {\bibinfo
  {journal} {Astronomical Techniques and Instruments}\ }\textbf {\bibinfo
  {volume} {3}},\ \bibinfo {pages} {93} (\bibinfo {year} {2026})}\BibitemShut
  {NoStop}%
\bibitem [{\citenamefont {Barack}(2009)}]{Barack:2009ux}%
  \BibitemOpen
  \bibfield  {author} {\bibinfo {author} {\bibfnamefont {L.}~\bibnamefont
  {Barack}},\ }\bibfield  {title} {\bibinfo {title} {{Gravitational self force
  in extreme mass-ratio inspirals}},\ }\href
  {https://doi.org/10.1088/0264-9381/26/21/213001} {\bibfield  {journal}
  {\bibinfo  {journal} {Classical Quantum Gravity}\ }\textbf {\bibinfo {volume}
  {26}},\ \bibinfo {pages} {213001} (\bibinfo {year} {2009})},\ \Eprint
  {https://arxiv.org/abs/0908.1664} {arXiv:0908.1664 [gr-qc]} \BibitemShut
  {NoStop}%
\bibitem [{\citenamefont {Pound}\ \emph {et~al.}(2020)\citenamefont {Pound},
  \citenamefont {Wardell}, \citenamefont {Warburton},\ and\ \citenamefont
  {Miller}}]{Pound:2019lzj}%
  \BibitemOpen
  \bibfield  {author} {\bibinfo {author} {\bibfnamefont {A.}~\bibnamefont
  {Pound}}, \bibinfo {author} {\bibfnamefont {B.}~\bibnamefont {Wardell}},
  \bibinfo {author} {\bibfnamefont {N.}~\bibnamefont {Warburton}},\ and\
  \bibinfo {author} {\bibfnamefont {J.}~\bibnamefont {Miller}},\ }\bibfield
  {title} {\bibinfo {title} {{Second-order self-force calculation of
  gravitational binding energy in compact binaries}},\ }\href
  {https://doi.org/10.1103/PhysRevLett.124.021101} {\bibfield  {journal}
  {\bibinfo  {journal} {Phys. Rev. Lett.}\ }\textbf {\bibinfo {volume} {124}},\
  \bibinfo {pages} {021101} (\bibinfo {year} {2020})},\ \Eprint
  {https://arxiv.org/abs/1908.07419} {arXiv:1908.07419 [gr-qc]} \BibitemShut
  {NoStop}%
\bibitem [{\citenamefont {Warburton}\ \emph {et~al.}(2021)\citenamefont
  {Warburton}, \citenamefont {Pound}, \citenamefont {Wardell}, \citenamefont
  {Miller},\ and\ \citenamefont {Durkan}}]{Warburton:2021kwk}%
  \BibitemOpen
  \bibfield  {author} {\bibinfo {author} {\bibfnamefont {N.}~\bibnamefont
  {Warburton}}, \bibinfo {author} {\bibfnamefont {A.}~\bibnamefont {Pound}},
  \bibinfo {author} {\bibfnamefont {B.}~\bibnamefont {Wardell}}, \bibinfo
  {author} {\bibfnamefont {J.}~\bibnamefont {Miller}},\ and\ \bibinfo {author}
  {\bibfnamefont {L.}~\bibnamefont {Durkan}},\ }\bibfield  {title} {\bibinfo
  {title} {{Gravitational-wave energy flux for compact binaries through second
  order in the mass ratio}},\ }\href
  {https://doi.org/10.1103/PhysRevLett.127.151102} {\bibfield  {journal}
  {\bibinfo  {journal} {Phys. Rev. Lett.}\ }\textbf {\bibinfo {volume} {127}},\
  \bibinfo {pages} {151102} (\bibinfo {year} {2021})},\ \Eprint
  {https://arxiv.org/abs/2107.01298} {arXiv:2107.01298 [gr-qc]} \BibitemShut
  {NoStop}%
\bibitem [{\citenamefont {Wardell}\ \emph {et~al.}(2023)\citenamefont
  {Wardell}, \citenamefont {Pound}, \citenamefont {Warburton}, \citenamefont
  {Miller}, \citenamefont {Durkan},\ and\ \citenamefont
  {Le~Tiec}}]{Wardell:2021fyy}%
  \BibitemOpen
  \bibfield  {author} {\bibinfo {author} {\bibfnamefont {B.}~\bibnamefont
  {Wardell}}, \bibinfo {author} {\bibfnamefont {A.}~\bibnamefont {Pound}},
  \bibinfo {author} {\bibfnamefont {N.}~\bibnamefont {Warburton}}, \bibinfo
  {author} {\bibfnamefont {J.}~\bibnamefont {Miller}}, \bibinfo {author}
  {\bibfnamefont {L.}~\bibnamefont {Durkan}},\ and\ \bibinfo {author}
  {\bibfnamefont {A.}~\bibnamefont {Le~Tiec}},\ }\bibfield  {title} {\bibinfo
  {title} {{Gravitational waveforms for compact binaries from second-order
  self-force theory}},\ }\href {https://doi.org/10.1103/PhysRevLett.130.241402}
  {\bibfield  {journal} {\bibinfo  {journal} {Phys. Rev. Lett.}\ }\textbf
  {\bibinfo {volume} {130}},\ \bibinfo {pages} {241402} (\bibinfo {year}
  {2023})},\ \Eprint {https://arxiv.org/abs/2112.12265} {arXiv:2112.12265
  [gr-qc]} \BibitemShut {NoStop}%
\bibitem [{\citenamefont {Poisson}\ \emph {et~al.}(2011)\citenamefont
  {Poisson}, \citenamefont {Pound},\ and\ \citenamefont
  {Vega}}]{Poisson:2011nh}%
  \BibitemOpen
  \bibfield  {author} {\bibinfo {author} {\bibfnamefont {E.}~\bibnamefont
  {Poisson}}, \bibinfo {author} {\bibfnamefont {A.}~\bibnamefont {Pound}},\
  and\ \bibinfo {author} {\bibfnamefont {I.}~\bibnamefont {Vega}},\ }\bibfield
  {title} {\bibinfo {title} {{The Motion of point particles in curved
  spacetime}},\ }\href {https://doi.org/10.12942/lrr-2011-7} {\bibfield
  {journal} {\bibinfo  {journal} {Living Rev. Rel.}\ }\textbf {\bibinfo
  {volume} {14}},\ \bibinfo {pages} {7} (\bibinfo {year} {2011})},\ \Eprint
  {https://arxiv.org/abs/1102.0529} {arXiv:1102.0529 [gr-qc]} \BibitemShut
  {NoStop}%
\bibitem [{\citenamefont {Barack}\ and\ \citenamefont
  {Pound}(2019)}]{Barack:2018yvs}%
  \BibitemOpen
  \bibfield  {author} {\bibinfo {author} {\bibfnamefont {L.}~\bibnamefont
  {Barack}}\ and\ \bibinfo {author} {\bibfnamefont {A.}~\bibnamefont {Pound}},\
  }\bibfield  {title} {\bibinfo {title} {{Self-force and radiation reaction in
  general relativity}},\ }\href {https://doi.org/10.1088/1361-6633/aae552}
  {\bibfield  {journal} {\bibinfo  {journal} {Rept. Prog. Phys.}\ }\textbf
  {\bibinfo {volume} {82}},\ \bibinfo {pages} {016904} (\bibinfo {year}
  {2019})},\ \Eprint {https://arxiv.org/abs/1805.10385} {arXiv:1805.10385
  [gr-qc]} \BibitemShut {NoStop}%
\bibitem [{\citenamefont {Pound}\ and\ \citenamefont
  {Wardell}(2020)}]{Pound:2021qin}%
  \BibitemOpen
  \bibfield  {author} {\bibinfo {author} {\bibfnamefont {A.}~\bibnamefont
  {Pound}}\ and\ \bibinfo {author} {\bibfnamefont {B.}~\bibnamefont
  {Wardell}},\ }\bibfield  {title} {\bibinfo {title} {Black hole perturbation
  theory and gravitational self-force},\ }\bibfield  {booktitle} {\emph
  {\bibinfo {booktitle} {Handbook of Gravitational Wave Astronomy}},\
  }\href@noop {} {\ ,\ \bibinfo {pages} {1} (\bibinfo {year}
  {2020})}\BibitemShut {NoStop}%
\bibitem [{\citenamefont {Mino}(2003)}]{Mino:2003yg}%
  \BibitemOpen
  \bibfield  {author} {\bibinfo {author} {\bibfnamefont {Y.}~\bibnamefont
  {Mino}},\ }\bibfield  {title} {\bibinfo {title} {{Perturbative approach to an
  orbital evolution around a supermassive black hole}},\ }\href
  {https://doi.org/10.1103/PhysRevD.67.084027} {\bibfield  {journal} {\bibinfo
  {journal} {Phys. Rev. D}\ }\textbf {\bibinfo {volume} {67}},\ \bibinfo
  {pages} {084027} (\bibinfo {year} {2003})},\ \Eprint
  {https://arxiv.org/abs/gr-qc/0302075} {arXiv:gr-qc/0302075} \BibitemShut
  {NoStop}%
\bibitem [{\citenamefont {Drasco}\ \emph {et~al.}(2005)\citenamefont {Drasco},
  \citenamefont {Flanagan},\ and\ \citenamefont {Hughes}}]{Drasco:2005is}%
  \BibitemOpen
  \bibfield  {author} {\bibinfo {author} {\bibfnamefont {S.}~\bibnamefont
  {Drasco}}, \bibinfo {author} {\bibfnamefont {E.~E.}\ \bibnamefont
  {Flanagan}},\ and\ \bibinfo {author} {\bibfnamefont {S.~A.}\ \bibnamefont
  {Hughes}},\ }\bibfield  {title} {\bibinfo {title} {{Computing inspirals in
  Kerr in the adiabatic regime. I. The scalar case}},\ }\href
  {https://doi.org/10.1088/0264-9381/22/15/011} {\bibfield  {journal} {\bibinfo
   {journal} {Classical Quantum Gravity}\ }\textbf {\bibinfo {volume} {22}},\
  \bibinfo {pages} {S801} (\bibinfo {year} {2005})},\ \Eprint
  {https://arxiv.org/abs/gr-qc/0505075} {arXiv:gr-qc/0505075} \BibitemShut
  {NoStop}%
\bibitem [{\citenamefont {Sago}\ \emph {et~al.}(2006)\citenamefont {Sago},
  \citenamefont {Tanaka}, \citenamefont {Hikida}, \citenamefont {Ganz},\ and\
  \citenamefont {Nakano}}]{Sago:2005fn}%
  \BibitemOpen
  \bibfield  {author} {\bibinfo {author} {\bibfnamefont {N.}~\bibnamefont
  {Sago}}, \bibinfo {author} {\bibfnamefont {T.}~\bibnamefont {Tanaka}},
  \bibinfo {author} {\bibfnamefont {W.}~\bibnamefont {Hikida}}, \bibinfo
  {author} {\bibfnamefont {K.}~\bibnamefont {Ganz}},\ and\ \bibinfo {author}
  {\bibfnamefont {H.}~\bibnamefont {Nakano}},\ }\bibfield  {title} {\bibinfo
  {title} {{The Adiabatic evolution of orbital parameters in the Kerr
  spacetime}},\ }\href {https://doi.org/10.1143/PTP.115.873} {\bibfield
  {journal} {\bibinfo  {journal} {Prog. Theor. Phys.}\ }\textbf {\bibinfo
  {volume} {115}},\ \bibinfo {pages} {873} (\bibinfo {year} {2006})},\ \Eprint
  {https://arxiv.org/abs/gr-qc/0511151} {arXiv:gr-qc/0511151} \BibitemShut
  {NoStop}%
\bibitem [{\citenamefont {Isoyama}\ \emph {et~al.}(2019)\citenamefont
  {Isoyama}, \citenamefont {Fujita}, \citenamefont {Nakano}, \citenamefont
  {Sago},\ and\ \citenamefont {Tanaka}}]{Isoyama:2018sib}%
  \BibitemOpen
  \bibfield  {author} {\bibinfo {author} {\bibfnamefont {S.}~\bibnamefont
  {Isoyama}}, \bibinfo {author} {\bibfnamefont {R.}~\bibnamefont {Fujita}},
  \bibinfo {author} {\bibfnamefont {H.}~\bibnamefont {Nakano}}, \bibinfo
  {author} {\bibfnamefont {N.}~\bibnamefont {Sago}},\ and\ \bibinfo {author}
  {\bibfnamefont {T.}~\bibnamefont {Tanaka}},\ }\bibfield  {title} {\bibinfo
  {title} {{{\textquotedblleft}Flux-balance formulae{\textquotedblright} for
  extreme mass-ratio inspirals}},\ }\href {https://doi.org/10.1093/ptep/pty136}
  {\bibfield  {journal} {\bibinfo  {journal} {Prog. Theor. Exp. Phys.}\
  }\textbf {\bibinfo {volume} {2019}},\ \bibinfo {pages} {013E01} (\bibinfo
  {year} {2019})},\ \Eprint {https://arxiv.org/abs/1809.11118}
  {arXiv:1809.11118 [gr-qc]} \BibitemShut {NoStop}%
\bibitem [{\citenamefont {Grant}(2025)}]{Grant:2024ivt}%
  \BibitemOpen
  \bibfield  {author} {\bibinfo {author} {\bibfnamefont {A.~M.}\ \bibnamefont
  {Grant}},\ }\bibfield  {title} {\bibinfo {title} {{Flux-balance laws for
  spinning bodies under the gravitational self-force}},\ }\href
  {https://doi.org/10.1103/PhysRevD.111.084015} {\bibfield  {journal} {\bibinfo
   {journal} {Phys. Rev. D}\ }\textbf {\bibinfo {volume} {111}},\ \bibinfo
  {pages} {084015} (\bibinfo {year} {2025})},\ \Eprint
  {https://arxiv.org/abs/2406.10343} {arXiv:2406.10343 [gr-qc]} \BibitemShut
  {NoStop}%
\bibitem [{\citenamefont {Teukolsky}(1972)}]{Teukolsky:1972my}%
  \BibitemOpen
  \bibfield  {author} {\bibinfo {author} {\bibfnamefont {S.~A.}\ \bibnamefont
  {Teukolsky}},\ }\bibfield  {title} {\bibinfo {title} {{Rotating black holes -
  separable wave equations for gravitational and electromagnetic
  perturbations}},\ }\href {https://doi.org/10.1103/PhysRevLett.29.1114}
  {\bibfield  {journal} {\bibinfo  {journal} {Phys. Rev. Lett.}\ }\textbf
  {\bibinfo {volume} {29}},\ \bibinfo {pages} {1114} (\bibinfo {year}
  {1972})}\BibitemShut {NoStop}%
\bibitem [{\citenamefont {Teukolsky}(1973)}]{Teukolsky:1973ha}%
  \BibitemOpen
  \bibfield  {author} {\bibinfo {author} {\bibfnamefont {S.~A.}\ \bibnamefont
  {Teukolsky}},\ }\bibfield  {title} {\bibinfo {title} {{Perturbations of a
  rotating black hole. 1. Fundamental equations for gravitational
  electromagnetic and neutrino field perturbations}},\ }\href
  {https://doi.org/10.1086/152444} {\bibfield  {journal} {\bibinfo  {journal}
  {Astrophys. J.}\ }\textbf {\bibinfo {volume} {185}},\ \bibinfo {pages} {635}
  (\bibinfo {year} {1973})}\BibitemShut {NoStop}%
\bibitem [{\citenamefont {Press}\ and\ \citenamefont
  {Teukolsky}(1973)}]{Press:1973zz}%
  \BibitemOpen
  \bibfield  {author} {\bibinfo {author} {\bibfnamefont {W.~H.}\ \bibnamefont
  {Press}}\ and\ \bibinfo {author} {\bibfnamefont {S.~A.}\ \bibnamefont
  {Teukolsky}},\ }\bibfield  {title} {\bibinfo {title} {{Perturbations of a
  rotating black hole. II. Dynamical stability of the Kerr metric}},\ }\href
  {https://doi.org/10.1086/152445} {\bibfield  {journal} {\bibinfo  {journal}
  {Astrophys. J.}\ }\textbf {\bibinfo {volume} {185}},\ \bibinfo {pages} {649}
  (\bibinfo {year} {1973})}\BibitemShut {NoStop}%
\bibitem [{\citenamefont {Teukolsky}\ and\ \citenamefont
  {Press}(1974)}]{Teukolsky:1974yv}%
  \BibitemOpen
  \bibfield  {author} {\bibinfo {author} {\bibfnamefont {S.~A.}\ \bibnamefont
  {Teukolsky}}\ and\ \bibinfo {author} {\bibfnamefont {W.~H.}\ \bibnamefont
  {Press}},\ }\bibfield  {title} {\bibinfo {title} {{Perturbations of a
  rotating black hole. III - Interaction of the hole with gravitational and
  electromagnetic radiation}},\ }\href {https://doi.org/10.1086/153180}
  {\bibfield  {journal} {\bibinfo  {journal} {Astrophys. J.}\ }\textbf
  {\bibinfo {volume} {193}},\ \bibinfo {pages} {443} (\bibinfo {year}
  {1974})}\BibitemShut {NoStop}%
\bibitem [{\citenamefont {Wald}(1973)}]{Wald:1973wwa}%
  \BibitemOpen
  \bibfield  {author} {\bibinfo {author} {\bibfnamefont {R.~M.}\ \bibnamefont
  {Wald}},\ }\bibfield  {title} {\bibinfo {title} {{On perturbations of a Kerr
  black hole}},\ }\href {https://doi.org/10.1063/1.1666203} {\bibfield
  {journal} {\bibinfo  {journal} {J. Math. Phys.}\ }\textbf {\bibinfo {volume}
  {14}},\ \bibinfo {pages} {1453} (\bibinfo {year} {1973})}\BibitemShut
  {NoStop}%
\bibitem [{\citenamefont {Hinderer}\ and\ \citenamefont
  {Flanagan}(2008)}]{Hinderer:2008dm}%
  \BibitemOpen
  \bibfield  {author} {\bibinfo {author} {\bibfnamefont {T.}~\bibnamefont
  {Hinderer}}\ and\ \bibinfo {author} {\bibfnamefont {E.~E.}\ \bibnamefont
  {Flanagan}},\ }\bibfield  {title} {\bibinfo {title} {{Two timescale analysis
  of extreme mass ratio inspirals in Kerr. I. Orbital motion}},\ }\href
  {https://doi.org/10.1103/PhysRevD.78.064028} {\bibfield  {journal} {\bibinfo
  {journal} {Phys. Rev. D}\ }\textbf {\bibinfo {volume} {78}},\ \bibinfo
  {pages} {064028} (\bibinfo {year} {2008})},\ \Eprint
  {https://arxiv.org/abs/0805.3337} {arXiv:0805.3337 [gr-qc]} \BibitemShut
  {NoStop}%
\bibitem [{\citenamefont {Miller}\ and\ \citenamefont
  {Pound}(2021)}]{Miller:2020bft}%
  \BibitemOpen
  \bibfield  {author} {\bibinfo {author} {\bibfnamefont {J.}~\bibnamefont
  {Miller}}\ and\ \bibinfo {author} {\bibfnamefont {A.}~\bibnamefont {Pound}},\
  }\bibfield  {title} {\bibinfo {title} {{Two-timescale evolution of
  extreme-mass-ratio inspirals: Waveform generation scheme for quasicircular
  orbits in Schwarzschild spacetime}},\ }\href
  {https://doi.org/10.1103/PhysRevD.103.064048} {\bibfield  {journal} {\bibinfo
   {journal} {Phys. Rev. D}\ }\textbf {\bibinfo {volume} {103}},\ \bibinfo
  {pages} {064048} (\bibinfo {year} {2021})},\ \Eprint
  {https://arxiv.org/abs/2006.11263} {arXiv:2006.11263 [gr-qc]} \BibitemShut
  {NoStop}%
\bibitem [{\citenamefont {Mathews}\ \emph {et~al.}(2026)\citenamefont
  {Mathews}, \citenamefont {Wardell}, \citenamefont {Pound},\ and\
  \citenamefont {Warburton}}]{Mathews:2025txc}%
  \BibitemOpen
  \bibfield  {author} {\bibinfo {author} {\bibfnamefont {J.}~\bibnamefont
  {Mathews}}, \bibinfo {author} {\bibfnamefont {B.}~\bibnamefont {Wardell}},
  \bibinfo {author} {\bibfnamefont {A.}~\bibnamefont {Pound}},\ and\ \bibinfo
  {author} {\bibfnamefont {N.}~\bibnamefont {Warburton}},\ }\bibfield  {title}
  {\bibinfo {title} {{Postadiabatic self-force waveforms: Slowly spinning
  primary and precessing secondary}},\ }\href
  {https://doi.org/10.1103/ph3p-mscl} {\bibfield  {journal} {\bibinfo
  {journal} {Phys. Rev. D}\ }\textbf {\bibinfo {volume} {113}},\ \bibinfo
  {pages} {064034} (\bibinfo {year} {2026})},\ \Eprint
  {https://arxiv.org/abs/2510.16113} {arXiv:2510.16113 [gr-qc]} \BibitemShut
  {NoStop}%
\bibitem [{\citenamefont {Hughes}(2000)}]{Hughes:1999bq}%
  \BibitemOpen
  \bibfield  {author} {\bibinfo {author} {\bibfnamefont {S.~A.}\ \bibnamefont
  {Hughes}},\ }\bibfield  {title} {\bibinfo {title} {{The Evolution of
  circular, nonequatorial orbits of Kerr black holes due to gravitational wave
  emission}},\ }\href {https://doi.org/10.1103/PhysRevD.65.069902} {\bibfield
  {journal} {\bibinfo  {journal} {Phys. Rev. D}\ }\textbf {\bibinfo {volume}
  {61}},\ \bibinfo {pages} {084004} (\bibinfo {year} {2000})},\ \bibinfo {note}
  {[Erratum: Phys.Rev.D 63, 049902 (2001), Erratum: Phys.Rev.D 65, 069902
  (2002), Erratum: Phys.Rev.D 67, 089901 (2003), Erratum: Phys.Rev.D 78, 109902
  (2008), Erratum: Phys.Rev.D 90, 109904 (2014)]},\ \Eprint
  {https://arxiv.org/abs/gr-qc/9910091} {arXiv:gr-qc/9910091} \BibitemShut
  {NoStop}%
\bibitem [{\citenamefont {Hughes}(2001)}]{Hughes:2001jr}%
  \BibitemOpen
  \bibfield  {author} {\bibinfo {author} {\bibfnamefont {S.~A.}\ \bibnamefont
  {Hughes}},\ }\bibfield  {title} {\bibinfo {title} {{Evolution of circular,
  nonequatorial orbits of Kerr black holes due to gravitational wave emission.
  II. Inspiral trajectories and gravitational wave forms}},\ }\href
  {https://doi.org/10.1103/PhysRevD.64.064004} {\bibfield  {journal} {\bibinfo
  {journal} {Phys. Rev. D}\ }\textbf {\bibinfo {volume} {64}},\ \bibinfo
  {pages} {064004} (\bibinfo {year} {2001})},\ \bibinfo {note} {[Erratum:
  Phys.Rev.D 88, 109902 (2013)]},\ \Eprint
  {https://arxiv.org/abs/gr-qc/0104041} {arXiv:gr-qc/0104041} \BibitemShut
  {NoStop}%
\bibitem [{\citenamefont {Drasco}\ and\ \citenamefont
  {Hughes}(2004)}]{Drasco:2003ky}%
  \BibitemOpen
  \bibfield  {author} {\bibinfo {author} {\bibfnamefont {S.}~\bibnamefont
  {Drasco}}\ and\ \bibinfo {author} {\bibfnamefont {S.~A.}\ \bibnamefont
  {Hughes}},\ }\bibfield  {title} {\bibinfo {title} {{Rotating black hole orbit
  functionals in the frequency domain}},\ }\href
  {https://doi.org/10.1103/PhysRevD.69.044015} {\bibfield  {journal} {\bibinfo
  {journal} {Phys. Rev. D}\ }\textbf {\bibinfo {volume} {69}},\ \bibinfo
  {pages} {044015} (\bibinfo {year} {2004})},\ \Eprint
  {https://arxiv.org/abs/astro-ph/0308479} {arXiv:astro-ph/0308479}
  \BibitemShut {NoStop}%
\bibitem [{\citenamefont {Drasco}\ and\ \citenamefont
  {Hughes}(2006)}]{Drasco:2005kz}%
  \BibitemOpen
  \bibfield  {author} {\bibinfo {author} {\bibfnamefont {S.}~\bibnamefont
  {Drasco}}\ and\ \bibinfo {author} {\bibfnamefont {S.~A.}\ \bibnamefont
  {Hughes}},\ }\bibfield  {title} {\bibinfo {title} {{Gravitational wave
  snapshots of generic extreme mass ratio inspirals}},\ }\href
  {https://doi.org/10.1103/PhysRevD.73.024027} {\bibfield  {journal} {\bibinfo
  {journal} {Phys. Rev. D}\ }\textbf {\bibinfo {volume} {73}},\ \bibinfo
  {pages} {024027} (\bibinfo {year} {2006})},\ \bibinfo {note} {[Erratum:
  Phys.Rev.D 88, 109905 (2013), Erratum: Phys.Rev.D 90, 109905 (2014)]},\
  \Eprint {https://arxiv.org/abs/gr-qc/0509101} {arXiv:gr-qc/0509101}
  \BibitemShut {NoStop}%
\bibitem [{\citenamefont {Hughes}\ \emph {et~al.}(2021)\citenamefont {Hughes},
  \citenamefont {Warburton}, \citenamefont {Khanna}, \citenamefont {Chua},\
  and\ \citenamefont {Katz}}]{Hughes:2021exa}%
  \BibitemOpen
  \bibfield  {author} {\bibinfo {author} {\bibfnamefont {S.~A.}\ \bibnamefont
  {Hughes}}, \bibinfo {author} {\bibfnamefont {N.}~\bibnamefont {Warburton}},
  \bibinfo {author} {\bibfnamefont {G.}~\bibnamefont {Khanna}}, \bibinfo
  {author} {\bibfnamefont {A.~J.~K.}\ \bibnamefont {Chua}},\ and\ \bibinfo
  {author} {\bibfnamefont {M.~L.}\ \bibnamefont {Katz}},\ }\bibfield  {title}
  {\bibinfo {title} {{Adiabatic waveforms for extreme mass-ratio inspirals via
  multivoice decomposition in time and frequency}},\ }\href
  {https://doi.org/10.1103/PhysRevD.103.104014} {\bibfield  {journal} {\bibinfo
   {journal} {Phys. Rev. D}\ }\textbf {\bibinfo {volume} {103}},\ \bibinfo
  {pages} {104014} (\bibinfo {year} {2021})},\ \bibinfo {note} {[Erratum:
  Phys.Rev.D 107, 089901 (2023)]},\ \Eprint {https://arxiv.org/abs/2102.02713}
  {arXiv:2102.02713 [gr-qc]} \BibitemShut {NoStop}%
\bibitem [{\citenamefont {Mano}\ \emph
  {et~al.}(1996{\natexlab{a}})\citenamefont {Mano}, \citenamefont {Suzuki},\
  and\ \citenamefont {Takasugi}}]{Mano:1996mf}%
  \BibitemOpen
  \bibfield  {author} {\bibinfo {author} {\bibfnamefont {S.}~\bibnamefont
  {Mano}}, \bibinfo {author} {\bibfnamefont {H.}~\bibnamefont {Suzuki}},\ and\
  \bibinfo {author} {\bibfnamefont {E.}~\bibnamefont {Takasugi}},\ }\bibfield
  {title} {\bibinfo {title} {{Analytic solutions of the Regge-Wheeler equation
  and the postMinkowskian expansion}},\ }\href
  {https://doi.org/10.1143/PTP.96.549} {\bibfield  {journal} {\bibinfo
  {journal} {Prog. Theor. Phys.}\ }\textbf {\bibinfo {volume} {96}},\ \bibinfo
  {pages} {549} (\bibinfo {year} {1996}{\natexlab{a}})},\ \Eprint
  {https://arxiv.org/abs/gr-qc/9605057} {arXiv:gr-qc/9605057} \BibitemShut
  {NoStop}%
\bibitem [{\citenamefont {Mano}\ \emph
  {et~al.}(1996{\natexlab{b}})\citenamefont {Mano}, \citenamefont {Suzuki},\
  and\ \citenamefont {Takasugi}}]{Mano:1996vt}%
  \BibitemOpen
  \bibfield  {author} {\bibinfo {author} {\bibfnamefont {S.}~\bibnamefont
  {Mano}}, \bibinfo {author} {\bibfnamefont {H.}~\bibnamefont {Suzuki}},\ and\
  \bibinfo {author} {\bibfnamefont {E.}~\bibnamefont {Takasugi}},\ }\bibfield
  {title} {\bibinfo {title} {{Analytic solutions of the Teukolsky equation and
  their low frequency expansions}},\ }\href
  {https://doi.org/10.1143/PTP.95.1079} {\bibfield  {journal} {\bibinfo
  {journal} {Prog. Theor. Phys.}\ }\textbf {\bibinfo {volume} {95}},\ \bibinfo
  {pages} {1079} (\bibinfo {year} {1996}{\natexlab{b}})},\ \Eprint
  {https://arxiv.org/abs/gr-qc/9603020} {arXiv:gr-qc/9603020} \BibitemShut
  {NoStop}%
\bibitem [{\citenamefont {Fujita}\ \emph {et~al.}(2009)\citenamefont {Fujita},
  \citenamefont {Hikida},\ and\ \citenamefont {Tagoshi}}]{Fujita:2009us}%
  \BibitemOpen
  \bibfield  {author} {\bibinfo {author} {\bibfnamefont {R.}~\bibnamefont
  {Fujita}}, \bibinfo {author} {\bibfnamefont {W.}~\bibnamefont {Hikida}},\
  and\ \bibinfo {author} {\bibfnamefont {H.}~\bibnamefont {Tagoshi}},\
  }\bibfield  {title} {\bibinfo {title} {{An efficient numerical method for
  computing gravitational waves induced by a particle moving on eccentric
  inclined orbits around a Kerr black hole}},\ }\href
  {https://doi.org/10.1143/PTP.121.843} {\bibfield  {journal} {\bibinfo
  {journal} {Prog. Theor. Phys.}\ }\textbf {\bibinfo {volume} {121}},\ \bibinfo
  {pages} {843} (\bibinfo {year} {2009})},\ \Eprint
  {https://arxiv.org/abs/0904.3810} {arXiv:0904.3810 [gr-qc]} \BibitemShut
  {NoStop}%
\bibitem [{\citenamefont {Fujita}(2012)}]{Fujita:2012cm}%
  \BibitemOpen
  \bibfield  {author} {\bibinfo {author} {\bibfnamefont {R.}~\bibnamefont
  {Fujita}},\ }\bibfield  {title} {\bibinfo {title} {{Gravitational waves from
  a particle in circular orbits around a Schwarzschild black hole to the 22nd
  post-Newtonian order}},\ }\href {https://doi.org/10.1143/PTP.128.971}
  {\bibfield  {journal} {\bibinfo  {journal} {Prog. Theor. Phys.}\ }\textbf
  {\bibinfo {volume} {128}},\ \bibinfo {pages} {971} (\bibinfo {year}
  {2012})},\ \Eprint {https://arxiv.org/abs/1211.5535} {arXiv:1211.5535
  [gr-qc]} \BibitemShut {NoStop}%
\bibitem [{\citenamefont {Sago}\ and\ \citenamefont
  {Fujita}(2015)}]{Sago:2015rpa}%
  \BibitemOpen
  \bibfield  {author} {\bibinfo {author} {\bibfnamefont {N.}~\bibnamefont
  {Sago}}\ and\ \bibinfo {author} {\bibfnamefont {R.}~\bibnamefont {Fujita}},\
  }\bibfield  {title} {\bibinfo {title} {{Calculation of radiation reaction
  effect on orbital parameters in Kerr spacetime}},\ }\href
  {https://doi.org/10.1093/ptep/ptv092} {\bibfield  {journal} {\bibinfo
  {journal} {Prog. Theor. Exp. Phys.}\ }\textbf {\bibinfo {volume} {2015}},\
  \bibinfo {pages} {073E03} (\bibinfo {year} {2015})},\ \Eprint
  {https://arxiv.org/abs/1505.01600} {arXiv:1505.01600 [gr-qc]} \BibitemShut
  {NoStop}%
\bibitem [{\citenamefont {Munna}(2020)}]{Munna:2020iju}%
  \BibitemOpen
  \bibfield  {author} {\bibinfo {author} {\bibfnamefont {C.}~\bibnamefont
  {Munna}},\ }\bibfield  {title} {\bibinfo {title} {{Analytic post-Newtonian
  expansion of the energy and angular momentum radiated to infinity by
  eccentric-orbit nonspinning extreme-mass-ratio inspirals to the 19th
  order}},\ }\href {https://doi.org/10.1103/PhysRevD.102.124001} {\bibfield
  {journal} {\bibinfo  {journal} {Phys. Rev. D}\ }\textbf {\bibinfo {volume}
  {102}},\ \bibinfo {pages} {124001} (\bibinfo {year} {2020})},\ \Eprint
  {https://arxiv.org/abs/2008.10622} {arXiv:2008.10622 [gr-qc]} \BibitemShut
  {NoStop}%
\bibitem [{\citenamefont {Sago}\ \emph {et~al.}(2025)\citenamefont {Sago},
  \citenamefont {Fujita},\ and\ \citenamefont {Nakano}}]{Sago:2024mgh}%
  \BibitemOpen
  \bibfield  {author} {\bibinfo {author} {\bibfnamefont {N.}~\bibnamefont
  {Sago}}, \bibinfo {author} {\bibfnamefont {R.}~\bibnamefont {Fujita}},\ and\
  \bibinfo {author} {\bibfnamefont {H.}~\bibnamefont {Nakano}},\ }\bibfield
  {title} {\bibinfo {title} {{Post-Newtonian templates for phase evolution of
  spherical extreme mass ratio inspirals}},\ }\href
  {https://doi.org/10.1103/PhysRevD.111.064043} {\bibfield  {journal} {\bibinfo
   {journal} {Phys. Rev. D}\ }\textbf {\bibinfo {volume} {111}},\ \bibinfo
  {pages} {064043} (\bibinfo {year} {2025})},\ \Eprint
  {https://arxiv.org/abs/2411.09147} {arXiv:2411.09147 [gr-qc]} \BibitemShut
  {NoStop}%
\bibitem [{\citenamefont {Castillo}\ \emph {et~al.}(2025)\citenamefont
  {Castillo}, \citenamefont {Evans}, \citenamefont {Kavanagh}, \citenamefont
  {Neef}, \citenamefont {Ottewill},\ and\ \citenamefont
  {Wardell}}]{Castillo:2024isq}%
  \BibitemOpen
  \bibfield  {author} {\bibinfo {author} {\bibfnamefont {J.~C.}\ \bibnamefont
  {Castillo}}, \bibinfo {author} {\bibfnamefont {C.~R.}\ \bibnamefont {Evans}},
  \bibinfo {author} {\bibfnamefont {C.}~\bibnamefont {Kavanagh}}, \bibinfo
  {author} {\bibfnamefont {J.}~\bibnamefont {Neef}}, \bibinfo {author}
  {\bibfnamefont {A.}~\bibnamefont {Ottewill}},\ and\ \bibinfo {author}
  {\bibfnamefont {B.}~\bibnamefont {Wardell}},\ }\bibfield  {title} {\bibinfo
  {title} {{Post-Newtonian expansion of gravitational energy and angular
  momentum fluxes: Inclined spherical orbits about a Kerr black hole}},\ }\href
  {https://doi.org/10.1103/PhysRevD.111.084004} {\bibfield  {journal} {\bibinfo
   {journal} {Phys. Rev. D}\ }\textbf {\bibinfo {volume} {111}},\ \bibinfo
  {pages} {084004} (\bibinfo {year} {2025})},\ \Eprint
  {https://arxiv.org/abs/2411.09700} {arXiv:2411.09700 [gr-qc]} \BibitemShut
  {NoStop}%
\bibitem [{\citenamefont {Sago}\ \emph {et~al.}(2026)\citenamefont {Sago},
  \citenamefont {Fujita}, \citenamefont {Isoyama},\ and\ \citenamefont
  {Nakano}}]{Sago:2026gxb}%
  \BibitemOpen
  \bibfield  {author} {\bibinfo {author} {\bibfnamefont {N.}~\bibnamefont
  {Sago}}, \bibinfo {author} {\bibfnamefont {R.}~\bibnamefont {Fujita}},
  \bibinfo {author} {\bibfnamefont {S.}~\bibnamefont {Isoyama}},\ and\ \bibinfo
  {author} {\bibfnamefont {H.}~\bibnamefont {Nakano}},\ }\bibfield  {title}
  {\bibinfo {title} {{Secular evolution of orbital parameters for general bound
  orbits in Kerr spacetime}},\ }\href@noop {} {\  (\bibinfo {year} {2026})},\
  \Eprint {https://arxiv.org/abs/2603.27941} {arXiv:2603.27941 [gr-qc]}
  \BibitemShut {NoStop}%
\bibitem [{\citenamefont {Chua}\ \emph {et~al.}(2021)\citenamefont {Chua},
  \citenamefont {Katz}, \citenamefont {Warburton},\ and\ \citenamefont
  {Hughes}}]{Chua:2020stf}%
  \BibitemOpen
  \bibfield  {author} {\bibinfo {author} {\bibfnamefont {A.~J.~K.}\
  \bibnamefont {Chua}}, \bibinfo {author} {\bibfnamefont {M.~L.}\ \bibnamefont
  {Katz}}, \bibinfo {author} {\bibfnamefont {N.}~\bibnamefont {Warburton}},\
  and\ \bibinfo {author} {\bibfnamefont {S.~A.}\ \bibnamefont {Hughes}},\
  }\bibfield  {title} {\bibinfo {title} {{Rapid generation of fully
  relativistic extreme-mass-ratio-inspiral waveform templates for LISA data
  analysis}},\ }\href {https://doi.org/10.1103/PhysRevLett.126.051102}
  {\bibfield  {journal} {\bibinfo  {journal} {Phys. Rev. Lett.}\ }\textbf
  {\bibinfo {volume} {126}},\ \bibinfo {pages} {051102} (\bibinfo {year}
  {2021})},\ \Eprint {https://arxiv.org/abs/2008.06071} {arXiv:2008.06071
  [gr-qc]} \BibitemShut {NoStop}%
\bibitem [{\citenamefont {Katz}\ \emph {et~al.}(2021)\citenamefont {Katz},
  \citenamefont {Chua}, \citenamefont {Speri}, \citenamefont {Warburton},\ and\
  \citenamefont {Hughes}}]{Katz:2021yft}%
  \BibitemOpen
  \bibfield  {author} {\bibinfo {author} {\bibfnamefont {M.~L.}\ \bibnamefont
  {Katz}}, \bibinfo {author} {\bibfnamefont {A.~J.~K.}\ \bibnamefont {Chua}},
  \bibinfo {author} {\bibfnamefont {L.}~\bibnamefont {Speri}}, \bibinfo
  {author} {\bibfnamefont {N.}~\bibnamefont {Warburton}},\ and\ \bibinfo
  {author} {\bibfnamefont {S.~A.}\ \bibnamefont {Hughes}},\ }\bibfield  {title}
  {\bibinfo {title} {{Fast extreme-mass-ratio-inspiral waveforms: New tools for
  millihertz gravitational-wave data analysis}},\ }\href
  {https://doi.org/10.1103/PhysRevD.104.064047} {\bibfield  {journal} {\bibinfo
   {journal} {Phys. Rev. D}\ }\textbf {\bibinfo {volume} {104}},\ \bibinfo
  {pages} {064047} (\bibinfo {year} {2021})},\ \Eprint
  {https://arxiv.org/abs/2104.04582} {arXiv:2104.04582 [gr-qc]} \BibitemShut
  {NoStop}%
\bibitem [{\citenamefont {Chapman-Bird}\ \emph {et~al.}(2025)\citenamefont
  {Chapman-Bird} \emph {et~al.}}]{Chapman-Bird:2025xtd}%
  \BibitemOpen
  \bibfield  {author} {\bibinfo {author} {\bibfnamefont {C.~E.~A.}\
  \bibnamefont {Chapman-Bird}} \emph {et~al.},\ }\bibfield  {title} {\bibinfo
  {title} {{Efficient waveforms for asymmetric-mass eccentric equatorial
  inspirals into rapidly spinning black holes}},\ }\href
  {https://doi.org/10.1103/scbp-75pf} {\bibfield  {journal} {\bibinfo
  {journal} {Phys. Rev. D}\ }\textbf {\bibinfo {volume} {112}},\ \bibinfo
  {pages} {104023} (\bibinfo {year} {2025})},\ \Eprint
  {https://arxiv.org/abs/2506.09470} {arXiv:2506.09470 [gr-qc]} \BibitemShut
  {NoStop}%
\bibitem [{\citenamefont {{Rico K. L. Lo}}()}]{GeneralizedSasakiNakamura}%
  \BibitemOpen
  \bibfield  {author} {\bibinfo {author} {\bibnamefont {{Rico K. L. Lo}}},\
  }\href@noop {} {\bibinfo {title} {{GeneralizedSasakiNakamura.jl}}},\ \bibinfo
  {howpublished}
  {\url{https://github.com/ricokaloklo/GeneralizedSasakiNakamura.jl}}\BibitemShut
  {NoStop}%
\bibitem [{\citenamefont {Lo}(2024)}]{Lo:2023fvv}%
  \BibitemOpen
  \bibfield  {author} {\bibinfo {author} {\bibfnamefont {R.~K.~L.}\
  \bibnamefont {Lo}},\ }\bibfield  {title} {\bibinfo {title} {{Recipes for
  computing radiation from a Kerr black hole using a generalized
  Sasaki-Nakamura formalism: Homogeneous solutions}},\ }\href
  {https://doi.org/10.1103/PhysRevD.110.124070} {\bibfield  {journal} {\bibinfo
   {journal} {Phys. Rev. D}\ }\textbf {\bibinfo {volume} {110}},\ \bibinfo
  {pages} {124070} (\bibinfo {year} {2024})},\ \Eprint
  {https://arxiv.org/abs/2306.16469} {arXiv:2306.16469 [gr-qc]} \BibitemShut
  {NoStop}%
\bibitem [{\citenamefont {Yin}\ \emph {et~al.}(2025)\citenamefont {Yin},
  \citenamefont {Lo},\ and\ \citenamefont {Chen}}]{Yin:2025kls}%
  \BibitemOpen
  \bibfield  {author} {\bibinfo {author} {\bibfnamefont {Y.}~\bibnamefont
  {Yin}}, \bibinfo {author} {\bibfnamefont {R.~K.~L.}\ \bibnamefont {Lo}},\
  and\ \bibinfo {author} {\bibfnamefont {X.}~\bibnamefont {Chen}},\ }\bibfield
  {title} {\bibinfo {title} {{Gravitational radiation from Kerr black holes
  using the Sasaki-Nakamura formalism: waveforms and fluxes at infinity}},\
  }\href@noop {} {\  (\bibinfo {year} {2025})},\ \Eprint
  {https://arxiv.org/abs/2511.08673} {arXiv:2511.08673 [gr-qc]} \BibitemShut
  {NoStop}%
\bibitem [{\citenamefont {Lo}\ and\ \citenamefont {Yin}(2025)}]{Lo:2025lpo}%
  \BibitemOpen
  \bibfield  {author} {\bibinfo {author} {\bibfnamefont {R.~K.~L.}\
  \bibnamefont {Lo}}\ and\ \bibinfo {author} {\bibfnamefont {Y.}~\bibnamefont
  {Yin}},\ }\bibfield  {title} {\bibinfo {title} {{Near-horizon gravitational
  perturbations of rotating black holes}},\ }\href@noop {} {\  (\bibinfo {year}
  {2025})},\ \Eprint {https://arxiv.org/abs/2512.07937} {arXiv:2512.07937
  [gr-qc]} \BibitemShut {NoStop}%
\bibitem [{\citenamefont {Nasipak}(2022)}]{Nasipak:2022xjh}%
  \BibitemOpen
  \bibfield  {author} {\bibinfo {author} {\bibfnamefont {Z.}~\bibnamefont
  {Nasipak}},\ }\bibfield  {title} {\bibinfo {title} {{Adiabatic evolution due
  to the conservative scalar self-force during orbital resonances}},\ }\href
  {https://doi.org/10.1103/PhysRevD.106.064042} {\bibfield  {journal} {\bibinfo
   {journal} {Phys. Rev. D}\ }\textbf {\bibinfo {volume} {106}},\ \bibinfo
  {pages} {064042} (\bibinfo {year} {2022})},\ \Eprint
  {https://arxiv.org/abs/2207.02224} {arXiv:2207.02224 [gr-qc]} \BibitemShut
  {NoStop}%
\bibitem [{\citenamefont {Nasipak}(2024)}]{Nasipak:2023kuf}%
  \BibitemOpen
  \bibfield  {author} {\bibinfo {author} {\bibfnamefont {Z.}~\bibnamefont
  {Nasipak}},\ }\bibfield  {title} {\bibinfo {title} {{Adiabatic gravitational
  waveform model for compact objects undergoing quasicircular inspirals into
  rotating massive black holes}},\ }\href
  {https://doi.org/10.1103/PhysRevD.109.044020} {\bibfield  {journal} {\bibinfo
   {journal} {Phys. Rev. D}\ }\textbf {\bibinfo {volume} {109}},\ \bibinfo
  {pages} {044020} (\bibinfo {year} {2024})},\ \Eprint
  {https://arxiv.org/abs/2310.19706} {arXiv:2310.19706 [gr-qc]} \BibitemShut
  {NoStop}%
\bibitem [{\citenamefont {{Z. Nasipak}}()}]{pybhpt}%
  \BibitemOpen
  \bibfield  {author} {\bibinfo {author} {\bibnamefont {{Z. Nasipak}}},\
  }\href@noop {} {\bibinfo {title} {{pybhpt}}},\ \bibinfo {howpublished}
  {\url{https://github.com/znasipak/pybhpt}}\BibitemShut {NoStop}%
\bibitem [{BHP()}]{BHPToolkit}%
  \BibitemOpen
  \href@noop {} {\bibinfo {title} {{Black Hole Perturbation Toolkit}}},\
  \bibinfo {howpublished}
  {(\href{http://bhptoolkit.org/}{bhptoolkit.org})}\BibitemShut {NoStop}%
\bibitem [{\citenamefont {Cook}\ and\ \citenamefont
  {Zalutskiy}(2014)}]{Cook:2014cta}%
  \BibitemOpen
  \bibfield  {author} {\bibinfo {author} {\bibfnamefont {G.~B.}\ \bibnamefont
  {Cook}}\ and\ \bibinfo {author} {\bibfnamefont {M.}~\bibnamefont
  {Zalutskiy}},\ }\bibfield  {title} {\bibinfo {title} {{Gravitational
  perturbations of the Kerr geometry: High-accuracy study}},\ }\href
  {https://doi.org/10.1103/PhysRevD.90.124021} {\bibfield  {journal} {\bibinfo
  {journal} {Phys. Rev. D}\ }\textbf {\bibinfo {volume} {90}},\ \bibinfo
  {pages} {124021} (\bibinfo {year} {2014})},\ \Eprint
  {https://arxiv.org/abs/1410.7698} {arXiv:1410.7698 [gr-qc]} \BibitemShut
  {NoStop}%
\bibitem [{\citenamefont {Slavyanov}\ and\ \citenamefont
  {Lay}(2000)}]{slavyanov2000unified}%
  \BibitemOpen
  \bibfield  {author} {\bibinfo {author} {\bibfnamefont {S.~Y.}\ \bibnamefont
  {Slavyanov}}\ and\ \bibinfo {author} {\bibfnamefont {W.}~\bibnamefont
  {Lay}},\ }\href@noop {} {\emph {\bibinfo {title} {A Unified Theory Based on
  Singularities}}}\ (\bibinfo  {publisher} {Oxford Mathematical Monographs},\
  \bibinfo {year} {2000})\BibitemShut {NoStop}%
\bibitem [{\citenamefont {Minucci}\ and\ \citenamefont
  {Panosso~Macedo}(2025)}]{Minucci:2024qrn}%
  \BibitemOpen
  \bibfield  {author} {\bibinfo {author} {\bibfnamefont {M.}~\bibnamefont
  {Minucci}}\ and\ \bibinfo {author} {\bibfnamefont {R.}~\bibnamefont
  {Panosso~Macedo}},\ }\bibfield  {title} {\bibinfo {title} {{The confluent
  Heun functions in black hole perturbation theory: a spacetime
  interpretation}},\ }\href {https://doi.org/10.1007/s10714-025-03364-7}
  {\bibfield  {journal} {\bibinfo  {journal} {Gen. Rel. Grav.}\ }\textbf
  {\bibinfo {volume} {57}},\ \bibinfo {pages} {33} (\bibinfo {year} {2025})},\
  \Eprint {https://arxiv.org/abs/2411.19740} {arXiv:2411.19740 [gr-qc]}
  \BibitemShut {NoStop}%
\bibitem [{\citenamefont {London}(2026)}]{London:2023aeo}%
  \BibitemOpen
  \bibfield  {author} {\bibinfo {author} {\bibfnamefont {L.~T.}\ \bibnamefont
  {London}},\ }\bibfield  {title} {\bibinfo {title} {{Radial scalar product for
  Kerr quasinormal modes}},\ }\href {https://doi.org/10.1103/vy3q-nx3w}
  {\bibfield  {journal} {\bibinfo  {journal} {Phys. Rev. D}\ }\textbf {\bibinfo
  {volume} {113}},\ \bibinfo {pages} {044008} (\bibinfo {year} {2026})},\
  \Eprint {https://arxiv.org/abs/2312.17678} {arXiv:2312.17678 [gr-qc]}
  \BibitemShut {NoStop}%
\bibitem [{\citenamefont {London}\ and\ \citenamefont
  {Foucoin}(2026)}]{London:2023idh}%
  \BibitemOpen
  \bibfield  {author} {\bibinfo {author} {\bibfnamefont {L.}~\bibnamefont
  {London}}\ and\ \bibinfo {author} {\bibfnamefont {M.}~\bibnamefont
  {Foucoin}},\ }\bibfield  {title} {\bibinfo {title} {{Natural polynomials for
  Kerr quasinormal modes}},\ }\href {https://doi.org/10.1103/nn8t-p14d}
  {\bibfield  {journal} {\bibinfo  {journal} {Phys. Rev. D}\ }\textbf {\bibinfo
  {volume} {113}},\ \bibinfo {pages} {044009} (\bibinfo {year} {2026})},\
  \Eprint {https://arxiv.org/abs/2312.17680} {arXiv:2312.17680 [gr-qc]}
  \BibitemShut {NoStop}%
\bibitem [{\citenamefont {Chen}\ and\ \citenamefont
  {Jing}(2023)}]{Chen:2023ese}%
  \BibitemOpen
  \bibfield  {author} {\bibinfo {author} {\bibfnamefont {C.}~\bibnamefont
  {Chen}}\ and\ \bibinfo {author} {\bibfnamefont {J.}~\bibnamefont {Jing}},\
  }\bibfield  {title} {\bibinfo {title} {{Radiation fluxes of gravitational,
  electromagnetic, and scalar perturbations in type-D black holes: an exact
  approach}},\ }\href {https://doi.org/10.1088/1475-7516/2023/11/070}
  {\bibfield  {journal} {\bibinfo  {journal} {JCAP}\ }\textbf {\bibinfo
  {volume} {11}},\ \bibinfo {pages} {070}},\ \Eprint
  {https://arxiv.org/abs/2307.14616} {arXiv:2307.14616 [gr-qc]} \BibitemShut
  {NoStop}%
\bibitem [{\citenamefont {Chen}\ and\ \citenamefont
  {Jing}(2024)}]{Chen:2023lsa}%
  \BibitemOpen
  \bibfield  {author} {\bibinfo {author} {\bibfnamefont {C.}~\bibnamefont
  {Chen}}\ and\ \bibinfo {author} {\bibfnamefont {J.}~\bibnamefont {Jing}},\
  }\bibfield  {title} {\bibinfo {title} {{Gravitational wave fluxes on generic
  orbits in near-extreme Kerr spacetime: Higher spin and large eccentricity}},\
  }\href {https://doi.org/10.1007/s11433-024-2431-0} {\bibfield  {journal}
  {\bibinfo  {journal} {Sci. China Phys. Mech. Astron.}\ }\textbf {\bibinfo
  {volume} {67}},\ \bibinfo {pages} {110411} (\bibinfo {year} {2024})},\
  \Eprint {https://arxiv.org/abs/2311.15295} {arXiv:2311.15295 [gr-qc]}
  \BibitemShut {NoStop}%
\bibitem [{\citenamefont {Bonelli}\ \emph {et~al.}(2023)\citenamefont
  {Bonelli}, \citenamefont {Iossa}, \citenamefont {Panea~Lichtig},\ and\
  \citenamefont {Tanzini}}]{Bonelli:2022ten}%
  \BibitemOpen
  \bibfield  {author} {\bibinfo {author} {\bibfnamefont {G.}~\bibnamefont
  {Bonelli}}, \bibinfo {author} {\bibfnamefont {C.}~\bibnamefont {Iossa}},
  \bibinfo {author} {\bibfnamefont {D.}~\bibnamefont {Panea~Lichtig}},\ and\
  \bibinfo {author} {\bibfnamefont {A.}~\bibnamefont {Tanzini}},\ }\bibfield
  {title} {\bibinfo {title} {{Irregular liouville correlators and connection
  formulae for Heun functions}},\ }\href
  {https://doi.org/10.1007/s00220-022-04497-5} {\bibfield  {journal} {\bibinfo
  {journal} {Commun. Math. Phys.}\ }\textbf {\bibinfo {volume} {397}},\
  \bibinfo {pages} {635} (\bibinfo {year} {2023})},\ \Eprint
  {https://arxiv.org/abs/2201.04491} {arXiv:2201.04491 [hep-th]} \BibitemShut
  {NoStop}%
\bibitem [{\citenamefont {Bautista}\ \emph {et~al.}(2024)\citenamefont
  {Bautista}, \citenamefont {Bonelli}, \citenamefont {Iossa}, \citenamefont
  {Tanzini},\ and\ \citenamefont {Zhou}}]{Bautista:2023sdf}%
  \BibitemOpen
  \bibfield  {author} {\bibinfo {author} {\bibfnamefont {Y.~F.}\ \bibnamefont
  {Bautista}}, \bibinfo {author} {\bibfnamefont {G.}~\bibnamefont {Bonelli}},
  \bibinfo {author} {\bibfnamefont {C.}~\bibnamefont {Iossa}}, \bibinfo
  {author} {\bibfnamefont {A.}~\bibnamefont {Tanzini}},\ and\ \bibinfo {author}
  {\bibfnamefont {Z.}~\bibnamefont {Zhou}},\ }\bibfield  {title} {\bibinfo
  {title} {{Black hole perturbation theory meets CFT2: Kerr-Compton amplitudes
  from Nekrasov-Shatashvili functions}},\ }\href
  {https://doi.org/10.1103/PhysRevD.109.084071} {\bibfield  {journal} {\bibinfo
   {journal} {Phys. Rev. D}\ }\textbf {\bibinfo {volume} {109}},\ \bibinfo
  {pages} {084071} (\bibinfo {year} {2024})},\ \Eprint
  {https://arxiv.org/abs/2312.05965} {arXiv:2312.05965 [hep-th]} \BibitemShut
  {NoStop}%
\bibitem [{\citenamefont {Aminov}\ \emph {et~al.}(2022)\citenamefont {Aminov},
  \citenamefont {Grassi},\ and\ \citenamefont {Hatsuda}}]{Aminov:2020yma}%
  \BibitemOpen
  \bibfield  {author} {\bibinfo {author} {\bibfnamefont {G.}~\bibnamefont
  {Aminov}}, \bibinfo {author} {\bibfnamefont {A.}~\bibnamefont {Grassi}},\
  and\ \bibinfo {author} {\bibfnamefont {Y.}~\bibnamefont {Hatsuda}},\
  }\bibfield  {title} {\bibinfo {title} {{Black hole quasinormal modes and
  seiberg\textendash{}witten theory}},\ }\href
  {https://doi.org/10.1007/s00023-021-01137-x} {\bibfield  {journal} {\bibinfo
  {journal} {Annales Henri Poincare}\ }\textbf {\bibinfo {volume} {23}},\
  \bibinfo {pages} {1951} (\bibinfo {year} {2022})},\ \Eprint
  {https://arxiv.org/abs/2006.06111} {arXiv:2006.06111 [hep-th]} \BibitemShut
  {NoStop}%
\bibitem [{\citenamefont {Bonelli}\ \emph {et~al.}(2022)\citenamefont
  {Bonelli}, \citenamefont {Iossa}, \citenamefont {Lichtig},\ and\
  \citenamefont {Tanzini}}]{Bonelli:2021uvf}%
  \BibitemOpen
  \bibfield  {author} {\bibinfo {author} {\bibfnamefont {G.}~\bibnamefont
  {Bonelli}}, \bibinfo {author} {\bibfnamefont {C.}~\bibnamefont {Iossa}},
  \bibinfo {author} {\bibfnamefont {D.~P.}\ \bibnamefont {Lichtig}},\ and\
  \bibinfo {author} {\bibfnamefont {A.}~\bibnamefont {Tanzini}},\ }\bibfield
  {title} {\bibinfo {title} {{Exact solution of Kerr black hole perturbations
  via CFT2 and instanton counting: Greybody factor, quasinormal modes, and Love
  numbers}},\ }\href {https://doi.org/10.1103/PhysRevD.105.044047} {\bibfield
  {journal} {\bibinfo  {journal} {Physical Review D}\ }\textbf {\bibinfo
  {volume} {105}},\ \bibinfo {pages} {044047} (\bibinfo {year} {2022})},\
  \Eprint {https://arxiv.org/abs/2105.04483} {arXiv:2105.04483 [hep-th]}
  \BibitemShut {NoStop}%
\bibitem [{\citenamefont {Casals}\ and\ \citenamefont
  {Longo~Micchi}(2019)}]{Casals:2019vdb}%
  \BibitemOpen
  \bibfield  {author} {\bibinfo {author} {\bibfnamefont {M.}~\bibnamefont
  {Casals}}\ and\ \bibinfo {author} {\bibfnamefont {L.~F.}\ \bibnamefont
  {Longo~Micchi}},\ }\bibfield  {title} {\bibinfo {title} {{Spectroscopy of
  extremal and near-extremal Kerr black holes}},\ }\href
  {https://doi.org/10.1103/PhysRevD.99.084047} {\bibfield  {journal} {\bibinfo
  {journal} {Phys. Rev. D}\ }\textbf {\bibinfo {volume} {99}},\ \bibinfo
  {pages} {084047} (\bibinfo {year} {2019})},\ \Eprint
  {https://arxiv.org/abs/1901.04586} {arXiv:1901.04586 [gr-qc]} \BibitemShut
  {NoStop}%
\bibitem [{\citenamefont {Ablowitz}\ and\ \citenamefont
  {Segur}(1981)}]{ablowitz1981solitons}%
  \BibitemOpen
  \bibfield  {author} {\bibinfo {author} {\bibfnamefont {M.~J.}\ \bibnamefont
  {Ablowitz}}\ and\ \bibinfo {author} {\bibfnamefont {H.}~\bibnamefont
  {Segur}},\ }\href {https://doi.org/10.1137/1.9781611970883} {\emph {\bibinfo
  {title} {Solitons and the inverse scattering transform}}}\ (\bibinfo
  {publisher} {Society for Industrial and Applied Mathematics},\ \bibinfo
  {year} {1981})\BibitemShut {NoStop}%
\bibitem [{\citenamefont {Its}\ and\ \citenamefont
  {Novokshenov}(2006)}]{its2006isomonodromic}%
  \BibitemOpen
  \bibfield  {author} {\bibinfo {author} {\bibfnamefont {A.~R.}\ \bibnamefont
  {Its}}\ and\ \bibinfo {author} {\bibfnamefont {V.~Y.}\ \bibnamefont
  {Novokshenov}},\ }\href {https://link.springer.com/book/10.1007/BFb0076661}
  {\emph {\bibinfo {title} {The isomonodromic deformation method in the theory
  of Painlev{\'e} equations}}},\ Vol.\ \bibinfo {volume} {1191}\ (\bibinfo
  {publisher} {Springer},\ \bibinfo {year} {2006})\BibitemShut {NoStop}%
\bibitem [{\citenamefont {Cavalcante}\ \emph
  {et~al.}(2024{\natexlab{a}})\citenamefont {Cavalcante}, \citenamefont
  {Richartz},\ and\ \citenamefont {da~Cunha}}]{Cavalcante:2024swt}%
  \BibitemOpen
  \bibfield  {author} {\bibinfo {author} {\bibfnamefont {J.~a.~P.}\
  \bibnamefont {Cavalcante}}, \bibinfo {author} {\bibfnamefont
  {M.}~\bibnamefont {Richartz}},\ and\ \bibinfo {author} {\bibfnamefont
  {B.~C.}\ \bibnamefont {da~Cunha}},\ }\bibfield  {title} {\bibinfo {title}
  {{Exceptional point and hysteresis in perturbations of Kerr black holes}},\
  }\href {https://doi.org/10.1103/PhysRevLett.133.261401} {\bibfield  {journal}
  {\bibinfo  {journal} {Phys. Rev. Lett.}\ }\textbf {\bibinfo {volume} {133}},\
  \bibinfo {pages} {261401} (\bibinfo {year} {2024}{\natexlab{a}})},\ \Eprint
  {https://arxiv.org/abs/2407.20850} {arXiv:2407.20850 [gr-qc]} \BibitemShut
  {NoStop}%
\bibitem [{\citenamefont {Cavalcante}\ \emph
  {et~al.}(2024{\natexlab{b}})\citenamefont {Cavalcante}, \citenamefont
  {Richartz},\ and\ \citenamefont {da~Cunha}}]{Cavalcante:2024kmy}%
  \BibitemOpen
  \bibfield  {author} {\bibinfo {author} {\bibfnamefont {J.~a.~P.}\
  \bibnamefont {Cavalcante}}, \bibinfo {author} {\bibfnamefont
  {M.}~\bibnamefont {Richartz}},\ and\ \bibinfo {author} {\bibfnamefont
  {B.~C.}\ \bibnamefont {da~Cunha}},\ }\bibfield  {title} {\bibinfo {title}
  {{Massive scalar perturbations in Kerr black holes: Near extremal
  analysis}},\ }\href {https://doi.org/10.1103/PhysRevD.110.124064} {\bibfield
  {journal} {\bibinfo  {journal} {Phys. Rev. D}\ }\textbf {\bibinfo {volume}
  {110}},\ \bibinfo {pages} {124064} (\bibinfo {year} {2024}{\natexlab{b}})},\
  \Eprint {https://arxiv.org/abs/2408.13964} {arXiv:2408.13964 [gr-qc]}
  \BibitemShut {NoStop}%
\bibitem [{\citenamefont {Motygin}(2018)}]{Motygin2018}%
  \BibitemOpen
  \bibfield  {author} {\bibinfo {author} {\bibfnamefont {O.~V.}\ \bibnamefont
  {Motygin}},\ }\bibfield  {title} {\bibinfo {title} {{On evaluation of the
  confluent Heun functions}},\ }in\ \href
  {https://doi.org/10.1109/DD.2018.8553032} {\emph {\bibinfo {booktitle} {2018
  Days on Diffraction (DD)}}}\ (\bibinfo  {publisher} {{IEEE}},\ \bibinfo
  {address} {New York},\ \bibinfo {year} {2018})\BibitemShut {NoStop}%
\bibitem [{\citenamefont {McMaken}\ and\ \citenamefont
  {Hamilton}(2023)}]{McMaken:2023tft}%
  \BibitemOpen
  \bibfield  {author} {\bibinfo {author} {\bibfnamefont {T.}~\bibnamefont
  {McMaken}}\ and\ \bibinfo {author} {\bibfnamefont {A.~J.~S.}\ \bibnamefont
  {Hamilton}},\ }\bibfield  {title} {\bibinfo {title} {{Hawking radiation
  inside a charged black hole}},\ }\href
  {https://doi.org/10.1103/PhysRevD.107.085010} {\bibfield  {journal} {\bibinfo
   {journal} {Phys. Rev. D}\ }\textbf {\bibinfo {volume} {107}},\ \bibinfo
  {pages} {085010} (\bibinfo {year} {2023})},\ \Eprint
  {https://arxiv.org/abs/2301.12319} {arXiv:2301.12319 [gr-qc]} \BibitemShut
  {NoStop}%
\bibitem [{\citenamefont {Ronveaux}\ and\ \citenamefont
  {Arscott}(1995)}]{Ronveaux:1995}%
  \BibitemOpen
  \bibfield  {author} {\bibinfo {author} {\bibfnamefont {A.}~\bibnamefont
  {Ronveaux}}\ and\ \bibinfo {author} {\bibfnamefont {F.~M.}\ \bibnamefont
  {Arscott}},\ }\href@noop {} {\emph {\bibinfo {title} {{Heun's Differential
  Equations}}}}\ (\bibinfo  {publisher} {Oxford University Press},\ \bibinfo
  {address} {New York},\ \bibinfo {year} {1995})\BibitemShut {NoStop}%
\bibitem [{\citenamefont {Olver}\ \emph {et~al.}(2010)\citenamefont {Olver},
  \citenamefont {Lozier}, \citenamefont {Boisvert},\ and\ \citenamefont
  {Clark}}]{olver2010nist}%
  \BibitemOpen
  \bibfield  {author} {\bibinfo {author} {\bibfnamefont {F.~W.}\ \bibnamefont
  {Olver}}, \bibinfo {author} {\bibfnamefont {D.~W.}\ \bibnamefont {Lozier}},
  \bibinfo {author} {\bibfnamefont {R.~F.}\ \bibnamefont {Boisvert}},\ and\
  \bibinfo {author} {\bibfnamefont {C.~W.}\ \bibnamefont {Clark}},\ }\href@noop
  {} {\emph {\bibinfo {title} {NIST handbook of mathematical functions hardback
  and CD-ROM}}}\ (\bibinfo  {publisher} {Cambridge university press},\ \bibinfo
  {year} {2010})\BibitemShut {NoStop}%
\bibitem [{\citenamefont {Chen}\ \emph {et~al.}(2025)\citenamefont {Chen},
  \citenamefont {Jing}, \citenamefont {Cao},\ and\ \citenamefont
  {Wang}}]{Chen:2025sbz}%
  \BibitemOpen
  \bibfield  {author} {\bibinfo {author} {\bibfnamefont {C.}~\bibnamefont
  {Chen}}, \bibinfo {author} {\bibfnamefont {J.}~\bibnamefont {Jing}}, \bibinfo
  {author} {\bibfnamefont {Z.}~\bibnamefont {Cao}},\ and\ \bibinfo {author}
  {\bibfnamefont {M.}~\bibnamefont {Wang}},\ }\bibfield  {title} {\bibinfo
  {title} {{Complete quasinormal modes of type-D black holes}},\ }\href
  {https://doi.org/10.1103/f8m8-vr4l} {\bibfield  {journal} {\bibinfo
  {journal} {Phys. Rev. D}\ }\textbf {\bibinfo {volume} {112}},\ \bibinfo
  {pages} {103036} (\bibinfo {year} {2025})},\ \Eprint
  {https://arxiv.org/abs/2506.14635} {arXiv:2506.14635 [gr-qc]} \BibitemShut
  {NoStop}%
\bibitem [{\citenamefont {Fujita}\ and\ \citenamefont
  {Hikida}(2009)}]{Fujita:2009bp}%
  \BibitemOpen
  \bibfield  {author} {\bibinfo {author} {\bibfnamefont {R.}~\bibnamefont
  {Fujita}}\ and\ \bibinfo {author} {\bibfnamefont {W.}~\bibnamefont
  {Hikida}},\ }\bibfield  {title} {\bibinfo {title} {{Analytical solutions of
  bound timelike geodesic orbits in Kerr spacetime}},\ }\href
  {https://doi.org/10.1088/0264-9381/26/13/135002} {\bibfield  {journal}
  {\bibinfo  {journal} {Classical Quantum Gravity}\ }\textbf {\bibinfo {volume}
  {26}},\ \bibinfo {pages} {135002} (\bibinfo {year} {2009})},\ \Eprint
  {https://arxiv.org/abs/0906.1420} {arXiv:0906.1420 [gr-qc]} \BibitemShut
  {NoStop}%
\bibitem [{\citenamefont {O'Sullivan}\ and\ \citenamefont
  {Hughes}(2014)}]{OSullivan:2014ywd}%
  \BibitemOpen
  \bibfield  {author} {\bibinfo {author} {\bibfnamefont {S.}~\bibnamefont
  {O'Sullivan}}\ and\ \bibinfo {author} {\bibfnamefont {S.~A.}\ \bibnamefont
  {Hughes}},\ }\bibfield  {title} {\bibinfo {title} {{Strong-field tidal
  distortions of rotating black holes: Formalism and results for circular,
  equatorial orbits}},\ }\href {https://doi.org/10.1103/PhysRevD.91.109901}
  {\bibfield  {journal} {\bibinfo  {journal} {Phys. Rev. D}\ }\textbf {\bibinfo
  {volume} {90}},\ \bibinfo {pages} {124039} (\bibinfo {year} {2014})},\
  \bibinfo {note} {[Erratum: Phys.Rev.D 91, 109901 (2015)]},\ \Eprint
  {https://arxiv.org/abs/1407.6983} {arXiv:1407.6983 [gr-qc]} \BibitemShut
  {NoStop}%
\bibitem [{\citenamefont {Cipriani}\ \emph {et~al.}(2025)\citenamefont
  {Cipriani}, \citenamefont {Di~Russo}, \citenamefont {Fucito}, \citenamefont
  {Morales}, \citenamefont {Poghosyan},\ and\ \citenamefont
  {Poghossian}}]{Cipriani:2025ikx}%
  \BibitemOpen
  \bibfield  {author} {\bibinfo {author} {\bibfnamefont {A.}~\bibnamefont
  {Cipriani}}, \bibinfo {author} {\bibfnamefont {G.}~\bibnamefont {Di~Russo}},
  \bibinfo {author} {\bibfnamefont {F.}~\bibnamefont {Fucito}}, \bibinfo
  {author} {\bibfnamefont {J.~F.}\ \bibnamefont {Morales}}, \bibinfo {author}
  {\bibfnamefont {H.}~\bibnamefont {Poghosyan}},\ and\ \bibinfo {author}
  {\bibfnamefont {R.}~\bibnamefont {Poghossian}},\ }\bibfield  {title}
  {\bibinfo {title} {{Resumming post-Minkowskian and post-Newtonian
  gravitational waveform expansions}},\ }\href
  {https://doi.org/10.21468/SciPostPhys.19.2.057} {\bibfield  {journal}
  {\bibinfo  {journal} {SciPost Phys.}\ }\textbf {\bibinfo {volume} {19}},\
  \bibinfo {pages} {057} (\bibinfo {year} {2025})},\ \Eprint
  {https://arxiv.org/abs/2501.19257} {arXiv:2501.19257 [gr-qc]} \BibitemShut
  {NoStop}%
\bibitem [{\citenamefont {Stein}\ and\ \citenamefont
  {Warburton}(2020)}]{Stein:2019buj}%
  \BibitemOpen
  \bibfield  {author} {\bibinfo {author} {\bibfnamefont {L.~C.}\ \bibnamefont
  {Stein}}\ and\ \bibinfo {author} {\bibfnamefont {N.}~\bibnamefont
  {Warburton}},\ }\bibfield  {title} {\bibinfo {title} {{Location of the last
  stable orbit in Kerr spacetime}},\ }\href
  {https://doi.org/10.1103/PhysRevD.101.064007} {\bibfield  {journal} {\bibinfo
   {journal} {Phys. Rev. D}\ }\textbf {\bibinfo {volume} {101}},\ \bibinfo
  {pages} {064007} (\bibinfo {year} {2020})},\ \Eprint
  {https://arxiv.org/abs/1912.07609} {arXiv:1912.07609 [gr-qc]} \BibitemShut
  {NoStop}%
\bibitem [{\citenamefont {Berti}\ and\ \citenamefont
  {Cardoso}(2006)}]{Berti:2006wq}%
  \BibitemOpen
  \bibfield  {author} {\bibinfo {author} {\bibfnamefont {E.}~\bibnamefont
  {Berti}}\ and\ \bibinfo {author} {\bibfnamefont {V.}~\bibnamefont
  {Cardoso}},\ }\bibfield  {title} {\bibinfo {title} {{Quasinormal ringing of
  Kerr black holes. I. The excitation factors}},\ }\href
  {https://doi.org/10.1103/PhysRevD.74.104020} {\bibfield  {journal} {\bibinfo
  {journal} {Phys. Rev. D}\ }\textbf {\bibinfo {volume} {74}},\ \bibinfo
  {pages} {104020} (\bibinfo {year} {2006})},\ \Eprint
  {https://arxiv.org/abs/gr-qc/0605118} {arXiv:gr-qc/0605118} \BibitemShut
  {NoStop}%
\bibitem [{\citenamefont {Zhang}\ \emph {et~al.}(2013)\citenamefont {Zhang},
  \citenamefont {Berti},\ and\ \citenamefont {Cardoso}}]{Zhang:2013ksa}%
  \BibitemOpen
  \bibfield  {author} {\bibinfo {author} {\bibfnamefont {Z.}~\bibnamefont
  {Zhang}}, \bibinfo {author} {\bibfnamefont {E.}~\bibnamefont {Berti}},\ and\
  \bibinfo {author} {\bibfnamefont {V.}~\bibnamefont {Cardoso}},\ }\bibfield
  {title} {\bibinfo {title} {{Quasinormal ringing of Kerr black holes. II.
  Excitation by particles falling radially with arbitrary energy}},\ }\href
  {https://doi.org/10.1103/PhysRevD.88.044018} {\bibfield  {journal} {\bibinfo
  {journal} {Phys. Rev. D}\ }\textbf {\bibinfo {volume} {88}},\ \bibinfo
  {pages} {044018} (\bibinfo {year} {2013})},\ \Eprint
  {https://arxiv.org/abs/1305.4306} {arXiv:1305.4306 [gr-qc]} \BibitemShut
  {NoStop}%
\bibitem [{\citenamefont {Della~Rocca}\ \emph {et~al.}(2025)\citenamefont
  {Della~Rocca}, \citenamefont {Pezzella}, \citenamefont {Berti}, \citenamefont
  {Gualtieri},\ and\ \citenamefont {Maselli}}]{DellaRocca:2025zbe}%
  \BibitemOpen
  \bibfield  {author} {\bibinfo {author} {\bibfnamefont {M.}~\bibnamefont
  {Della~Rocca}}, \bibinfo {author} {\bibfnamefont {L.}~\bibnamefont
  {Pezzella}}, \bibinfo {author} {\bibfnamefont {E.}~\bibnamefont {Berti}},
  \bibinfo {author} {\bibfnamefont {L.}~\bibnamefont {Gualtieri}},\ and\
  \bibinfo {author} {\bibfnamefont {A.}~\bibnamefont {Maselli}},\ }\bibfield
  {title} {\bibinfo {title} {{Quasinormal ringing of Kerr black holes. III.
  Excitation coefficients for equatorial inspirals from the innermost stable
  circular orbit}},\ }\href@noop {} {\  (\bibinfo {year} {2025})},\ \Eprint
  {https://arxiv.org/abs/2512.07959} {arXiv:2512.07959 [gr-qc]} \BibitemShut
  {NoStop}%
\bibitem [{\citenamefont {Berens}\ \emph {et~al.}(2024)\citenamefont {Berens},
  \citenamefont {Gravely},\ and\ \citenamefont {Lupsasca}}]{Berens:2024czo}%
  \BibitemOpen
  \bibfield  {author} {\bibinfo {author} {\bibfnamefont {R.}~\bibnamefont
  {Berens}}, \bibinfo {author} {\bibfnamefont {T.}~\bibnamefont {Gravely}},\
  and\ \bibinfo {author} {\bibfnamefont {A.}~\bibnamefont {Lupsasca}},\
  }\bibfield  {title} {\bibinfo {title} {{Gravitational waves on Kerr black
  holes: I. Reconstruction of linearized metric perturbations}},\ }\href
  {https://doi.org/10.1088/1361-6382/ad6c9c} {\bibfield  {journal} {\bibinfo
  {journal} {Classical Quantum Gravity}\ }\textbf {\bibinfo {volume} {41}},\
  \bibinfo {pages} {195004} (\bibinfo {year} {2024})},\ \Eprint
  {https://arxiv.org/abs/2403.20311} {arXiv:2403.20311 [gr-qc]} \BibitemShut
  {NoStop}%
\bibitem [{\citenamefont {{Changkai Chen}}()}]{ChenGWFlux}%
  \BibitemOpen
  \bibfield  {author} {\bibinfo {author} {\bibnamefont {{Changkai Chen}}},\
  }\href@noop {} {\bibinfo {title} {{GWFluxHeunC}}},\ \bibinfo {howpublished}
  {\url{https://github.com/IronChen1/GWFluxHeunC}}\BibitemShut {NoStop}%
\bibitem [{\citenamefont {{O. V.
  Motygin}}()}]{motygin_confluent_Heun_functions}%
  \BibitemOpen
  \bibfield  {author} {\bibinfo {author} {\bibnamefont {{O. V. Motygin}}},\
  }\href@noop {} {\bibinfo {title} {{confluent\_Heun\_functions: Octave/Matlab
  code for evaluation of the confluent Heun functions}}},\ \bibinfo
  {howpublished}
  {\url{https://github.com/motygin/confluent_Heun_functions}}\BibitemShut
  {NoStop}%
\bibitem [{\citenamefont {Zwillinger}\ \emph {et~al.}(2015)\citenamefont
  {Zwillinger}, \citenamefont {Moll}, \citenamefont {Gradshteyn},\ and\
  \citenamefont {Ryzhik}}]{book2015249}%
  \BibitemOpen
  \bibinfo {editor} {\bibfnamefont {D.}~\bibnamefont {Zwillinger}}, \bibinfo
  {editor} {\bibfnamefont {V.}~\bibnamefont {Moll}}, \bibinfo {editor}
  {\bibfnamefont {I.}~\bibnamefont {Gradshteyn}},\ and\ \bibinfo {editor}
  {\bibfnamefont {I.}~\bibnamefont {Ryzhik}},\ eds.,\ \href
  {https://doi.org/https://doi.org/10.1016/B978-0-12-384933-5.00003-5} {\emph
  {\bibinfo {title} {Table of Integrals, Series, and Products (Eighth
  Edition)}}},\ \bibinfo {edition} {eighth edition}\ ed.\ (\bibinfo
  {publisher} {Academic Press},\ \bibinfo {address} {Boston},\ \bibinfo {year}
  {2015})\ pp.\ \bibinfo {pages} {249--519}\BibitemShut {NoStop}%
\bibitem [{\citenamefont {Becker}(1997)}]{Becker1997}%
  \BibitemOpen
  \bibfield  {author} {\bibinfo {author} {\bibfnamefont {P.~A.}\ \bibnamefont
  {Becker}},\ }\bibfield  {title} {\bibinfo {title} {{Normalization integrals
  of orthogonal Heun functions}},\ }\href {https://doi.org/10.1063/1.532062}
  {\bibfield  {journal} {\bibinfo  {journal} {Journal of Mathematical Physics}\
  }\textbf {\bibinfo {volume} {38}},\ \bibinfo {pages} {3692} (\bibinfo {year}
  {1997})}\BibitemShut {NoStop}%
\end{thebibliography}%

\end{document}